\documentstyle[12pt,equation]{article}
\begin{document}
\def\to{\rightarrow}
\def\ub{\underbar}
\def\rs{\mbox{$\sqrt{s}$}}
\def\phad{\mbox{$P_{\rm had}$}}
\def\nav{\mbox{$\langle n_{\rm jet} \rangle $}}
\def\wgg{\mbox{$W_{\gamma\gamma}$}}
\def\gga{\mbox{$G^{\gamma}$}}
\def\jpsi{\mbox{$J/\psi$}}
\def\lqcd{\mbox{$\Lambda_{\rm QCD}$}}
\def\ppbar{\mbox{$p \bar{p}$}}
\def\qqbar{\mbox{$q \bar{q}$}}
\def\QQbar{\mbox{$Q \bar{Q}$}}
\def\bbbar{\mbox{$b \bar{b}$}}
\def\ccbar{\mbox{$c \bar{c}$}}
\def\mcc{\mbox{$M_{c \bar{c}}$}}
\def\ttbar{\mbox{$t \bar{t}$}}
\def\ben{\begin{subequations}}
\def\be{\begin{equation}}
\def\een{\end{subequations}}
\def\ee{\end{equation}}
\def\beq{\begin{eqalignno}}
\def\eeq{\end{eqalignno}}
\def\bea{\begin{eqnarray}}
\def\eea{\end{eqnarray}}
\def\qsq{\mbox{$Q^2$}}
\def\xgam{\mbox{$x_{\gamma}$}}
\def\sigtot{${\sigma^{tot}_W}$}
\def\gamg{${ \gamma g }$}
\def\pbarp{\mbox{$ \bar{p}p$}}
\def\ptmin{\mbox{$ p_{T,{\rm min}} $}}
\def\egam{\mbox{$e \gamma $}}
\def\gaga{\mbox{$\gamma \gamma $}}
\def\epem{\mbox{$e^+ e^- $}}
\def\alphas{\mbox{$\alpha_s$}}
\def\aem{\mbox{$\alpha_{\rm em}$}}
\def\rts{\mbox{$ \sqrt{s} $ }}
\def\gamp{\mbox{$\gamma p$}}
\def\totwo{$2 \rightarrow 2$}
\def\glph{\mbox{$ G^{\gamma}(x,Q^2)$}}
\def\Gg{\mbox{$ G^{\gamma}$}}
\def\ug{\mbox{$ u^{\gamma}$}}
\def\dg{\mbox{$ d^{\gamma}$}}
\def\sg{\mbox{$ s^{\gamma}$}}
\def\cg{\mbox{$ c^{\gamma}$}}
\def\qvph{\mbox{$\vec q^{\gamma} $}}
\def\qv0{\mbox{$\vec q^{\gamma}_0 $}}
\def\qph{\mbox{${q^{\gamma} } $}}
\def\vqxqsq{\mbox{$ \vec q^{\gamma} (x,Q^2)  $}}
\def\qisq{\mbox{$ q_i^{\gamma} (x,Q^2) $}}
\def\qolsq{ (Q^2 / \Lambda ^2)}
\def\fgme{\mbox{$f_{\gamma|e}$}}
\def\fgmebeam{\mbox{$ f_{\gamma|e}^{beam}$ }}
\def\fgmer{\mbox{$ f_{\gamma|e}^{res} $}}
\def\sigh{ \hat \sigma}
\def\xtsq{\mbox{$ x{_T}{^2}$}}
\def\f2gam{\mbox{$ F_2^\gamma $}}
\renewcommand{\thefootnote}{\fnsymbol{footnote}}
\hyphenation{brems-strahl-ung beam-strahl-ung}
\setcounter{page}{0}
\setcounter{footnote}{0}
\begin{flushright}
BU--TH--95/2\\
MADPH--95--898\\
July 1995\\
\end{flushright}
\vspace{1cm}
\begin{center}
{\Large \bf Resolved Photon Processes} \\
\vspace{1cm}
Manuel Drees\footnote{Heisenberg Fellow}\\
{\it University of Wisconsin, Dept. of Physics, 1150 University Ave., Madison,
WI53706, USA} \\
\vspace*{0.5cm}
Rohini M. Godbole\\
{\it Physics Department, University of Bombay, Vidyanagari, Bombay,400 098,
India.}\\
\vspace{1cm}
\end{center}
\begin{abstract}
We review the present level of knowledge of the hadronic structure of the
photon, as revealed in interactions involving quarks and gluons ``in" the
photon. The concept of photon structure functions is introduced in the
description of deep--inelastic $e \gamma$ scattering, and existing
parametrizations of the parton densities in the photon are reviewed. We then
turn to hard \gamp\ and \gaga\ collisions, where we treat the production of
jets, heavy quarks, hard (direct) photons, \jpsi\ mesons, and lepton pairs. We
also comment on issues that go beyond perturbation theory, including recent
attempts at a comprehensive description of both hard and soft \gamp\ and
\gaga\ interactions. We conclude with a list of open problems.

\end{abstract}
\newpage
\begin{center}
{ \large \bf Table of Contents}
\begin{enumerate}
\item Introduction
\item Photon Structure Functions.
\begin{enumerate}
\item Deep--inelastic $e \gamma$ Scattering
\item Parametrizations of Photonic Parton Densities
\end{enumerate}

\item Resolved Photon Reactions in $\gamma p $ Scattering
\begin{enumerate}
\item $ep \rightarrow $ jets $ + \ X $
\item Heavy Quark Production
\item Direct Photon Production
\item $J/\Psi $ Production
\item Lepton Pair Production
\end{enumerate}

\item Real \gaga\ Scattering at \epem\ Colliders
\begin{enumerate}
\item $\epem \rightarrow $ jets $ + \ X $
\item Heavy Quark Production
\item $ J/\Psi$ Production
\item Direct Photon Production
\end{enumerate}

\item Beyond Perturbation Theory
\item Outlook
\item Acknowledgements
\item References
\end{enumerate}
\end{center}
\newpage
\pagestyle{plain}

\section*{1) Introduction}
The photon is the simplest of all bosons. It is the gauge boson of QED, which
implies that it is massless and structureless (i.e., pointlike). Predictions
for $\gamma e$ interactions can be made with impressive accuracy, in some
cases (e.g., $g-2$ \cite{1}) to better than one part in $10^8$. Indeed, for
many physicists the great quantitative success of QED is one of the strongest
arguments in favour of Quantum Field Theories in general.

At first glance it might therefore be surprising, perhaps even a bit
embarassing, that many reactions involving (quasi--)real photons are much less
well understood, both theoretically and experimentally. However, a moment's
reflection will show that this is really not astonishing at all. The
uncertainty principle tells us that for a short period of time a photon can
fluctuate into a pair of charged particles. Fluctuations into virtual
two--lepton states are well understood, and are in fact a crucial ingredient
of the quantitative success of QED. Fluctuations into quark--antiquark pairs
are much more problematic, however. Whenever the lifetime of the virtual
state exceeds about $10^{-25}$ sec, corresponding to a characteristic energy
or momentum scale of about 1 GeV or less, the (virtual) \qqbar\ pair has
sufficient time to evolve into a complicated hadronic state that cannot be
described by perturbative methods only. Even if the lifetime is shorter, i.e.
the energy--momentum scale is larger, hard gluon emission and related
processes complicate the picture substantially.

The understanding of such virtual hadronic states becomes particularly
important when they are ``kicked on the mass shell" by an interaction of the
photon. The most thoroughly studied reactions of this type involve
interactions of a real and a virtual photon (e.g. in $e \gamma$ scattering);
of two real photons (\gaga\ scattering at \epem\ colliders); and of a real
photon with a hadron (e.g. \gamp\ scattering). All these reactions allow to
probe some aspects of the hadronic nature of the photon, and we will discuss
them in turn.

Why is it important to study such reactions? There are at least two answers.
First, we can use data taken at present colliders to sharpen our predictions
for (background) processes at future colliders. We will see that in some cases
this may contribute significantly to the assessment of the potential of future
colliders; this is true for high--energy linear \epem\ colliders that are now
being discussed, and especially for the so--called \gaga\ colliders. Secondly,
while the hadronic structure of the photon certainly has nonperturbative
aspects, we expect the photon to be a simpler system than any real hadron,
like the proton. After all, no matter how complicated it is, the hadronic
structure of the photon certainly originated from the $\gamma \qqbar$ vertex;
if this vertex vanished, the photon would not have any ``hadronic nature" to
speak of. No such (in principle) simple starting point can be defined for
protons, constituent quarks themselves being highly complicated objects. One
can therefore expect photonic reactions to be particularly well suited to
study both perturbative and nonperturbative aspects of strong interactions,
and everything in between. We should mention right away that the present
theoretical description of, say, \gamp\ scattering is, if anything, even more
complex than that of \pbarp\ scattering; however, we feel that this does not
refute our argument that \gamp\ collisions are simpler at some fundamental
level. At the very least reactions probing the hadronic nature of the photon
give us an additional handle on most aspects of perturbative and
nonperturbative QCD that are of relevance for collider physics.

The remainder of this article is organized as follows. We start in Sec.~2 with
a discussion of the scattering of a highly virtual (probing) photon off an
almost real (target) photon, $\gamma^* \gamma$ or $e \gamma$ scattering. These
were the first photonic reactions for which predictions were made in the
framework of the quark parton model (QPM) \cite{2} and within QCD \cite{3}.
$e \gamma$ scattering was also among the first of the ``hard'' photonic
reactions, which can at least partly be described by perturbation theory, to
be studied experimentally \cite{4}. Finally, the conceptual simplicity of
these processes makes them ideal for introducing the notion of photon
structure functions and the parton content of the photon.

The commissioning of the $ep$ collider HERA has opened a new era in the
experimental study of hard $\gamma p$ collisions, increasing the available
\gamp\ centre--of--mass energy by about a factor of ten compared to previous
experiments. In Sec.~3 we attempt to cover both the recent intense theoretical
activity \cite{5} that has been triggered by this prospect, and the relevant
experimental data.

Lately there has also been much progress in the field of hard scattering of
two quasi--real photons. An important milestone was set \cite{6} by the AMY
collaboration at the TRISTAN collider, who for the first time analyzed their
data using an essentially complete leading order QCD event generator. Previous
generators had omitted important classes of diagrams, and had therefore not
been able to reproduce data taken at the older PEP and PETRA colliders.
Recent developments in this area are summarized in Sec.~4.

The emphasis in Secs.~2 to 4 is on hard processes, which are amenable to a
perturbative treatment. In Sec.~5 we loosen this restriction and comment on
issues that go beyond standard perturbative QCD. In particular, we discuss
\gamp\ and \gaga\ cross--sections, and related quantities, in the minijet
model \cite{7}. This model has been introduced to describe purely hadronic
($pp, \  \pbarp$) reactions; some modifications are necessary \cite{8}
before it can be used for \gamp\ and \gaga\ scattering. Finally, Sec.~6
contains some concluding remarks, as well as a list of open problems and
experimental challenges.

This field is still very much in flux. Hardly a week goes by without a new
experimental study of some reaction that probes the hadronic structure of the
photon, a new or refined calculation of a relevant cross--section for given
photon structure, and/or a new model for or parametrization of this structure.
We nevertheless think it worthwhile to summarize the present status, partly
to celebrate what has already been achieved, but mostly to highlight open
questions and how to address them. We hope that this review, which updates
ref.\cite{9}, will be of some use both for those who have already worked on
some aspects of the hadronic nature of the photon, and for those who
contemplate starting such work.
\setcounter{footnote}{0}
\section*{2) Photon Structure Functions}
In this section we introduce the concept of photon structure functions. To
this end we first discuss in Sec.~2a deep--inelastic \egam\ scattering, which
is theoretically very clean, being fully inclusive; it is thus well suited to
serve as the defining process for photon structure functions and the parton
content of the photon. For reasons of space we have to be rather brief here;
we refer the reader to ref.\cite{9a} for a more pedagogical introduction to
photon structure functions. In Sec.~2b we then describe existing
parametrizations of the photonic parton distribution functions.

\subsection*{2a) Deep--inelastic \egam\ Scattering}
Studies of deep--inelastic electron nucleon scattering,
\be \label{e2.1}
e N \to e X,
\ee
where $X$ is any hadronic system and the squared four momentum transfer $Q^2
\equiv - q^2 \geq 1$ GeV$^2$, have contributed much to our understanding of
strong interactions. The pioneering SLAC experiments \cite{10} played a key
role in the development of both the quark parton model (QPM) \cite{11}, and,
due to the observation of scaling violations and the concept of asymptotic
freedom \cite{12}, of QCD. Since then our understanding of QCD in general and
the structure of the proton in particular has improved greatly. The most
recent progress in this area has come from the analysis of data taken at the
$ep$ collider HERA; see ref.\cite{13} for a study of the impact of these data
on the determination of the parton content of the proton.

Formally deep--inelastic \egam\ scattering is quite similar to $ep$
scattering. The basic kinematics is explained in Fig.~1. The differential
cross section can be written in terms of the scaling variables $x \equiv Q^2 /
(2 p \cdot q)$ and $y \equiv Q^2 / (s x)$, where \rs\ is the total available
centre--of--mass (cms) energy:
\be \label{e2.2}
\frac {d^2 \sigma (e \gamma \to e X)} {dx dy} =
\frac {2 \pi \alpha^2_{\rm em} s} {Q^4}
\left\{ \left[ 1 + \left( 1-y \right)^2 \right] \f2gam(x,Q^2) - y^2
F_L^{\gamma}(x,Q^2) \right\};
\ee
this expression is completely analogous to the equation defining the protonic
structure functions $F_2$ and $F_L$ in terms of the differential
cross--section for $ep$ scattering via the exchange of a virtual
photon.\footnote{The $Z$ exchange contribution to eq.(\ref{e2.2}) is
negligible for $Q^2$ values that can be achieved in the foreseeable future.}
The special significance \cite{2} of \egam\ scattering lies in the fact that,
while (at present) the $x-$dependence of the nucleonic structure functions can
only be parametrized from data, the structure functions appearing in
eq.(\ref{e2.2}) can be {\em computed} in the QPM from the diagram shown in
Fig.~2a:
\be \label{e2.3}
F^{\gamma,{\rm QPM}}_2(x,Q^2) = \frac {6 \aem}{\pi} x \sum_q e_q^4 \left\{
\left[ x^2 + \left( 1-x \right)^2 \right] \log \frac {W^2} {m_q^2} + 8 x
\left( 1-x \right) - 1 \right\},
\ee
where we have introduced the squared cms energy of the $\gamma^* \gamma$
system
\be \label{e2.4}
W^2 = Q^2 \left( \frac {1}{x} - 1 \right).
\ee
The sum in eq.(\ref{e2.3}) runs over all quark flavours, and $e_q$ is the
electric charge of quark $q$ in units of the proton charge. An experimental
test of this equation was thought to not only allow to confirm the existence
of pointlike quarks, but also to measure their charges through the $e_q^4$
factor; both were topics of interest in the early 1970's when the study of
\egam\ scattering was first proposed \cite{2}.

Unfortunately, eq.(\ref{e2.3}) depends on the quark masses $m_q$. If this
ansatz is to describe data \cite{14} even approximately, one has to use
constituent quark masses of a few hundred MeV here; constituent quarks are not
very well defined in field theory. Moreover, we now know that QPM predictions
can be modified substantially by QCD effects. In case of proton structure
functions these lead to, among other things, scaling violations and a nonzero
$F_L$. In case of \egam\ scattering, QCD corrections are described by the kind
of diagrams shown in Figs.~2b,c. Diagrams of the type 2b leave the flavour
structure unchanged and are therefore part of the (flavour) nonsinglet
contribution to \f2gam, while diagrams with several disconnected quark lines,
as in Fig.~2c, contribute to the (flavour) singlet part of \f2gam.

The interest in photon structure functions received a boost in 1977, when
Witten showed \cite{3} that such diagrams can be computed exactly, at least
in the so--called ``asymptotic" limit of infinite $Q^2$. Including
next--to--leading order (NLO) corrections \cite{15}, the result can be
written as
\be \label{e2.5}
F_2^{\gamma,{\rm asymp}}(x,Q^2) = \aem \left[ \frac {1} {\alpha_s(Q^2)}
a(x) + b(x) \right],
\ee
where $a$ and $b$ are {\em calculable} functions of $x$. The absolute
normalization of this ``aymptotic" solution is therefore given uniquely by
$\alpha_s(Q^2)$, i.e. by the value of the QCD scale parameter $\Lambda_{\rm
QCD}$. It was therefore hoped that eq.(\ref{e2.5}) might be exploited for a
very precise measurement of $\Lambda_{\rm QCD}$.

Unfortunately this no longer appears feasible. One problem is that, in order
to derive eq.(\ref{e2.5}), one has to neglect terms of the form
${\displaystyle \left( \frac {\alpha_s(Q^2)} {\alpha_s(Q_0^2)} \right)^P}$,
where $Q_0^2$ is some input scale (see below). Neglecting such terms is
formally justified {\em if} $\alpha_s(Q^2) \ll \alpha_s(Q_0^2)$ {\em} and $P$
is positive. Unfortunately the first inequality is usually not satisfied at
experimentally accessible values of $Q^2$, assuming $Q_0^2$ is chosen in the
region of applicability of perturbative QCD, i.e. $\alpha_s(Q_0^2)/\pi \ll 1$.
Worse yet, $P$ can be zero or even negative! In this case ignoring such terms
is obviously a bad approximation. Indeed, one finds that eq.(\ref{e2.5})
contains divergencies as $x \to 0$ \cite{3,15}:
\be \label{e2.6}
a(x) \sim x^{-0.59}, \ \ \ \ \ b(x) \sim x^{-1}.
\ee
The coefficient of the $1/x$ pole in $b$ is {\em negative}; eq.(\ref{e2.5})
therefore predicts negative counting rates at small $x$. Notice that the
divergence is worse in the NLO contribution $b$ than in the LO term $a$. It
can be shown \cite{16} that this trend continues in yet higher orders, i.e.
the ``asymptotic" prediction for \f2gam\ rapidly becomes more and more
divergent for $x \to 0$ as more higher order corrections are included:
$\f2gam \sim x^{-4.3}$ in 3rd order, $\sim x^{-25.6}$ in 4th order, and so on.
Clearly the ``asymptotic" solution is not a very useful concept, having a
violently divergent perturbative expansion.\footnote{Notice that in the same
``asymptotic" limit, nucleonic structure functions collapse to a
$\delta-$function at $x=0$. While this is formally correct for infinite $Q^2$,
it gives obviously a poor description of the true proton structure at {\em
any} finite $Q^2$.}

The worst divergencies in $F_2^{\gamma,{\rm asymp}}$ occur in the singlet
sector, i.e. originate from diagrams of the type shown in Fig.~2c. It has been
speculated \cite{17} that this hints at a nonperturbative solution. For
example, if the invariant mass of the lowest \qqbar\ pair in Fig.~2c is small,
they might form a bound state. Traditionally, however, nonperturbative
contributions have been estimated using the vector dominance model (VDM)
\cite{18}, from the diagrams shown in Fig.~2d: The target photon undergoes a
transition into a nearly on--shell vector meson $(\rho, \ \omega, \ \phi,
\dots)$, so that \egam\ scattering is ``really" $e \rho, \ e \omega, \dots$
scattering, which should look qualitatively like $ep$ scattering. In
particular, the contribution of Fig.~2d should itself be well--behaved, i.e.
non--singular; it {\em cannot} cancel the divergencies of the ``asymptotic"
solution.\footnote{Of course, one can always {\em define} \f2gam\ to be the
sum of the (divergent) ``asymptotic" solution and a (divergent)
``nonperturbative contribution". However, nothing is gained by this as long
as one cannot at least estimate the nonperturbative contribution. Our
argument shows that the VDM is of no help here.}

Indeed, the existence of the contribution shown in Fig.~2d demonstrates that
we cannot hope to compute $\f2gam(x,Q^2)$ from perturbation theory alone.
Moreover, even if we assume that the VDM correctly describes {\em all}
nonperturbative contributions to \f2gam, it seems essentially impossible to
estimate them without making further assumptions. The problem is that the
vector mesons $\rho, \ \omega, \ \phi, \dots$ are much too short--lived to
allow an independent measurement of their parton distribution
functions.\footnote{Assuming that such distributions can even be defined for
resonance states; e.g., should a $\rho$ be treated as a \qqbar\ resonance with
two valence quarks, or a $\pi \pi$ resonance with four valence quarks?} It
therefore seems to us that the only meaningful approach is that suggested by
Gl\"uck and Reya \cite{19}. That is, one {\em formally} sums the contributions
from Figs.~2a--d into the single diagram of Fig.~2e, where we have introduced
quark densities in the photon \qisq\ such that (in LO)
\be \label{e2.7}
\f2gam(x,Q^2) = 2 x \sum_i e^2_{q_i} \qisq,
\ee
where the sum runs over flavours, $e_{q_i}$ is the electric charge of quark
$q_i$ in units of the proton charge, and the factor of 2 takes care or
antiquarks. This is merely a definition. In the approach of ref.\cite{19}
one does not attempt to compute the absolute size of the quark densities
inside the photon. Rather, one introduces input distribution functions
$q^{\gamma}_{i,0}(x) \equiv q_i^{\gamma}(x,Q_0^2)$ at some scale $Q_0^2$. This
scale is in principle arbitrary, as long as $\alpha_s(Q_0^2)$ is sufficiently
small to allow for a meaningful perturbative expansion. In practice, $Q_0^2$
is usually chosen as the {\em smallest} value where this criterion is assumed
to be satisfied. We will come back to this point in the following
subsection.

Given these input distributions, the photonic parton densities, and thus
\f2gam, at different values of $Q^2$ can be computed using the inhomogeneous
evolution equations. In LO, they read \cite{3,20}:
\ben \label{e2.8} \beq
\frac { d q^{\gamma}_{\rm NS} (x,Q^2)} { d \log Q^2} &= \frac {\aem}{2 \pi}
k^{\gamma}_{\rm NS}(x) + \frac {\alpha_s(Q^2)} {2 \pi} \left( P^0_{qq} \otimes
q^{\gamma}_{\rm NS} \right) (x,Q^2) ; \label{e2.8a} \\
\frac { d \Sigma^{\gamma} (x,Q^2)} { d \log Q^2} &= \frac {\aem}{2 \pi}
k^{\gamma}_{\Sigma}(x) + \frac {\alpha_s(Q^2)} {2 \pi} \left[ \left( P^0_{qq}
\otimes \Sigma^{\gamma} \right) (x,Q^2) + \left( P^0_{qG} \otimes G^{\gamma}
\right) (x,Q^2) \right] ; \label{e2.8b} \\
\frac { d G^{\gamma} (x,Q^2)} { d \log Q^2} &= \frac {\alpha_s(Q^2)} {2 \pi}
\left[ \left( P^0_{Gq}
\otimes \Sigma^{\gamma} \right) (x,Q^2) + \left( P^0_{GG} \otimes G^{\gamma}
\right) (x,Q^2) \right], \label{e2.8c}
\eeq \een
where we have used the notation
\be \label{e2.9}
\left( P \otimes q \right) (x,Q^2) \equiv \int_x^1 \frac {dy}{y} P(y)
q(\frac{x}{y},Q^2).
\ee
The $P^0_{ij}$ are the usual (LO) $j \to i$ splitting functions \cite{21}.
The inhomogeneous terms $k_i^{\gamma}$ describe $\gamma \to \qqbar$ splitting,
i.e. the diagram of Fig.~2a; for one quark flavour, one has
\be \label{e2.1n}
k_{q_i}^{\gamma} (x) = 3 e^2_{q_i} \left[ x^2 + (1-x)^2 \right].
\ee
The $k_i^{\gamma}$ of eqs.(\ref{e2.8}) follow from eq.(\ref{e2.1n}) by taking
appropriate sums or differences of quark flavours. Eq.(\ref{e2.8a}) describes
the evolution of the nonsinglet distributions (differences of quark
densities), i.e. resums only diagrams of the type shown in Fig.~2b, while
eqs.(\ref{e2.8b},\ref{e2.8c}) describe the evolution of the singlet sector
($\Sigma^{\gamma} \equiv \sum_i q_i^{\gamma} + \bar{q}_i^{\gamma}$), which
includes diagrams of the kind shown in Fig.~2c. Notice that this necessitates
the introduction of a gluon density inside the photon \glph, with its
corresponding input distribution $G_0^{\gamma}(x) \equiv G^{\gamma}(x,Q_0^2)$.

It is crucial to note that, given non--singular input distributions, the
solutions of eqs.(\ref{e2.8}) will also remain \cite{19} well--behaved at all
finite values of $Q^2$. This is true both in LO and in NLO \cite{22}. On the
other hand, by introducing a priori unknown input distributions, one clearly
abandons the hope to make an absolute prediction of $\f2gam(x,Q^2)$ in terms
of $\Lambda_{\rm QCD}$ alone. The solutions of eqs.(\ref{e2.8}) still show an
approximately linear growth with $\log Q^2$; in this sense eq.(\ref{e2.5})
remains approximately correct, but the functions $a$ and $b$ now do depend
weakly on $Q^2$ (approximately like $\log \log Q^2$), and the $x-$dependence
of $b$ is {\em not} computable. In fact, only the nonleading $Q^2$ dependence
(contained in the functions $a$ and $b$), which corresponds formally to the
scaling violations in the case of $ep$ scattering\footnote{Recall that in case
of \egam\ scattering, the QPM also predicts a logarithmic growth of \f2gam\
with $Q^2$, see eq.(\ref{e2.3}).}, can in this approach be used to determine
$\Lambda_{\rm QCD}$; a change of the normalization $1/\alpha_s$ multiplying
the first term can always be compensated by adding a constant to the input
distributions. Notice that {\em no} momentum sum rule applies for the parton
densities in the photon as defined here. The reason is that these densities
are all of first order in the fine structure constant \aem. Even a relatively
large change in these densities can therefore always be compensated by a small
change of the ${\cal O}(\alpha^0_{\rm em})$ term in the decomposition of the
physical photon, which is simply the ``bare" photon [with distribution
function $\delta(1-x)$]. Formally this would manifest itself by the addition
of ${\cal O}(\alpha^2_{\rm em})$ contributions to the inhomogeneous terms in
eqs.(\ref{e2.8}), which are numerically negligible.\footnote{It is sometimes
argued that one can still use the ``asymptotic" NLO prediction for \f2gam\ to
determine $\Lambda_{\rm QCD}$, if one sticks to the region of large $x$ where
the influence of the $x \to 0$ pole is supposed to be weak. However, this
rests on the {\em assumption} that the nonperturbative contributions to
\f2gam\ are small at large $x$, and the {\em second assumption} that the terms
that regularize the pole \cite{23} vanish as $x \to 1$. The only way to test
these assumptions seems to be to compare the value of $\Lambda_{\rm QCD}$
extracted in this way with other measurements \cite{24} of $\alpha_s$. In our
opinion this shows that a measurement of \f2gam\ at fixed $Q^2$ {\em cannot}
be used to determine $\Lambda_{\rm QCD}$ unambiguously. Recall also that the
perturbative expansion of the ``asymptotic" prediction in yet higher orders in
QCD is extremely problematic.}

Before discussing our present knowledge of and parametrizations for the
parton densities in the photon, we briefly address a few issues related to the
calculation of \f2gam. As mentioned above, eqs.(\ref{e2.7}),(\ref{e2.8})
have been extended to NLO quite early, although a mistake in the two--loop
$\gamma \to G$ splitting function was found \cite{25} only fairly recently. A
full NLO treatment of massive quarks is now also available \cite{26} for both
\f2gam\ and $F_L^{\gamma}$. A first treatment of small$-x$ effects in the
photon structure functions, i.e. $\log 1/x$ resummation and parton
recombination, has been presented in ref.\cite{27}; however, the predicted
steep increase of \f2gam\ at small $x$ has not been observed experimentally
\cite{28}. Finally, nonperturbative contributions to \f2gam\ are expected to
be greatly suppressed if the target photon is also far off--shell. One can
therefore derive unambiguous QCD predictions \cite{29} in the region $Q^2 \gg
P^2 \gg \Lambda^2$, where the first strong inequality has been imposed to
allow for a meaningful definition of structure functions.\footnote{If $Q^2
\simeq P^2$, the use of fixed--order perturbation theory is more appropriate,
since it includes terms $\propto (P^2/Q^2)^n$, which are not included in the
usual structure functions.} However, it has recently been pointed out
\cite{30} that nonperturbative effects might survive longer than previously
expected; an unambiguous prediction would then only be possible for very large
$P^2$, and even larger $Q^2$, where the cross--section is very small.

\setcounter{footnote}{0}
\subsection*{2b) Parametrizations of Photonic Parton Densities}
As discussed in the previous subsection, the $Q^2$ evolution of the photonic
parton densities $\qvph(x,Q^2) \equiv (q_i^{\gamma}, G^{\gamma})(x,Q^2)$ is
uniquely determined by perturbative QCD, eqs.(\ref{e2.8}) and their NLO
extension. However, given the flaws of the ``asymptotic" prediction for \qvph,
one has to specify input distributions \qv0\ at a fixed $Q^2 = Q_0^2$. The
situation is therefore in principle quite similar to the case of nucleonic
parton densities. In practice, however, the determination of these input
distributions is much more difficult in case of the photon, for a variety of
reasons.

To begin with, no momentum sum rule applies for \qv0, as discussed above. This
means that it will be difficult to derive reliable information on
$G_0^{\gamma}$ from measurements of $F_2^{\gamma}$ alone: in LO, the gluon
density only enters via the (subleading) $Q^2$ evolution in $F_2^{\gamma}$. In
contrast, analyses of $eN$ scattering revealed very early that gluons must
carry about 50\% of the proton's momentum, thereby fixing the overall scale of
the gluon density in the nucleon. We will see below that existing
parametrizations for $G^{\gamma}$ still differ by sizable factors over the
entire $x$ range unless $Q^2$ is very large.

Secondly, so far deep--inelastic $e \gamma$ scattering could only be studied
at \epem\ colliders, where the target photon is itself radiated off one of the
incoming leptons. Its spectrum is given by the well--known
Weizs\"acker--Williams function \cite{31}
\be \label{e2.10}
\fgme(z) = \frac {\aem} {2 \pi z} \left\{ \left[ 1 + \left( 1-z \right)^2
\right] \log \frac {2 E^2 \left( 1 - z \right)^2 \left( 1 -
\cos \theta_{\rm max} \right) } { m_e^2 z^2 } - 2 \left( 1-z \right) \right\},
\ee
where $E$ is the electron beam energy and $zE$ the energy of the target
photon. In eq.(\ref{e2.10}) we have assumed that there is no experimentally
imposed lower bound on the virtuality $p^2 \equiv -P^2$ of the target photon,
so that $P^2_{\rm min} = m_e^2 z^2 / (1-z)$; however, we have introduced an
upper limit $\theta_{\rm max}$ on the scattering angle of the electron
emitting the target photon, in order to allow for antitagging.

Eq.(\ref{e2.10}) implies that the cross section from the measurement of which
$F_2^{\gamma}$ is to be determined is of order $d \sigma / d Q^2 \sim
\alpha^4_{\rm em}/ \left( \pi Q^4 \right) \log \left( E/m_e \right)$, see
eq.(\ref{e2.2}). [Recall that $F_2^{\gamma}$ is itself ${\cal O}(\aem)$.]
The event rate is therefore quite small; the most recent measurements
\cite{14,28,32} typically have around 1,000 events at $Q^2 \simeq 5$ GeV$^2$,
and the statistics rapidly gets worse at higher $Q^2$. This is to be compared
with millions of events in, for example, deep--inelastic $\nu N$ scattering
\cite{33}.

Another problem is that the $e^{\pm}$ emitting the target photon is usually
not detected, since it emerges at too small an angle. This means that the
energy of the target photon, and hence the Bjorken variable $x$, can {\em only}
be determined from the hadronic system. All existing analyses try to determine
$x$ from the invariant mass $W$, using eq.(\ref{e2.4}). This is problematic,
since at least some of the produced hadrons usually also escape undetected,
e.g. in the beam pipes; the measured value of $W$, usually denoted by $W_{\rm
vis}$, is therefore generally {\em smaller} than the true value. (It can
exceed the true $W$ due to the finite energy resolution of real--world
detectors.) One has to correct for this by ``unfolding" the measured $W_{\rm
vis}$ distribution to arrive at the true $W$ (or $x$) distribution. However,
in order to do this one has to model the hadronic system $X$. In other words,
a ``measurement" of $F_2^{\gamma}$ only seems possible if we already know many
details about multi--hadron production in $\gamma^* \gamma$ scattering!

In practice the situation is not quite as bad, since one can check the
assumptions made by comparing various distributions predicted by the Monte
Carlo with real data. An iterative procedure can then be used to arrive at a
model for $X$ that should allow to do the unfolding reliably. However, one
should be aware that this might lead to large uncertainties at the boundaries
of the accessible range of $x$ values. The reason is that in the event
selection one usually imposes upper and lower cuts on $W_{\rm vis}$,
corresponding to small and large $x$, respectively. Relatively minor changes
of the model for $X$ can therefore significantly change the predicted
efficiency for accepting events with $x$ close to one of the kinematic
boundaries. The region of large $x$ (small $W$) is in any case plagued by
higher twist uncertainties (e.g., production of resonances); however, the
region of small $x$ is very interesting, since here $F_2^{\gamma}$ is
dominated by sea quarks whose density is closely related to the gluon
distribution. It has been shown explicitly \cite{34} that different
ans\"atze for $X$ can lead to quite different ``measurements" of
$F_2^{\gamma}$ at small $x$.

Most recent analyses use the ``FKP model" of refs.\cite{35} as starting point
of the unfolding procedure. Here $F_2^{\gamma}$ is split into ``soft" and
``hard" components, depending on the virtuality of the quark exchanged in the
$t-$ or $u-$channel, see figs.~2a--c. The cut--off parameter $t_0$ separating
these two regions is to be fitted from the data. Contributions with $|t| \leq
t_0$ are modelled using VDM ideas; in practice this means that a scaling
($Q^2-$independent) ansatz is used, fitted from low$-Q^2$ data \cite{14}.
Contributions with $|t| > t_0$ can be computed from an evolution equation in
$\log |t|$, with the boundary condition that this ``pointlike" part vanish at
$t = t_0$. This procedure is formally equivalent to the one suggested by
Gl\"uck and Reya \cite{19}, {\em if} we identify $Q_0^2 = t_0$ and impose the
upper limit $|t|_{\rm max} = Q^2$ for the $t-$evolution. However, as presently
used \cite{36,28,32}, the unfolding procedure has several weaknesses:
\begin{itemize}
\item It uses a parametrization \cite{36} of the ``pointlike" part of
$F_2^{\gamma}$ which includes some terms which are of next--to--leading order
in a formal operator product expansion; however, not all such terms are
included. In particular, it uses the actual kinematical maximum for the
$t-$evolution, $|t|_{\rm max} \simeq Q^2/x$; as a result, the predicted
$F_2^{\gamma}$ tends to be larger at small and median $x$ than what one would
get from the usual $Q^2-$evolution.\footnote{Very roughly, one replaces the
overall growth $\propto \log Q^2$ by $\log \left( Q^2/x \right)$.} On the
other hand, this expression ignores sea quarks, i.e. $g \to \qqbar$ splitting;
it therefore underestimates the true result for very small $x$.
\item The $Q^2$ (or $t$) evolution of the ``hadronic part" is ignored; at high
$Q^2$, this over--estimates the soft contribution to $F_2^{\gamma}$ at median
and large $x$, and underestimates it at small $x$.
\item While this procedure treats the effect of gluon radiation on the shape
of $F_2^{\gamma}$ (approximately) correctly, it does {\em not} include any
parton showering in the MC event generator. Rather, the generator produces
\qqbar\ pairs whose $p_T$ distribution in the $\gamma^* \gamma$ cms is either
exponential (for the ``hadronic part") or follows the QPM prediction (for the
``pointlike part"). This \qqbar\ pair is then fed into a string--based
fragmentation program. While string fragmentation can mimick the effects of
parton showering to some extent, it is not able to produce additional jets.
It is therefore not surprising that a recent analysis \cite{32} found larger
jet rates in $e \gamma$ scattering than the MC predicted.
\end{itemize}
Some of these points have also been raised in ref.\cite{36a}. We would like to
emphasize that none of these problems is intrinsic to the formalism of
ref.\cite{35}. Moreover, the input model used for the unfolding always turned
out to be consistent within errors with the extracted $F_2^{\gamma}$, if $t_0$
is chosen in the vicinity of 0.5 GeV$^2$.

The point of this lengthy discussion is not to criticize our experimental
colleagues. Rather, we hope that it might serve as a starting point for
further work by people familiar with the design of MC event generators, which
we are not. However, this discussion shows that present data on $F_2^{\gamma}$
have to be taken with a grain of salt. In particular, the shortcomings listed
above are usually {\em not} addressed in the estimates of the systematic error
due to the unfolding procedure; this error only includes things like the
choice of binning \cite{28}. This might help to explain the apparent
discrepancy between different data sets \cite{14}. Fortunately, new ideas
for improved unfolding algorithms \cite{36b} are now under investigation; this
should facilitate the measurement of $F_2^{\gamma}$ at small $x$, especially
at high energy (LEP2).

In spite of this, measurements of $F_2^{\gamma}$ probably still provide the
most reliable constraints on the input distributions $\qv0(x)$; they are
certainly the only data that have been taken into account when constructing
existing parametrizations of \vqxqsq. In the remainder of this subsection we
will briefly describe these parametrizations.

The simplest and oldest parametrizations \cite{37,37a} are based on the
``asymptotic" LO prediction \cite{3,37b}. While the theoretical basis for this
prediction is weak, as described in the previous subsection, their simplicity
makes these parametrizations useful for first estimates of reaction rates, as
long as one stays away from very small $x$. Refs.\cite{37,37a} only give
parametrizations for $N_f=4$ active flavours. More recently, Gordon and
Storrow \cite{38} provided different, more accurate fits for $N_f = 3, \ 4$
and $5$.

All other parametrizations involve some amount of data fitting. However, due
to the rather large experimental errors of data on $F_2^{\gamma}$, additional
{\em assumptions} always had to be made. In particular, quark and antiquark
distributions (of the same flavour) are always assumed to be identical, which
guarantees that the photon carries no flavour. This assumption is eminently
reasonable. The $\gamma \qqbar$ vertex treats quarks and antiquarks
symmetrically, and we do not know of any effect that could destroy this
symmetry.

The \ub{DG parametrization} \cite{39} was the first to start from input
distributions. At the time of this fit, only a single measurement of
$F_2^{\gamma}$ at fixed $Q^2\simeq 5.3$ GeV$^2$ existed. In order to determine
three, in principle independent, input distributions (for nonsinglet and
singlet quarks as well as gluons), two assumptions were made: All input quark
densities were assumed to be proportional to the squared quark charges, i.e.
$\ug = 4 \dg = 4 \sg$ at $Q_0^2 = 1$ GeV$^2$; and the gluon input was
generated purely radiatively, i.e.
\be \label{e2.11}
G^{\gamma}_{0, {\rm DG}} = \frac {2} {\beta_0} \Sigma^{\gamma}_{0, {\rm DG}}
\otimes P_{Gq},
\ee
where $\beta_0$ = 9 is the coefficient of the 1--loop QCD $\beta$ function.
This parametrization only exists in LO. Moreover, it treats flavour thresholds
by introducing three independent sets of distributions for $N_f=3, \ 4, \ 5$,
so that the transition across a threshold is not automatically smooth.
Nevertheless it continues to describe data reasonably well, although a
combined fit would probably give a higher $\chi^2$ than for more recent
parametrizations.

The \ub{LAC parametrizations} \cite{40} are based on a much larger data set.
The main point of these fits was to demonstrate that data on $F_2^{\gamma}$
constrain \Gg\ very poorly. In particular, they allow a very hard gluon,
with $x \Gg$ having a maximum at $x \simeq 0.9$ (LAC3), as well as very soft
gluon distributiuons, with $x\Gg$ rising very steeply at low $x$ (LAC1, LAC2).
The LAC parametrizations only exist for $N_f=4$ massless flavours and in LO.
{\em No} assumptions about the relative sizes of the four input quark
densities were made in the fit. LAC3 has been clearly excluded by data on jet
production in $ep$ scattering (Sec.~3a) as well as in real \gaga\ scattering
(Sec.~4a); the experimental status of LAC1,2 is less clear.

The recent \ub{WHIT parametrizations} \cite{41} follow a similar philosophy
as LAC, at least regarding the gluon input; however, their choices for
$G_0^{\gamma}$ are much less extreme. In the WHIT1,2,3 parametrizations,
gluons carry about half as much of the photon's momentum as quarks do (at the
input scale $Q_0^2 = 4$ GeV$^2$), while in WHIT4,5,6 gluons and quarks carry
about the same momentum fraction. These two groups of fits also have slightly
different valence quark densities.\footnote{The definition of ``valence" and
``sea" quarks used here differs from the more common ``nonsinglet" and
"singlet" distributions. The $Q^2$ evolution of the valence density is
independent of \Gg, i.e. obeys eq.(\ref{e2.8a}), while a non--zero sea
quark density is produced only through $g \to \qqbar$ splitting. In the
absence of mass effects, the valence distributions are proportional to the
squared quark charges, while the sea distributions are independent of the
quark charge.} The input gluon density is assumed to have the simple shape $x
G_0^\gamma \propto \left( 1-x \right)^{c_g}$, with $c_g=3$ for WHIT1,4; $c_g=9$
for WHIT2,5; and $c_g=15$ for WHIT3,6. Recall that the normalization is
adjusted such that $\int x G_0^\gamma(x) dx$ is constant within each group of
parametrizations; a larger $c_g$ therefore also means a larger \Gg\ at small
$x$. The input distributions for the sea quarks are computed from the cross
section for $\gamma^* g \to \qqbar$, regularized by light quark masses $m=0.5$
GeV. The WHIT parametrizations only exist in LO, but great care has been taken
to treat the ($x-$dependent) charm threshold correctly. This is much more
important here than for nucleonic parton densities, since the photon very
rapidly develops an ``intrinsic charm" component from $\gamma \to \ccbar$
splitting.

The \ub{GRV parametrization} \cite{42} is the first NLO fit of \qvph; a LO
version is also available. This parametrization is based on the same
``dynamical" philosophy as the earlier fits of protonic \cite{43} and pionic
\cite{44} parton densities by the same authors. The idea is to start from a
very simple input at a very low $Q_0^2$ (0.25 GeV$^2$ in LO, 0.3 GeV$^2$ in
NLO); this scale is assumed to be the same for $p, \ \pi$ and $\gamma$
targets. The observed, more complex structure is then generated dynamically by
the evolution equations. In case of the proton, only valence quarks were
originally assumed to be present at scale $Q_0^2$. While the gluon density
does evolve fast enough to carry approximately half the proton's momentum at
$Q^2$ of a few GeV$^2$, it was found to be too soft in shape. The ansatz
therefore had to be modified \cite{45} by introducing ``valence--like" gluon
and even sea quark densities already at the input scale, thereby giving up
much of the original simplicity of the idea. Their pionic input distributions
also include a ``valence--like" gluon density, which is in fact strictly
proportional to the valence quark density, but no sea quarks at scale $Q_0^2$.

In the photonic case,
\be \label{e2.12}
\vec{q}^{\gamma}_{0,{\rm GRV}}(x) = \kappa \frac {4 \pi \aem} {f^2_{\rho}}
\vec{q}^{\pi}_{0,{\rm GRV}}(x).
\ee
This input is motivated from VDM ideas, where $f_{\rho}^2$ determines the
$\gamma \to \rho$ transition probability ($f^2_{\rho}/4\pi \simeq 2.2$), and
$\kappa$ has been introduced to describe contributions from  heavier vector
mesons ($\omega, \phi, \dots$). In fact, $\kappa$ is the {\em only} free
parameter in this ansatz; it was determined to be $\kappa=2$ (1.6) in LO (NLO).
However, it should be clear that eq.(\ref{e2.12}) is an {\em assumption} that
has to be tested experimentally; in particular, it is not obvious that the
parton densities in a pion resemble those in a vector meson, or that QCD is
applicable at such small input scales.\footnote{It has been shown previously
\cite{45a} that an ansatz like eq.(\ref{e2.12}) cannot be brought into
agreement with the data if one insists on $Q_0^2 \geq 1$ GeV$^2$. One way out
\cite{27} would be to multiply the resulting $F_2^{\gamma}$ with something
like $Q^2/(Q^2 + \mu^2)$, but this goes beyond the leading--twist partonic
contributions.} On the positive side, the GRV parametrization ensures a smooth
onset of the charm density, using an $x-$independent threshold. Moreover, care
has been taken to split the NLO parametrizations for \qvph\ into LO and NLO
pieces; only the former should be multiplied with NLO pieces of hard cross
sections.\footnote{This distinction is more important in the photonic case,
since possible $\log ( 1-x)$ divergencies are {\em not} always regularized by
parton densities falling like a power of $1-x$. However, even in the photonic
case this problem is greatly ameliorated if one uses the ``DIS$_{\gamma}$"
scheme introduced \cite{46} by the same authors.}

The \ub{GS parametrizations} \cite{38} were developed shortly after GRV, but
follow a quite different strategy. Problems with low input scales \cite{38a}
are avoided by choosing $Q_0^2 = 5.3$ GeV$^2$. This is certainly in the
perturbative region, but necessitates a rather complicated ansatz for the
input distributions:
\be \label{e2.13}
\vec{q}^{\gamma}_{0,{\rm GS}}(x) = \kappa \frac {4 \pi \aem} {f^2_{\rho}}
\vec{q}^{\pi}_0(x,Q_0^2) + \vec{q}^{\gamma}_{\rm QPM}(x,Q_0^2).
\ee
The free parameters in the fit are the momentum fractions carried by gluons
and sea--quarks in the pion, the parameter $\kappa$, and the light quark
masses. In the GS2 parametrization, $G_0^{\gamma}$ is assumed to come entirely
from the first term in eq.(\ref{e2.13}), while in GS1 the second term also
contributes via radiation, see eq.(\ref{e2.11}). The fit gives $m_u = m_d =
0.29$ GeV and $\kappa = 1.96$. These values are not unreasonable. However, the
ansatz (\ref{e2.13}) might be somewhat suspect: It holds in the perturbative
region of $Q^2$, but its form is not invariant under the evolution equations.
On the other hand, for practical purposes it includes sufficiently many free
parameters to allow a decent description of data on $F_2^{\gamma}$. This
parametrization is now being updated \cite{38b}. The new version uses a
slightly reduced input scale $Q_0^2=3$ GeV$^2$, and for the first time
includes data on jet production in two--photon collisions (see Sec.~4a) in the
fit; unfortunately this still does not allow to pin down \gga\ with any
precision.

The \ub{AGF parametrization} \cite{47} is (in its ``standard" form) quite
similar to GRV. In particular, they also assume that at a low input scale
$Q_0^2 = 0.25$ GeV$^2$ the photonic parton densities are described by the VDM.
There are some differences, however. First, AGF point out that in NLO the
input densities are scheme--dependent, if physical quantities like
$F_2^{\gamma}$ are to be scheme--independent. GRV use their ansatz
(\ref{e2.12}) in their DIS$_{\gamma}$ scheme, since it is perturbatively more
stable than the more commonly used $\overline{\rm MS}$ scheme. AGF point out
that this treatment includes certain process--dependent terms in \qvph;
they therefore prefer to use the $\overline{\rm MS}$ scheme, and define their
input distributions to be the difference of a (regular) ``VDM" term and a
process--independent term containing a $\log(1-x)$ divergence, so that
$F_2^{\gamma}(x,Q_0^2)$ is well--behaved in the limit $x \to 1$. Secondly,
they include $\rho-\omega-\phi$ interference when specifying the ``VDM" input,
so that
\beq \label{e2.14}
u^{\gamma}_{0,{\rm AGF}}(x) &= K \aem \left[ \frac{4}{9}
u^{\pi}_{\rm valence}(x,Q_0^2) + \frac{2}{3} u^{\pi}_{\rm sea}(x,Q_0^2)
\right]; \nonumber \\
d^{\gamma}_{0,{\rm AGF}}(x) = s^{\gamma}_{0,{\rm AGF}}(x) & =
K \aem \left[  \frac{1}{9} u^{\pi}_{\rm valence}(x,Q_0^2)
+ \frac{2}{3} u^{\pi}_{\rm sea}(x,Q_0^2) \right]; \nonumber \\
G^{\gamma}_{0,{\rm AGF}}(x) &= K \aem \frac{2}{3} G^{\pi}(x,Q_0^2),
\eeq
where the pion distribution functions are taken from ref.\cite{48}. (Recall
that in NLO the input quark densities have to be modified by a subtraction
term.) Perhaps the for practical purposes most important difference from the
GRV parametrization is that the coefficient $K$ is left free, i.e. separate
fits are provided for the ``anomalous" (or ``pointlike") and
``nonperturbative" contributions to \qvph, allowing the user to specify the
absolute normalization (although not the shape) of the latter.

Finally, two of the \ub{SaS parametrizations} \cite{36a} are based on a
similar philosophy as the GRV and AGF parametrizations, by assuming that at a
low $Q_0 \simeq 0.6$ GeV the perturbative component vanishes (SaS1). However,
while the normalization of the nonperturbative contribution is taken from
the VDM (including $\rho, \ \omega$ and $\phi$ contributions with fixed
normalization), the shapes of the quark and gluon distributions are fitted
from data. Although the SaS parametrizations are available in LO only, the
authors attempt to estimate the scheme dependence (formally an NLO effect)
providing a parametrization (SaS1M) where the non leading--log part of the QPM
prediction for $F_2^{\gamma}$ has been added to eq.(\ref{e2.7}), while SaS1D
is based on eq.(\ref{e2.7}) alone. There are also two parametrizations with
$Q_0=2$ GeV; however, in this case the normalization of the fitted ``soft"
contribution had to be left free, and its shape is much harder than one what
expects from hadronic parton densities. Again two sets of parametrizations are
available, using different schemes (SaS2D, SaS2M).

The SaS1 sets preferred by the authors are quite similar to AGF; the real
significance of ref.\cite{36a} is that it carefully describes the properties
of the hadronic state $X$ for both the hadronic and ``anomalous"
contributions, as needed for a full event characterization. We will come back
to this aspect of their work in Sec.~5.

In Fig.~3 we compare various LO parametrizations of $F_2^{\gamma}$ at
$Q^2=15$ GeV$^2$ with recent data taken by the OPAL \cite{28} and TOPAZ
\cite{32} collaborations; present data are not able to distinguish between LO
and NLO fits. In order to allow for a meaningful comparison, we have added a
charm contribution to the OPAL data, as estimated from the QPM; this
contribution had been subtracted in their analysis. We have used the DG and GRV
parametrizations with $N_f=3$ flavours, since their parametrizations of
$c^\gamma$ are meant to be used only if $\log Q^2 / m_c^2 \gg 1$; the charm
contribution has again been estimated from the QPM.\footnote{We have ignored
the small contribution \cite{26} from $\gamma^* g \to \ccbar$ in this figure.}
As discussed earlier, WHIT provides a parametrization of \cg\ that includes
the correct kinematical threshold, while LAC treat the charm as massless at
all $Q^2$.

We see that most parametrizations give quite similar results for
$F_2^{\gamma}$ over most of the relevant $x-$range; the exception is LAC1,
which exceeds the other parametrizations both at large and at very small $x$.
It should be noted that the data points represent averages over the
respective $x$ bins; the lowest bin starts at $x=0.006 \ (0.02)$ for the OPAL
(TOPAZ) data. The first OPAL point is therefore in conflict \cite{34} with the
LAC1 prediction. Unfortunately there is also some discrepancy between the
TOPAZ and OPAL data at low $x$. As discussed above, one is sensitive to the
unfolding procedure here; for this reason, WHIT chose not to use these (and
similar) points in their fit. (The other fits predate the data shown in
Fig.~3.) This ambiguity in present low$-x$ data is to be regretted, since
in principle these data have the potential to discriminate between different
ans\"atze for $G_0^{\gamma}$. This can most clearly be seen by comparing the
curves for WHIT4 (long dashed) and WHIT6 (long--short dashed), which have the
{\em same} valence quark input, and even the same $\int x G_0^{\gamma} dx$:
WHIT4 has a harder gluon input distribution, and therefore predicts a larger
$F_2^{\gamma}$ at $x \simeq 0.1$; WHIT6 has many more soft gluons, and
therefore a very rapid increase of $F_2^{\gamma}$ for $x \leq 0.05$, not
unlike LAC1. Finally, we should mention that the GS, AGF and SaS
parametrizations also reproduce these data quite well.

Discriminating between these parametrizations would be much easier if one
could measure the gluon density directly. This is demonstrated in Fig.~4a,
where we show results for $x \Gg$ at the same value of $Q^2$; we have
chosen the same LO parametrizations as in Fig.~3, and included the LAC3
parametrization with its extremely hard gluon density. Note that, for
example, WHIT4 and WHIT6 now differ by a factor of 5 for $x$ around 0.3.
The gluon distribution of WHIT6 is rather similar in shape to the one of LAC1,
but significantly smaller in magnitude. Indeed, in all three LAC
parametrizations, gluons carry significantly more momentum than quarks for
$Q^2 \leq 20$ GeV$^2$; this is counter--intuitive \cite{38}, since in known
hadrons, and hence presumably in a VMD--like low$-Q^2$ photon, gluons and
quarks carry about equal momentum fractions, while at very high $Q^2$ the
inhomogeneous evolution equations (\ref{e2.8}) predict that quarks in the
photon carry about three times more momentum than gluons. Notice finally that
GRV predicts a relatively flat gluon distribution. This results partly from the
low value of the input scale $Q_0^2=0.25$ GeV$^2$, compared to 1 GeV$^2$ for
DG and 4 GeV$^2$ for WHIT and LAC1; a larger $Q^2/Q_0^2$ allows for more
radiation of relatively hard gluons off large$-x$ quarks. Recall also that
their (pionic) input distribution includes a valence--like (hard) gluon
density.

Finally, Fig.~4b shows that some parametrizations also differ substantially in
the flavour structure. Both DG and WHIT assume $\dg=\sg$ at all $Q^2$.
Moreoever, DG assumes $q^{\gamma}_{i,0} \propto e_{q_i}^2$ for the entire
input quark distribution, while WHIT assumes this only for the valence input.
Nevertheless the smaller value of $Q_0^2$ assumed by DG leads to a strangeness
content quite similar to that predicted by WHIT: At very low $x$, sea quarks
dominate, which have $\ug=\dg=\sg$, so that $\sg/(\ug+\dg) \simeq
1/2$; at high $x$, $q^{\gamma}_i \propto e^2_{q_i}$, so that $\sg/(\ug+\dg)
\simeq 1/5$.\footnote{The ratio of 4:1 between large$-x$ (valence) $u$ and
$d$ quark densities is implicit to the structure of the WHIT fits, while in
DG, \ug\ and \dg\ are parametrized independently; this explains the $\sim 2\%$
deviation between the two parametrizations shown in Fig.~4b at large $x$.}
Since sea quarks are produced by gluon splitting, the transition between the
sea--dominated and valence--dominated regions depends on \Gg.

In contrast, GRV assumes $u_0^{\gamma}=d_0^{\gamma}$ and $s_0^{\gamma}=0$ at
the input scale; for given $F_2^{\gamma}$, the former assumption increases
$\ug+\dg$ and the latter reduces \sg, compared to the ansatz $q_i^{\gamma}
\propto e^2_{q_i}$. This explains the smallness of the strangeness content
of the photon predicted by GRV, which persists to surprisingly large values
of $x$. It is worth mentioning that AGF, which is otherwise quite similar to
GRV [at least for the standard normalization of the nonperturbative
contribution, $K=1$ in eqs.(\ref{e2.14})], also assumes the input valence
quark distributions to be proportional to the squared quark charges; its
predictions for $\sg/(\ug+\dg)$ are therefore quite similar to those of the
WHIT group. GS falls between WHIT and GRV, since here a small but nonzero
$s_0^{\gamma}$ is assumed, from the pionic sea quarks as well as the QPM part
in eq.(\ref{e2.13}). On the other hand, LAC treats all four quark input
distributions as completely independent quantities, without imposing any
constraints between them. This results in the erratic behaviour of the
strangeness content depicted in Fig.~4b; the ratio even exceeds unity for
$x \simeq 0.1$. Clearly LAC should therefore not be used when the flavour
structure of the photon is important, e.g. in $W$ and $Z$ boson production at
HERA.
\setcounter{footnote}{0}
\section*{3) Resolved Photon Processes in \gamp\ Scattering}
In this section we discuss \gamp\ scattering reactions that are sensitive to
the hadronic structure of the photon. Most of our numerical results will be
for the $ep$ collider HERA, with appropriate (anti--)tagging conditions for
the outgoing electron, in order to make sure that the virtuality of the
exchanged photon is small; we will also make a few comments on fixed target
\gamp\ scattering. We follow the terminology of ref.\cite{49} in
distinguishing between {\em direct} and {\em resolved photon} contributions to
a given process; the existence of these physically distinct contributions had
already been emphasized in ref.\cite{37b}. The former are defined as reactions
where the incident photon participates directly in the relevant hard
scattering process; some examples are shown in Fig.~5a. In contrast, in
resolved photon contributions the incoming photon takes part in the hard
scattering via one of its constituents, a quark or gluon, as shown in Fig.~5b.
Notice that in direct processes the entire photon energy goes into the hard
partonic final state, while in resolved photon processes only a fraction
$x_{\gamma}$ of the photon energy is available for the hard scattering. This
also leads to different topologies for the two classes of contributions. In
LO, the direct contribution to jet production will give rise to two high$-p_T$
jets and the remnant jet from the proton, see Fig.~6a; there will be little or
no hadronic activity in the direction of the incident photon. In contrast,
resolved photon contributions are characterized by a second spectator jet,
which is formed when a coloured parton is ``taken out" of the photon; this
photonic remnant or spectator jet will usually go approximately in the
direction of the incident photon, which coincides with the electron direction
at HERA. This additional jet in principle allows  distinction between direct
and resolved photon contributions on an event--by--event basis.

Occasionally one sees the erroneous statement in the literature that resolved
photon contributions are NLO corrections to the corresponding direct process.
However, by counting powers of coupling constants in Figs.~5a and 5b one can
easily convince oneself that this is not the case. The direct contribution to
di--jet production is obviously of order $\aem \alphas$. The resolved photon
contribution is of order $\qvph \cdot \alpha^2_s$, where \qvph\ stands for any
photonic parton density. We have seen early in Sec.~2 that these densities are
of first order in \aem. Moreover, we also saw that they grow logarithmically
with the relevant scale $Q^2$ of the given process; in leading--log summed
perturbation theory this $\log Q^2$ has to be counted as a factor $1/\alphas$,
see eq.(\ref{e2.5}).\footnote{Recall that this expression remains
approximately correct even beyond the ``asymptotic" approximation, if we allow
a weak $Q^2$ dependence of the functions $a$ and $b$.} Altogether the photonic
parton densities therefore have to be counted as ${\cal O} \left( \aem /
\alphas \right)$, so that $\qvph \cdot \alpha^2_s$ is again of order $\aem
\alphas$, just like the direct contribution. It should be emphasized that LO
QCD is {\em always} leading log summed, i.e. $\alphas(Q^2) \cdot \log Q^2$ is
always counted as ${\cal O}(1)$, not ${\cal O}(\alphas)$. To give one example,
the resummation of such leading logarithms leads to the scaling violations
of nucleonic structure functions; it should be clear that one is {\em not}
performing an NLO analysis by simply using $Q^2-$dependent parton densities
when estimating \ppbar\ cross--sections.

Part of the confusion is caused by the fact that in NLO, direct and resolved
photon contributions mix. This can, e.g., be seen from the Feynman diagram
of Fig.~7, which shows an NLO contribution to direct jet production. If no
restrictions are imposed on the transverse momentum of the outgoing
antiquark, it could go in the photon direction and thus form a ``remnant
jet". In fact, the total contribution from the diagram of Fig.~7 to the
inclusive jet (pair) cross--section will be dominated by configurations where
the transverse momentum of the antiquark is small, due to the $1/t$ pole
associated with the exchanged quark. The crucial point is that the
contribution from this pole has already been included in the resolved LO $q q'$
scattering process, where the exchanged quark is treated as being on--shell.
In order to avoid double--counting, the contribution of the $1/t$ pole
therefore has to be subtracted from the NLO contribution of Fig.~7; in other
words, the corresponding collinear divergence is absorbed in the quark
distribution function in the photon. We emphasize that this treatment is
completely analogous to the calculation of NLO corrections to jet production
in \ppbar\ scattering. For example, the incident photon in Fig.~7 could be
replaced by a gluon coming from the $\bar{p}$. The above argument then tells
us that $qq'$ and $gq$ scattering mix in NLO, i.e. a part of the NLO
contribution from $gq$ scattering has to be absorbed in $q q'$ scattering;
nevertheless nobody would consider one to be an NLO correction to the
other.\footnote{The diagram of Fig.~7 also has a divergence when the exchanged
gluon goes on--shell; this is absorbed in the gluon density in the proton,
i.e. in direct $\gamma g$ scattering. Finally, in principle both the exchanged
quark and the gluon can be (nearly) on--shell. However, in this case none of
the final state partons has large transverse momentum; this configuration
therefore only contributes to soft processes, which cannot be treated
perturbatively.}

We do not attempt to split resolved photon contributions into those coming
from the ``anomalous", ``pointlike" or ``perturbative" part of \qvph\ and
those due to the ``hadronic" or ``nonperturbative" part. We have seen in the
previous section that it is not easy to separate these parts consistently;
indeed, it should be clear that in reality there is a smooth transition from
the one to the other. Nevertheless we will see later (in Sec.~5) that the
existence of the ``pointlike" contribution may have some impact on overall
event characteristics.

After these preliminaries, we are ready to discuss various hard \gamp\
reactions. We start with jet production in Sec.~3a, where both NLO
calculations and high--energy data from the $ep$ collider HERA are available.
Open heavy quark production (Sec.~3b) has also been treated in NLO, and first
HERA data have started to appear. Direct photon production (Sec.~3c) also also
been treated in NLO, but no HERA data have yet been published. We then discuss
the production of \jpsi\ mesons in Sec.~3d, and lepton--pair (Drell--Yan)
production in Sec.~3e.

\setcounter{footnote}{0}
\subsection*{3a) Jet Production in \gamp\ Collisions}
The production of high$-p_T$ jets offers the largest cross--section of all
hard \gamp\ scattering reactions. It was therefore the first such process for
which resolved photon contributions were calculated \cite{50}, and also among
the first reactions to be studied experimentally at HERA \cite{51,52}.
Moreover, many properties of direct and resolved photon contributions to jet
production carry over to the production of heavy quarks and direct photons, to
be discussed in subsequent subsections. We therefore treat jet production in
somewhat more detail than other photoproduction processes.

Feynman diagrams contributing to jet production in \gamp\ scattering are
sketched in Fig.~5. In leading order (LO), direct contributions (Fig.~5a) come
from either the ``QCD Compton" process (left diagram), or from photon--gluon
fusion (right diagram); notice that, unlike DIS, this latter process is
sensitive to the gluon content of the proton already in LO. The resolved
photon contributions (Fig.~5b) involve the same matrix elements for $2
\rightarrow 2$ QCD scattering processes that appear in calculations of jet
production at purely hadronic (\ppbar\ or $pp$) scattering. Note that, in
contrast to the direct processes, many of these QCD scattering processes can
proceed via the exchange of a gluon in the $t-$ or $u-$channel; processes like
$gg \to \qqbar$ that only proceed via the exchange of a quark in the $t-$ or
$u-$channel, and/or the exchange of a gluon in the $s-$channel, contribute
only little to the inclusive jet cross--section.

In LO, the cross--section for the electroproduction of two (partonic) jets
with transverse momentum $p_T$ and (pseudo)rapidities\footnote{There is no
difference between the partonic rapidity and pseudorapidity in LO.} $\eta_1,
\ \eta_2$ can be written as:
\be \label{e3.1}
\frac {d^3 \sigma (e p \to e j_1 j_2 X)} {d p_T d \eta_1 d \eta_2} = 2 p_T x_e
x_p \sum_{i,j,k,l} f_{i|e}(x_e) f_{j|p}(x_p) \frac {d \hat{\sigma}_{ij \to kl}
(\hat{s},\hat{t},\hat{u})} {d \hat{t}}.
\ee
Here, $i$ stands for a photon, quark or gluon, and $j, \ k, \ l$ stand for a
quark or gluon. If $i=\gamma$ (direct contribution), the function $f_{i|e}$ is
just the Weizs\"acker--Williams photon flux (\ref{e2.10}); otherwise it is
given by
\be \label{e3.2}
f_{i|e}(x_e) = \int_{x_e}^1 \frac {dz}{z} \fgme(z) f_{i|\gamma} \left( \frac
{x_e}{z} \right) .
\ee
The resolved photon contribution to the cross--section (\ref{e3.1}) therefore
only depends on the product $x_e$ of the fraction $z$ of the electron energy
carried by the incident photon, and the fraction \xgam\ ($=x_e/z$) of the
photon energy carried by the parton in the photon. Of course, $f_{i|\gamma}$
and $f_{j|p}$ are nothing but the parton densities in the photon and proton,
respectively.

The (pseudo)rapidities of the jets are related to these Bjorken variables
by:
\ben \label{e3.3} \beq
x_p &= \frac {1}{2} x_T \sqrt{ \frac{E_e}{E_p} } \left( e^{\eta_1} +
e^{\eta_2} \right) ; \label{e3.3a} \\
x_e &= \frac{1}{2} x_T \sqrt{ \frac{E_p}{E_e} } \left( e^{-\eta_1} +
e^{-\eta_2} \right) \label{e3.3b},
\eeq \een
with $x_T = 2 p_T / \rs$, where \rs\ is the $ep$ centre--of--mass energy; note
that in our convention positive rapidities correspond to the direction of the
proton. Finally, the subprocess cross--sections $\hat{\sigma}$ for the direct
\cite{53,50} and resolved photon \cite{54} contributions depend on the
Mandelstam variables describing the hard partonic scattering, with $\hat{s} =
x_e x_p s$ and
\be \label{e3.4}
\hat{t} = - \frac{\hat{s}}{2} \left( 1 \pm \sqrt{ 1 - \frac {4 p_T^2}
{\hat{s}} } \right);
\ee
both solutions in eq.(\ref{e3.4}) have to be included when evaluating
eq.(\ref{e3.1}), which can be accomplished by simply symmetrizing the
subprocess cross--sections under $\hat{t} \leftrightarrow \hat{u}$.

Direct and resolved photon contributions to the photoproduction of jets were
first compared by Owens \cite{50} in 1979 for fixed target energies, where
the resolved photon contributions were found to be subdominant, except at very
small $p_T$. Although a few theoretical analyses \cite{55a,55} of the
photoproduction of jets appeared in the first half of the 1980's, the
importance of resolved photon contributions was fully appreciated only in
1987, when it was realized \cite{49,56b,56} that at HERA energies, they could
exceed the direct contribution by as much as a factor of ten at $p_T \simeq 5$
GeV, and remained dominant out to $p_T \simeq 35$ to 40 GeV.

An update \cite{56a} of this result is shown in Fig.~8, where the ratio of
resolved photon and direct contributions to the single jet inclusive
cross--section\footnote{Recall that by definition, events with two accepted
jets count twice here.} is plotted for the nominal HERA energy $\rs = 314$
GeV. Unlike in ref.\cite{56}, we have imposed some acceptance cuts, taken from
a recent ZEUS analysis \cite{57}: The jet has to fall in the pseudorapidity
range $-1 \leq \eta_{\rm jet} \leq 2$, and the ``antitag" requirement that
the outgoing electron is not seen in the main detector implies that the photon
virtuality $Q^2 \leq 4$ GeV$^2$. Most results in Fig.~8 have been obtained
using the MRSD-' parametrization \cite{58a} for the parton densities in the
proton; comparison between the solid and the dotted curve, which is for the
MRSD0' parametrization, shows that even the pre--HERA uncertainty of nucleon
densities only leads to an uncertainty of a few percent in the ratio of
Fig.~8, except at very small $p_T$. HERA data on deep--inelastic scattering
have since then improved our knowledge of the nucleon structure considerably;
the impact of these data on parametrizations of $G^p(x)$ is still under
investigation \cite{60}, but it is already clear that very soon the
uncertainty from the parton densities in the nucleon will be negligible
compared to the differences between predictions based on various
parametrization of \qvph. In particular, HERA data clearly favour MRSD-' over
MRSD0'; we therefore take the former as our standard choice.

We see that implementation of the acceptance cuts reduces the region where
resolved photon contributions are dominant to $p_T \leq 25$ to 30 GeV; this is
mostly due to the upper limit on $\eta_{\rm jet}$, which reduces the resolved
photon contribution much more than the direct one (see below). Nevertheless,
the former still exceeds the latter by a factor between 5 and 11 at $p_T = 5$
GeV; at present our lack of knowledge of \qvph\ does not allow us to predict
this ratio more precisely. This dominance of resolved photon contributions at
small $p_T$ can partly be explained by the fact that they get contributions
from gluon exchange in the $t-$channel, see Fig.~5b; this enhances the squared
matrix elements for resolved photon processes by a factor $\hat{s}/|\hat{t}| >
2$, compared to those for direct processes. Colour factors generally also
favour the former over the latter. Finally, we saw in Figs.~4 that for small
\xgam, the parton densities in the photon can actually exceed $1/\aem$
substantially; for small $x_e$, which contribute only at small $p_T$, the
integral in eq.(\ref{e3.2}) can therefore enhance resolved photon
contributions even further.\footnote{Recall that the factor of \aem\ contained
in \qvph\ is cancelled in the ratio by the explicit factor of \aem\ appearing
in the $\hat{\sigma}$ for direct processes.} On the other hand, this
convolution integral decreases more rapidly with increasing $x_e$ than the
photon flux factor \fgme\ does; this explains the more rapid decrease of
resolved photon contributions with increasing $p_T$, and thus the shape of the
curves in Fig.~8.

Obviously parametrizations with sizable \gga\ (LAC1, WHIT4) predict a
considerably larger resolved photon contribution at small $p_T$ than those
with smaller \gga\ [WHIT1, WHIT3; DG, GRV, AGF and GS2 (not shown) also belong
to this class]. Note that the prediction from the WHIT1 parametrization
exceeds that using WHIT3 even at the smallest $p_T$ shown; this indicates
that the $p_T$ distribution integrated over jet rapidities is not sensitive to
very small \xgam, where the WHIT3 gluon density exceeds that of WHIT1, see
Fig.~4b. Finally, we should warn the reader that additional experimental cuts
can change the ratio of direct and resolved photon contributions
substantially. This is demonstrated by the dot--dashed curve, which has been
obtained with the same parametrizations of parton densities in the proton and
photon as the solid curve, but where we have demanded that the outgoing
electron be detectable in the ZEUS luminosity monitor; this implies $Q^2 <
0.01$ GeV$^2$ and, more importantly, $0.2 \leq z \leq 0.75$. The lower cut on
the scaled photon energy $z$ greatly reduces direct contributions at low
$p_T$, while the upper cut reduces resolved photon contributions at high $p_T$
more than direct ones; as a result, the $p_T-$dependence of the ratio of the
two contributions becomes considerably steeper.

In Fig.~9 we further split the resolved photon (9a) and direct (9b)
contributions depending on whether the two high$-p_T$ partons in the final
states are of two quarks, two gluons, or a quark and a gluon. Since the
contributions from $\qqbar \to gg$ and $gg \to \qqbar$ are very small, the
curves in Fig.~9a can also be read as coming from $qq, \ gg$ and $qg$ initial
states, while in Fig.~9b the $qg$ and $qq$ final states come from $\gamma q$
and $\gamma g$ initial states, respectively. We have used the WHIT1
parametrization for \qvph, and applied the same acceptance cuts as in Fig.~8.

WHIT1 assumes a relatively small \gga; as a result, the $gg$ final state is
dominant only for $p_T \leq 5$ GeV. However, even the WHIT4 parametrization,
which assumes a two times larger input distribution $G_0^{\gamma}$, predicts
the cross--section for the $gg$ final state to be well below that for the $qg$
final state for $p_T > 10$ GeV. Note that in the majority of $qg$ events the
quark comes from the photon and the gluon from the proton; this is partly
because the photon is assumed to be relatively poor in gluons, and partly
because the cut $\eta_{\rm jet} \leq 2$ removes many events with a gluon in
the photon in the initial state, as discussed below. We saw in sec.~2 that the
quark densities in the photon are much better known than \gga, except at small
\xgam. Together with the result of Fig.~9a, this explains the rapid
convergence of the curves in Fig.~8, although some difference between the LAC1
and WHIT predictions persists even at large $p_T$ (see also
Fig.~3\footnote{Notice, however, that all quark flavours contribute equally to
the jet cross--section, while contributions to \f2gam\ are weighted by the
squared charge. Two parametrizations can therefore have very similar \f2gam\
and yet lead to different predictions for quark--initiated jet production at
HERA.}). Finally, the comparison of Figs.~9a and 8 shows that two quark final
states dominate resolved photon contributions only for values of $p_T$ where
the total jet cross section is already dominated by direct contributions.

As expected from Fig.~8, the direct contributions shown in Fig.~9b have a
considerably flatter transverse momentum spectrum than the resolved photon
contributions. Unfortunately, photon--gluon fusion (the two quark final state)
dominates the direct contribution only at relatively small $p_T$; this is
because the gluon density in the proton is softer (i.e., decreases more
rapidly with increasing $x_p$) than the valence quark distributions. Since at
small $p_T$ the inclusive jet cross section is dominated by resolved photon
contributions, a direct study of photon--gluon fusion, which might allow to
further constrain the gluon density in the proton even at rather small $x_p$
\cite{58}, will be difficult unless the resolved photon contribution can be
suppressed by additional cuts.

As shown in Figs.~10, the (pseudo)rapidity distribution of the jet can be
used to help disentangle direct and resolved photon contributions. In Fig.~10a
we have used the present HERA energy $\rs= 296$ GeV, and applied the same cuts
on the parton level that the ZEUS collaboration applied \cite{57} on their
reconstructed jets. Unfortunately the $p_T$ cut chosen is still too low to
allow a direct comparison between partonic and jet cross sections, since a
substantial part of the jets still comes from the ``underlying event" (beam
fragments, initial state radiation, and possibly multiple interactions
producing minijets \cite{59,210}; see sec.~5); these effects are expected to
become more important as one approaches the proton beam direction, which
corresponds to $\eta = + \infty$ in our convention, and can thus distort the
jet rapidity distribution compared to the parton--level distribution shown in
Figs.~10. We nevertheless expect that the {\em differences} between the curves
shown in these figures will not be washed out by the underlying event.

We see that in this single--differential cross section the direct contribution
is always subdominant; however, one can enhance its relative importance by
requiring $\eta_{\rm jet} < 0$. Of more interest for us is the opposite region
of positive, sizable $\eta_{\rm jet}$, which is sensitive to small $x_e$, see
eq.(\ref{e3.3b}), and thus allows to probe the parton densities in the photon
at small \xgam, see eq.(\ref{e3.2}). Indeed, the figure shows that in this
region one is most sensitive to differences between the various
parametrizations of \qvph. We also observe again that these differences are
much larger than even the pre--HERA uncertainty from nucleonic structure
functions.

We have already noted above that the jet rapidities only depend on the product
$x_e = z \cdot \xgam$; events with large $\eta_{\rm jet}$ can come from
soft partons in hard photons (small \xgam, large $z$), but also from hard
partons in soft photons (large \xgam, small $z$). Fortunately, ZEUS has
demonstrated \cite{57} the ability to (approximately) determine $z$ purely
from the longitudinal momenta of particles in the main detector, without
having to detect the outgoing electron in the luminosity monitor, which would
reduce the accepted event rate by about a factor of four. This should allow
one to increase the lower cut on $z$ from 0.2 to 0.5, which enhances the
relative importance of contributions with small \xgam, so that the differences
between predictions based on different ans\"atze for \qvph\ become larger,
as illustrated in Fig.~10b. (H1 prefers to only use photoproduction events
where the outgoing electron is tagged in the small--angle detector; $z$ can
then be determined from its energy.) Although this stronger cut on $z$ reduces
the resolved photon contribution by slightly more than a factor of two for
$\eta_{\rm jet} \geq 1$, it should still enhance the discriminative power of
this measurement, given that the ZEUS data sample \cite{57} with the looser
cuts contains almost 20,000 events with reconstructed jets.

The integration over the rapidity of the second jet in single--jet inclusive
cross sections leads to a substantial spread in $x_e$ and $x_p$, see
eqs.(\ref{e3.3})\footnote{The integration over $p_T$ is less important here,
since most events will have $p_T \simeq p_{T, {\rm cut}}$ anyway.}; in order
to further increase the sensitivity of the jet rate to the region of small
\xgam\ in general and the gluon density in the photon in particular, one
therefore has to study more differential cross--sections. In Figs.~11a,b we
show predictions for the triple--differential di--jet cross section, as given
by eq.(\ref{e3.1}), for $\rs=314 \ {\rm GeV}, \ p_T = 10$ GeV and $\eta_1 =
\eta_2 \equiv \eta$. As in Fig.~9a we display resolved photon contributions
with different final states separately, as predicted from the WHIT1
parametrization. In Fig.~11a we have applied the antitag cut $Q^2 \leq 4$
GeV$^2$ on the virtuality of the photon, but we have not restricted the
allowed range of the scaled photon energy $z$. We see that the direct
contribution now dominates at $\eta < 0$, and remains sizable even for $\eta =
2$; in the region of negative $\eta$ it is mostly due to photon--gluon fusion.
Note that our choice $\eta_1 = \eta_2$ implies $\hat{t}=\hat{u} = - \hat{s} /
2$, which minimizes the dynamical enhancement factor $\hat{s} / |\hat{t}|$ of
most resolved photon contributions. Together with the slightly larger value of
$p_T$ this explains why direct contributions appear more prominent in Fig.~10a
than in Fig.~9a.

Contributions from the gluon in the photon peak at $\eta \simeq +2$. The
reason is that a gluon usually only carries a rather small fraction of the
photon energy, see Fig.~4a; for given $\hat{s} \geq 4 p_T^2$, a large
contribution to the total energy in the hard sub--process then has to come
from the proton, leading to a rather strong boost of the high$-p_T$ partons in
the proton direction. The quarks in the photon can be substantially more
energetic, and therefore already contribute at $\eta \simeq 0$.

Although according to the WHIT1 parametrization, the contribution from gluons
in the photon is clearly enhanced at large $\eta$, at $\eta=2$ slightly more
than half of the total cross--section still comes from quarks in the photon or
direct processes. As before, the sensitivity to the region of small \xgam\ can
be further enhanced by imposing cuts on the scaled photon energy $z$. This is
illustrated in Fig.~11b, where we have required $0.3 \leq z \leq 0.8$. The
lower limit has been chosen such that the direct contribution vanishes
identically for $\eta > 0$, see eq.(\ref{e3.3b}). (Recall that $x_e=z$ for
direct contributions, while $x_e = z \xgam$ otherwise.) Moreover, the relative
importance of contributions from gluons in the photon is clearly enhanced
compared to Fig.~11a; in particular, the $gg$ final state is now dominant for
$\eta \simeq 2$, where only $\sim 20 \%$ of the total cross section is
predicted to come from quarks in the photon. Recall that \gga\ might well be
larger than assumed in the WHIT1 parametrization, so that the contributions
from gluons in the photon might be even larger. It should therefore be
possible to derive stringent constraints on \gga\ even using jets with
relatively large $p_T$, by focussing on events with large rapidity and large
$z$. Of course, the sensitivity to resolved photon contributions involving
soft partons in the photon is even larger \cite{60a} if one can measure a
cross--section for fixed $z$; however, this will need a rather large event
sample.

If $z$ and both jet rapidities have been measured, one can in principle
reconstruct both $x_p$ and \xgam, using eqs.(\ref{e3.3}). In practice, HERA
experiments use an estimator for \xgam\ to separate the events into ``direct''
and ``resolved'' samples; this estimator reduces to \xgam\ in LO QCD, but only
uses the measured rapidities and transverse energies of the jets, and can thus
also be computed in NLO (where it will differ from the partonic \xgam). The
ZEUS collaboration \cite{61} has used 284 reconstructed di--jet events from
the first run of HERA to show that the \xgam\ distribution has a peak near
$\xgam = 1$, as expected from direct contributions. The H1 collaboration
\cite{62} has gone one step further and subtracted the contribution from
quarks in the photon, estimated using the GRV parametrization. They find
evidence for a non--vanishing contribution from gluons in the photon only for
$\xgam \leq 0.2$, thereby ruling out the LAC3 parametrization (see Fig.~4a).
H1 reconstructs \gga\ in the range $0.02 \leq \xgam \leq 0.2$ at an effective
scale $\mu^2 \simeq 60$ GeV$^2$; the result is in good agreement with GRV, but
disfavours LAC1. However, a fairly sophisticated Monte Carlo analysis was
necessary to extract \gga\ even using the LO formalism. For example, we saw in
the discussion of Fig.~10a that parts of the ``underlying event" contribute to
the reconstructed jets, which obscures the relation between hard partons and
jets. The H1 analysis \cite{62} depends on such details quite sensitively. It
might therefore be somewhat premature to exclude the LAC1 parametrization on
the basis of this evidence alone.

The ZEUS collaboration has also published \cite{62b} an updated jet analysis,
based on 12,000 reconstructed di--jet events with $E_T($jet$) \geq 6$ GeV.
They study $d \sigma / d \overline{\eta}$  for 4,000 events with $|\eta_1 -
\eta_2| \leq 0.5$, where $\overline{\eta}$ is the average pseudorapidity of
the two jets; this quantity is closely related to the double differential
cross section $d^2 \sigma / (d \eta_1 d \eta_2)$ at $\eta_1 = \eta_2$. They
find reasonable agreement between their data and LO QCD predictions in the
``direct" sample (events with large measured \xgam), but most parametrizations
of \qph\ give too low predictions for the ``resolved" sample. This again
indicates that at this rather low value of $E_T$ the ``underlying event" plays
quite an important role; as we will discuss in more detail in Sec.~5, this
aspect is not well described by present standard QCD Monte Carlo event
generators.

\setcounter{footnote}{0}
Direct and resolved photon contributions are also expected to have different
distributions in $\cos \!\theta^*$, where $\theta^*$ is the cms scattering
angle \cite{62a}. Due to diagrams where a gluon is exchanged in the $t-$ or
$u-$channel, resolved photon contributions are more strongly peaked at small
$\theta^*$ than direct contributions. The increasing importance of the
latter over the former at higher $p_T$ means that the $\cos \! \theta^*$
distribution of di--jet events with large $p_T$, will be flatter than at low
$p_T$.

As remarked earlier, the most obvious distinction between direct and resolved
photon events is that only the latter contain a photonic spectator (or
remnant) jet, see Fig.~6. In Fig.~12 we show the average energy of this jet in
the lab frame, as predicted from the WHIT1 parametrization; on the parton
level, this energy is simply given by $E_e \cdot z \cdot (1-\xgam)$. For
negative $\eta$ (of the high$-p_T$ jets), the spectator jet is rather soft,
since both $z$ and \xgam\ have to be large, so that $1-\xgam$ is quite small.
For $\eta \simeq 0$, events with a gluon from the photon usually have quite
large $z$, but moderate \xgam, yielding a high spectator jet energy; it
declines at large $\eta$ since the average $z$ becomes smaller.
Quark--initiated events typically have considerably larger \xgam\ and hence a
softer photonic spectator jet. Obviously the average spectator jet energy will
increase (decrease) if a lower (upper) cut on $z$ is applied.

In principle the results of Fig.~12 offer another possibility to enhance the
contributions from gluons in the photon, by requiring the presence of an
energetic remnant jet in the electron direction. However, in practice the
energy of this jet cannot be measured very accurately, since some part of it
will usually be lost in the beam pipe. It should nevertheless be emphasized
that the first analysis of jet data \cite{51} taken during the HERA pilot run
found substantial energy deposition in the backward calorimeter even if all
high$-p_T$ jets have positive rapidities; this can be understood only if the
photonic remnant jet is included in the MC simulation. This jet has recently
been studied in more detail by the ZEUS collaboration \cite{62c}; we will
discuss their results in Sec.~5.

Recently new data from the Fermilab fixed target photoproduction experiment
E683 have been published \cite{63}. It uses a tagged photon beam with mean lab
energy of 260 GeV, giving a mean \rs\ of slightly over 20 GeV. A Monte Carlo
analysis suggest that their di--jet sample, required to have two reconstructed
jets with average $p_T > 4$ GeV, gets approximately equal contributions from
direct and resolved photon contributions, in agreement with theoretical
expectations. However, unlike at HERA, no direct experimental evidence for the
existence of resolved photon contributions could be established (other than
the overall event rate). In particular, the hadronic energy flow in the very
forward direction was found to be quite similar for direct and resolved photon
events, and even for di--jet events produced from a pion beam; this somewhat
counter--intuitive result can be explained in terms of jet fluctuations.

So far all our predictions have been computed in LO in QCD. As well known, the
overall normalization of such predictions is uncertain, since in the leading
log approximation used here, one cannot with certainty determine the values of
the factorization scales in the parton distribution functions and the
renormalization scale appearing in the running QCD coupling constant. These
scales have no physical significance; predictions would be independent of them
if all orders of perturbation theory could be summed. One therefore expects
reduced scale dependence already in next--to--leading order (NLO). Moreover,
many quantities can be predicted meaningfully only if one allows at least
three high$-p_T$ partons in the final state; these include the dependence of
jet cross--sections on the jet definition, and the distribution in the
transverse opening angle between the two (hardest) jets in events with (at
least) two jets.

The first step towards a full NLO calculation of jet photoproduction was taken
in 1980 with a calculation \cite{64} of direct $2 \rightarrow 3$ cross
sections (e.g. $\gamma q \rightarrow qgg$ and $\gamma g \rightarrow \qqbar
g$). A first complete NLO calculation of the direct contribution, including
virtual (1--loop) corrections, was performed \cite{65} in 1986. These results
were applied to jet production at fixed target energies in \cite{66}, and at
HERA energies in \cite{67}. An NLO prediction of resolved photon contributions
to jet production become possible only after corrections to the hard partonic
QCD cross--sections had been computed \cite{68}. These results were applied to
jet production from resolved photons at HERA in ref.\cite{69}. Finally, in
refs.\cite{70,71,72,73}, complete NLO calculations for single--jet inclusive
jet cross--sections were presented, including both direct and resolved photon
contributions.

Typical results are presented in Fig.~13, adapted from B\"odeker et al.
\cite{72}. We show the scale dependence of the predicted jet cross--section
at HERA for $E_T=25$ GeV and $\eta_{\rm jet} = 1.5$. Note that $E_T$ and $p_T$
are in general no longer identical in NLO, since now a jet might be made up of
two partons. In Fig.~13 two partons have been merged into a single jet if
$\Delta R \equiv \sqrt{ \left( \Delta \eta \right)^2 + \left( \Delta \phi
\right)^2 } \leq 0.7$.

The solid and short dashed curves have been obtained by setting all three
scales (the renormalization scale, and the factorization scales for the
photonic and nucleonic parton densities) equal to each other. As expected, the
LO prediction exhibits a much stronger scale dependence than the NLO result.
It is worth noting, however, that for the ``natural" choice $\mu = p_T$ (or
$E_T$ in NLO), LO and NLO predictions almost coincide. Generally the
difference between LO and NLO predictions was found to be quite small at
HERA energies if $\mu = p_T$ has been chosen and jets are defined with
$\Delta R$ between about 0.7 and 1.\footnote{There appears to be some
discrepancy between the results of ref.\cite{71} and the earlier calculations
\cite{69,70}; this is now being sorted out (M. Greco, private communication).}
This means that the results of Figs.~8--12 should not be affected too much by
NLO corrections.

As already discussed in the beginning of this section, in NLO the distinction
between direct and NLO contributions is blurred.  Diagrams like the one shown
in Fig.~7 contribute to direct jet production in NLO, but they also contain
a logarithmically divergent piece which has already been included in the LO
resolved photon contribution; this piece therefore has to be subtracted from
the NLO direct contribution. This subtraction term grows logarithmically with
the photonic factorization scale $M_{\gamma}$, which is also the scale
appearing in the photonic parton distribution functions. In NLO the direct
contribution (dotted curve) therefore decreases with increasing $M_{\gamma}$,
while the resolved photon contribution (dot--dashed curve) increases. The sum
of the two (long dashed) is nearly independent of $M_{\gamma}$; the dependence
does not cancel completely since in NLO the subtraction term is exactly
proportional to $\log M_{\gamma}$, while the photonic parton densities only
increase approximately like $\log M_{\gamma}$, as discussed in Sec.~2.

We saw above that di--jet cross sections that are differential in both jet
rapidities are more powerful discriminators of photonic parton densities than
single--jet inclusive cross sections. Unfortunately, in conventional NLO
calculations going from single--jet to di--jet cross sections introduces
considerable complications;\footnote{This step should be much easier using the
Monte Carlo method of ref.\cite{66}.} as a result, only the direct production
of jet pairs has been treated in NLO to date \cite{74}. The urgent need for a
full calculation is emphasized by the recent ZEUS data \cite{62b}, which have
the potential to discriminate between different parametrizations of photonic
parton densities once the cross section can be predicted with some confidence.

Recently the calculation of jet rates at HERA has been further refined by
including two additional effects. First, eq.(\ref{e3.3a}) shows that (direct)
jet production at negative rapidities probes the proton structure at quite low
values of $x_p$; for $p_T \leq 10$ GeV, partons with $x_p$ as low as $10^{-3}$
contribute. Small$-x$ effects may then become important, and one may have to
use \cite{75} the so--called $k_T$ factorization \cite{76}. Secondly, if one
defines the photoproduction sample with a rather moderate (no--tag) cut on the
outgoing electron, one includes contributions where the photon virtuality
$Q^2$ may not be entirely negligible; recall, for example, that the ZEUS cuts
\cite{57} include events with $Q^2$ up to 4 GeV$^2$. On the other hand, when
using the simple Weizs\"acker--Williams approximation (\ref{e3.2}), one
assumes $Q^2$ to be small compared to all other scales in the problem. This is
still a good approximation for the direct contribution in this case, but the
parton densities in the photon become suppressed \cite{29,77} once $Q^2 >
\Lambda^2_{\rm QCD}$. The simple factorization (\ref{e3.2}) then breaks down,
but one can still define a ``parton density in the electron", which will
depend on the experimental cut on $Q^2$ \cite{78}. For the ZEUS cuts, this
suppression only amounts to a few percent.

Finally, as mentioned in the discussion of Fig.~10, so far jet production at
HERA could only be investigated experimentally at rather moderate transverse
momenta, where the ``underlying event" can still contribute significantly to
the reconstructed jets. The influence of the underlying event might be less
problematic when one studies the production of high$-p_T$ particles
\cite{79,65} rather than jets. On the other hand, one now has to specify not
only parton distribution functions, but also fragmentation functions, before
a prediction can be made. Also, the cross--section falls off very rapidly with
$p_T$, forcing one to work at scales of only a few GeV where the convergence
of the perturbative expansion might be rather slow. Moreover, one either has
to experimentally identify different particle species (chiefly pions and
kaons, if only charged particles are counted), or make assumptions about the
relative abundances of these species. Recent measurements of the
cross--section for the production of charged particles with $p_T \geq 1.5$
GeV, by both the H1 and ZEUS collaborations \cite{80}, find good agreement
with a theoretical NLO prediction \cite{81} as far as the $p_T$ spectrum is
concerned; the pseudorapidity distribution measured by H1 is less well
described, but the experimental errors do not allow to draw a definite
conclusion at this point.

\setcounter{footnote}{0}
\subsection*{3b) The Photoproduction of Heavy Quarks}
The production of heavy quarks offers two theoretical advantages over the
production of light partons (jets) discussed in the previous subsection.
First, their large mass $m_Q \gg \lqcd$ ensures that QCD perturbation theory
is applicable in all of phase space, although nonperturbative corrections
$\propto \left( \lqcd / m_Q \right)^{n\geq 1}$ might not be negligible for
charm quarks. In particular, the total cross--section without any cuts could
now be predicted with some reliability if the values of certain parameters
($m_Q, \ \lqcd$ and the parton distribution functions) were known precisely.
Secondly, at least in leading order the number of contributing partonic
processes is much smaller, making heavy quark production easier to analyze.
Specifically, in LO only photon--gluon fusion contributes to \QQbar\
production from direct photons, while the relevant resolved photon processes
are $gg$ fusion and light \qqbar\ annihilation.\footnote{It can be argued that
at very high transverse momentum, $p_T \gg m_Q$, $\alpha_s \log p_T/m_Q$
should be counted as ${\cal O}(1)$, rather than ${\cal O}(\alphas)$. In this
case the ``excitation" processes $Qg \to Qg$ and $Qq \to Qq$ also contribute
in LO, since the $Q-$quark density in the photon grows logarithmically with
the hard scale of the process. In these reactions the heavy quark jet is
balanced by a light quark or gluon jet, while in \QQbar\ creation events two
heavy quark jets occur with equal and opposite $p_T$. These ``flavour
excitation" contributions are now under study \cite{82}.}

In LO, the \QQbar\ production cross--section in $ep$ scattering is still
given by eqs.(\ref{e3.1})--(\ref{e3.3}). However, $\eta_{1,2}$ now have to be
interpreted as true rapidities, which differ from the pseudorapidity for
massive particles; moreover, $x_T$ in eqs.(\ref{e3.3}) is now given by $2
\sqrt{ p_T^2 + m_Q^2 } / \rts$, and eq.(\ref{e3.4}) has to be replaced by
\be \label{e3.5}
\hat{t} = \frac {\hat{s}} {2} \left[ \frac {2 m_Q^2} {\hat{s}} - 1 \pm
\sqrt{ 1 - \frac {4 \left( m_Q^2 + p_T^2 \right)} {\hat{s}} } \right].
\ee
The partonic cross--sections $\hat{\sigma}$ for $\gamma g \to \QQbar$ and
$gg,\qqbar \to \QQbar$ can be found in refs.\cite{83} and \cite{84,84a},
respectively.

Resolved photon contributions to heavy quark production were first treated in
ref.\cite{39}, for the case of the top quark, whose mass was then believed to
be in the vicinity of 40 GeV. We now know \cite{85} that the top quark is too
heavy to be produced at HERA, but \ccbar\ and \bbbar\ pairs will be produced
copiously. It was pointed out in refs.\cite{49,86} that (for most
parametrizations of \qph) resolved photon contributions to the total
cross--sections are subdominant, but not negligible; e.g., they amount to
$\sim 20\%$ for the DG parametrization. Resolved photon contributions to the
production of heavy quark pairs are therefore considerably less important than
for high$-p_T$ jet production. The reason is that now the resolved photon
processes also only involve gluon exchange in the $s-$channel or (heavy) quark
exchange in the $t-$ or $u-$channel; there is no enhancement factor
$\hat{s}/|\hat{t}|$, unlike for jet production. Moreover, the two
sub--processes involving the parton content of the photon have rather small
colour factors, again in contrast to the matrix elements appearing in resolved
photon contributions to jet production. These analyses also showed that $gg$
fusion is predicted to dominate over \qqbar\ annihilation, by a factor of 10
(3.5) in case of the DG parametrization and \ccbar\ (\bbbar) production. The
resolved photon contribution therefore offers a good opportunity to constrain
\gga, while (in LO) the direct contribution is proportional to $G^p$.

We saw in the previous subsection that at HERA, contributions from gluons
in the photon are most important at sizable, positive rapidities, but are
suppressed at negative rapidity. The rapidity distribution therefore offers
a good handle for separating the two contributions to \QQbar\ production. In
Fig.~14 we show the $p_T$ spectrum of $c$ and $b$ quarks at central rapidity,
$y_1 = y_2 = 0$. We see that the resolved photon contribution (difference
between solid and dashed curves) only amounts to 10--15\% at small $p_T$, and
becomes entirely negligible for $p_T > 5$ GeV. In fact, as shown in
ref.\cite{56}, the resolved photon contribution to heavy quark production at
HERA can always be suppressed to an insignificant level by requiring (at
least) {\em one} of the two heavy quarks to emerge at $y<0$.\footnote{This
result does not hold for the LAC3 parametrization; fortunately, this
parametrization is excluded by other data, as discussed in secs.~3a and 4a.}
Note that we have chosen the WHIT4 parametrization in Fig.~14, which is
characterized by a rather large and hard \gga; it predicts that resolved
photon contributions amount to more than 20\% (30\%) of the total \ccbar\
(\bbbar) cross--section at HERA. This contribution is more important for
\bbbar\ production, since the direct contribution is suppressed by the small
charge of the $b$ quark; for parametrizations with rather hard \gga\ (WHIT1,4,
DG, GS2, GRV) this suppression is stronger than the relative reduction of the
resolved photon contributon with increasing $m_Q$, which is caused by the
additional convolution (\ref{e3.2}) with the gluon density in the photon.

Since resolved photon contributions to \QQbar\ production are insignificant at
high $p_T$ (unless their importance is enhanced by specific cuts, as discussed
below), the ratio of \ccbar\ to \bbbar\ cross--sections in the region $p_T
\geq 10$ GeV simply reflects the ratio of their squared charges. This relative
suppression of the \bbbar\ cross--section means that $b-$tagging at HERA will
be significantly more difficult than it is at \ppbar\ colliders, where \bbbar\
and \ccbar\ cross--sections become equal at high $p_T$. At \ppbar\ colliders
the harder fragmentation function of $b-$flavoured hadrons \cite{87} means
that inclusive high$-p_T$ muon production is dominated by \bbbar\ events; this
will not be true at HERA, however, except at very high $p_T$ where the
cross--section is quite small. Charm quarks will also be a serious background
to $b-$tagging by micro--vertex detectors at HERA.

As mentioned earlier, the resolved photon contribution is concentrated at
positive rapidities. Just as in the case of jet production it can be isolated
by a cut on the scaled incident photon energy $z$; e.g., requiring $z>0.3$ at
$p_T=10$ GeV removes all direct contributions with $y_1=y_2>0$. Fig.~15 shows
that the remaining resolved photon contribution is indeed very sensitive to
the gluon content of the photon. Even according to the WHIT1 parametrization
the \qqbar\ annihilation contribution (short dashed) is considerably below the
one from $gg$ fusion (long dashed). The WHIT4 parametrization therefore
predicts a considerably larger cross--section, but the shape of the rapidity
distribution is similar to that predicted from WHIT1. In contrast, the LAC1
prediction differs in both normalization and shape. However, we should warn
the reader that without the cut on $z$, even at $y_1 = y_2 = 2$ the direct
contribution would be at least ten times larger than the resolved one; the
experimental implementation of this cut therefore has to be very efficient. It
might even be necessary to require the presence of a photonic remnant jet to
extract the resolved photon contribution; recall that this jet is expected to
be quite energetic in events that originate from the gluon content of the
photon. Recall also that there will be a large contribution from ``charm
excitation" \cite{82} if only one of the two high$-p_T$ jets is tagged as a
heavy quark. Finally, the cross--section shown in Fig.~15 is quite small, even
though we have not yet required any specific charm signal (e.g., a hard muon
or reconstructed $D^*$ meson). Clearly HERA will have to accumulate
significantly more data than the present 6 pb$^{-1}$ (as of the end of 1994)
to measure such triple--differential cross--sections even at lower $p_T$.

The predictions shown in Figs.~14 and 15 were computed in LO. NLO calculations
of the photoproduction of heavy quarks exist \cite{86,88,89}; unlike for jet
production, even the fully differential cross--section is available in NLO
\cite{90}. It has been demonstrated \cite{91} that the direct contribution can
be extracted reliably also in NLO, i.e. the sensitivity to the gluon content
of the proton is not degraded. In view of their smaller size, extraction of
the resolved photon contribution might prove more difficult. In particular, in
direct events with an additional hard parton in the final state ($\gamma g \to
\QQbar g, \ \gamma q \to \QQbar q$) the heavy quarks can occur at large
positive rapidity even after a cut on $z$ has been applied; one may have to
veto the occurence of additional high$-p_T$ jets, and/or require the two
highest $p_T$ jets to be back--to--back in the transverse plane, in order to
efficiently suppress such backgrounds to the cross--section shown in Fig.~15.

The considerable body of data on photoproduction of charm at fixed target
energies ($\rts \leq 20$ GeV) is well described by NLO QCD calculations
\cite{92}. However, at these low energies the resolved photon contribution is
quite small; it is significant only on the backward direction (opposite to the
incident photon) \cite{90}, where the experimental acceptance is poor.

Very recently, first data on charm production at HERA have become available.
The ZEUS collaboration \cite{93} searched for fully reconstructed $D^{\pm*}$
mesons. They observe a signal of $48 \pm 11$ events within the acceptance
region $p_T(D^*) > 1.5$ GeV, $|\eta(D^*)| < 1.5$; this corresponds to
$\sigma(ep \to D^{\pm*} X) = (32 \pm 7 \ {}^{+4}_{-7})$ nb at $\rts=296.7$ GeV
with $Q^2 \leq 4$ GeV$^2$ and $0.15 \leq z \leq 0.86$. They then attempt to
estimate the {\em total} \ccbar\ cross--section from this measurement; this,
however, sensitively depends on the extrapolation of the cross--section into
kinematic regions where it has not been measured, which introduces a strong
dependence of the ``measured" \ccbar\ cross--section on the assumed parton
distribution functions in the proton and photon as well as on $m_c$. This not
only greatly increases the quoted (systematic) error; even the central value
depends on these assumptions. It does not make much sense to compare this
``measured" cross--section with different theoretical predictions, since
``measurement" and ``prediction" depend on the same quantities!  Notice also
that the prediction for the total \ccbar\ cross section suffers from large
uncertainties \cite{95}. Since $\alphas(m_c)$ is still quite large, the
perturbative expansion only converges slowly, which manifests itself in a
rather strong dependence of even the NLO prediction on the factorization and
renormalization scales. Moreover, the prediction is very sensitive to $m_c$,
decreasing by more than a factor of three when $m_c$ is increased from 1.2 to
1.8 GeV. Finally, small$-x$ effects might be sizable \cite{95,95a}. All these
sources of theoretical uncertainties are reduced once we require the charm
quarks, or their fragmentation and decay products, to have significant
transverse momentum. One should therefore directly compare QCD predictions for
the cross--section in the experimentally accessible region with the data.

The same remarks also apply to the as yet preliminary analysis of charm
production by the H1 collaboration \cite{94}, which is based on events with a
hard muon. They find 484 events where at least one muon satisfies $p_T(\mu) >
1.5$ GeV and $30^{\circ} \le \theta(\mu) \le 130^{\circ}$; some 280 of these
events are expected to contain fake muons, or muons from $\pi$ and $K$ decays.
This gives an accepted cross--section $\sigma(ep \to \mu^{\pm} X) = (2.03
\pm 0.43 \pm 0.7)$ nb; about 95\% of this signal is expected to come from
\ccbar\ events, the rest coming from \bbbar\ production.

We attempted to reproduce the cross--sections measured by the ZEUS and H1
collaborations with a parton--level MC generator based on LO QCD expressions.
We take the renormalization and factorizations scales to be $\sqrt{m_c^2 +
p_T^2}$ and $m_c=1.6$ GeV; as mentioned earlier, the $p_T$ cuts greatly
reduce the sensitivity to $m_c$. However, these cuts also introduce an
additional difficulty in the theoretical treatment. They are sufficiently high
so that fragmentation effects will play a role. On the other hand, $p_T$
cannot safely be assumed to be much larger than $m_c$ here, so factorizing the
result into a hard production cross section and a fragmentation function may
not yet be a good approximation. We therefore ran our MC programs with two
different options, using the standard Peterson et al. fragmentation functions
\cite{87} or no fragmentation at all. In the former case we also include
contributions from the charm in the photon ($qc \to qc$ and $gc \to
gc$);\footnote{There is also a very small contribution of this kind from the
charm in the proton, which we neglect.} after all, the use of both
fragmentation and structure functions rests on the factorization theorem, so
it seems reasonable to treat them symmetrically.

In our comparison with the ZEUS result we always include a factor of 0.26,
which is \cite{93} the probability for a charm quark to fragment into a
charged $D^*$ meson. We find \cite{82} that if we leave out both fragmentation
and the contribution from the charm in the photon, we can reproduce the
experimental cross section only if we assume a gluon distribution that
increases rapidly at small $x$; MRSD-' works well, while the prediction of
MRSD0' is too low by nearly a factor of two for all reasonable choices of
momentum scale. Due to the rapidity cut, the contribution from resolved photon
processes (chiefly $gg$ fusion) only amounts to 20\% or less. On the other
hand, if we include fragmentation effects in the standard way, the average
$p_T$ of the charm quarks in accepted events increases by nearly a factor of
two, thereby reducing the sensitivity to very small $x$; moreover, the total
cross section is now actually dominated by contributions involving the charm
content in the photon, chiefly $cg$ scattering. As a result, we can reproduce
the experimental cross section using either MRSD0' or MRSD-' partons in the
proton, and LAC1 or WHIT partons for the photon; the DG parametrization now
predicts a far too large cross section, since it assumes $c^{\gamma} =
u^{\gamma}$, which is manifestly a bad approximation at these rather low
momentum scales.

The H1 sample is less sensitive to the gluon in the proton at small $x$,
due to both the $p_T(\mu)$ cut (which leads to a considerably higher mean
transverse momentum for accepted charm quarks than in the ZEUS data sample),
and the requirement that $\theta(\mu) \le 130^{\circ}$. The data therefore do
not even allow to discriminate between MRSD0' and MRSD-' if we ignore both
fragmentation and the charm content of the photon. In this case the resolved
photon contribution is quite small, so that any (reasonable) combination of
photon and proton structure functions is in agreement with the data, yielding
a LO cross section of about 1.4 to 2.4 nb. If we include both charm
fragmentation and the charm content of the photon, the sensitivity to soft
gluons in the proton is reduced even further, and predictions using MRSD0'
differ from those using MRSD-' by less than 0.1 nb. The LAC1 and WHIT
parametrizations yield a predicted cross--section of about 1.5 nb, very close
to the MRSD0' prediction without fragmentation and without the contribution
from charm in the photon. The DG parametrization gives a prediction of about
2.9 nb; as already discussed, this parametrization over--estimates the charm
content of the photon, but even this high number is not inconsistent with the
experimental result.

\setcounter{footnote}{0}
\subsection*{3c) Direct Photon Production in \gamp\ Scattering}
Interest in the ``deep--inelastic Compton process" $\gamp \to \gamma X$ dates
back to the early days of the quark--parton model \cite{96}; among other
things, the cross--section for the simplest contributing sub--process, $\gamma
q \to \gamma q$, depends on the fourth power of the quark charge, and thus
allows an independent determination of these charges. However, very soon after
the introduction of the concept of the parton content of the photon it was
realized \cite{37b} that other subprocesses contribute to the production of
direct photons\footnote{As opposed to photons stemming from the decay of
hadrons, mostly $\pi^0$ and $\eta$ mesons.} at high $p_T$: Not only resolved
photon processes like $\qqbar \to g \gamma$ and $gq \to q \gamma$ have to be
included in a consistent LO QCD treatment, but also fragmentation processes,
where a high$-p_T$ parton fragments into a hard photon and an additional
(nearly collinear) jet; note that the $q \to \gamma$ and $g \to \gamma$
fragmentation functions \cite{37a} are ${\cal O}(\aem/\alphas)$, just like the
parton densities in the photon. A full LO treatment therefore has to include
direct processes like $\gamma q \to g q \to g q \gamma$ and resolved photon
processes like $q q' \to q q' \to q q' \gamma$; these processes involve the
same partonic matrix elements that appear in direct and resolved photon
contributions to the production of high$-p_T$ jets. Hence direct photon
production is in general actually more complicated to analyze theoretically
than jet production, contrary to statements often found in the literature. On
the other hand, photons are easier to study experimentally than jets; in
particular, their energy can be measured considerably more precisely.

In LO the triple differential cross--section for the production of a direct
photon in processes that do not involve parton $\to$ photon fragmentation is
still given by eq.(\ref{e3.1}); the cross--section for fragmentation processes
contain a convolution with a fragmentation function:
\be \label{e3.2n}
\frac{d^3 \sigma_{\rm frag}(ep \to e \gamma j X)}{d p_T d \eta_\gamma d \eta_j}
= \sum_{i,j,k,l} \int \frac {d z'}{z'} f_{i|e}(x_e)
f_{j|p}(x_p) D_{k \to \gamma}(z')
\frac { d \hat{\sigma}_{ij \to kl} } {d \hat{t}}.
\ee
Eqs.(\ref{e3.3}),(\ref{e3.4}) also still apply here, but for the fragmentation
contribution one has to replace $p_T$ by $p_T' = p_T/z'$. Moreover, unlike the
case of di--jet production, the final state particles are now distinguishable.
Note that eqs.(\ref{e3.3}) allow a two--fold ambiguity for $\eta_{1,2}$ in
terms of the Bjorken$-x$ variables; taking $\eta_1 \equiv \eta_\gamma$, one
has
\be \label{e3.1n}
\eta_\gamma = \ln \left[ \sqrt{ \frac{E_p}{E_e} } \frac {x_p}{x_T} \left(
1 \pm \sqrt{ 1 - \frac {x_T^2} {x_e x_p} } \right) \right].
\ee
In case of jet production one can arbitrarily fix the sign in
eq.(\ref{e3.1n}), since this only corresponds to the definition of which of
the two partons gives ``jet 1". However, in case of direct photon production
the sign in eq.(\ref{e3.1n}) is correlated with the choice of sign in
$\hat{t}$, eq.(\ref{e3.4}). In particular, if $\hat{t}$ is defined as the
momentum transfer from the incident photon, or parton in the incident photon,
to the final photon, or parton fragmenting into the final photon, taking a $+$
sign in eq.(\ref{e3.4}) means $|\hat{t}| > \hat{s}/2$, which implies that one
has to take the $+$ sign in eq.(\ref{e3.1n}); recall that we define the proton
direction has having positive rapidity.

The first quantitative estimates of cross--sections for the production of
direct photons in \gamp\ scattering, including all sub--processes listed
above, were presented in refs.\cite{97}, using very simple ans\"atze for
parton distribution and fragmentation functions. The first partial NLO
analysis for fixed--target energies was published in ref.\cite{37a}, where
only NLO corrections to $\gamma q \to \gamma q$, as well as the direct NLO
process $\gamma g \to \qqbar \gamma$, were included; all other contributions
were treated in LO. This was justified by the observation that at these (low)
energies a rather modest $p_T$ cut on the outgoing photon suffices to greatly
suppress contributions involving photonic parton densities and/or
fragmentation functions. Notice that only processes involving both the parton
content of the photon and parton to photon fragmentation can proceed via gluon
exchange in the $t-$ or $u-$channel; one needs to convolute the cross section
with {\em two} additional functions, compared to simple $\gamma q \to \gamma
q$ scattering, before the $\hat{s}/|\hat{t}|$ enhancement characteristic of
these gluon exchange processes becomes available. It is therefore not
surprising that resolved photon contributions are not quite as important in
the case of high$-p_T$ direct photon production as they are for jet
production.

The analysis of ref.\cite{37a} was repeated independently, as well as extended
to HERA energies, in ref.\cite{98}. However, here the resolved photon
contributions involving fragmentation (which have comparatively soft $p_T$
spectra) were omitted, while the contribution from $\gamma g \to \gamma g$
(via a box diagram \cite{99}) was included; this last contribution was shown
to be significant in certain regions of phase space (see below), even though
it is formally of next--to--next--to--leading order (NNLO).

In contrast, the emphasis of refs.\cite{100} was on processes involving the
parton content of the photon. In particular, it was shown that a clear signal
from the gluon content of the photon (largely from $gq \to \gamma q$) can be
extracted from a sample of events with {\em fixed energy} of the incident
photon. Due to the large Lorentz boost from the parton--parton cms to the lab
frame these photons can be quite energetic if they emerge at large rapidity,
even at rather small $p_T$.\footnote{The same is true for very forward jets,
of course. However, it should be significantly easier to detect a photon at
small angle than to reconstruct a jet just a few degrees away from the proton
beam.}

A further step towards a full NLO analysis was taken in ref.\cite{101}, where
corrections to the resolved photon contributions were included; however, all
fragmentation contributions were still treated in LO. Notice that inclusion of
NLO corrections to non--fragmentation contributions are necessary to reduce
the dependence on the scale appearing in the fragmentation functions. In NLO,
fragmentation and non--fragmentation contributions mix, just like direct and
resolved photon processes do. For example, $g q \to g q \gamma$ is an NLO
correction to $g q \to q \gamma$, but also part of the LO contribution
involving $q \to \gamma$ fragmentation {\em if} the final state $q$ and
$\gamma$ are (nearly) collinear. In order to avoid double counting, this
collinear (divergent) contribution therefore has to be subtracted from the NLO
correction. This subtraction term is proportional to the logarithm of the
scale appearing in the fragmentation function. Since the subtraction term
(obviously) appears with a negative sign in the final result it largely
cancels the dependence on this ``fragmentation scale"; the cancellation is not
perfect, since the fragmentation function resums all orders of this leading
logarithm, while the subtraction term does not. On the other hand, since the
fragmentation contributions have only been treated in LO in ref.\cite{101},
a rather strong dependence on the renormalization scale appearing in \alphas\
remains.

In Fig.~16 we show some LO predictions for the rapidity dependence of the
direct photon cross--section, adapted from numerical results of
ref.\cite{101}. We see that the ``Born" cross--section (from $\gamma q \to
\gamma q$) peaks at negative rapidity (in the direction of the incoming
photon); even though the $1/\hat{u}$ pole of the hard matrix element favours
configurations where the final photon is emitted in the proton direction,
negative rapidities are favoured since they probe the proton at small $x$,
where the (sea) quark densities increase quickly. In the resolved photon
process $g^{\gamma} q^p \to \gamma q$ the final photon is also preferentially
emitted in the proton direction; moreover, at large positive rapidity one
becomes sensitive to the gluon density in the photon at small $x$, where it is
(presumably) large. As a result, contributions $\propto \gga$ are expected to
dominate at large, positive $\eta_{\gamma}$, as first noted in refs.\cite{100}.
Notice that resolved photon contributions involving parton to photon
fragmentation have not been included in Fig.~16; however, we already know from
the discussion of Sec.~3a that at large positive rapidity they are also
dominated by contributions $\propto \gga$. Finally, the box contribution,
$\gamma g \to \gamma g$, is important only at very negative $\eta_{\gamma}$,
where one probes the gluon density in the proton at small $x$.

The results of Fig.~16 have been computed using LO expressions. While NLO
corrections are significant (e.g., they change the rapidity distribution of
the direct contributons \cite{37a,98}), they do not change the conclusion that
direct photon production offers a good handle for constraining \gga. More
recently this conclusion has been challenged on different grounds
\cite{102,103}: The integral over the photon spectrum (\ref{e2.10}) tends to
smear out the distributions, which in Fig.~16 were shown for a fixed energy
(10 GeV) of the incident photon. In particular, direct contributions from
rather soft initial photons also populate the region of positive rapidity.
These papers also complete the NLO calculations by including corrections to
the fragmentation processes, but these corrections have little bearing on the
question whether direct photon production is useful for constraining the gluon
content of the photon.

Ref.\cite{103} introduces two additional refinements. It makes use of an NLO
parametrization \cite{104} of the parton to photon fragmentation functions,
which updates the ``asymptotic" expressions of ref.\cite{37a}. More
importantly, this analysis for the first time introduces an isolation
requirement. At fixed target energies, backgrounds from $\pi^0$ and $\eta$
decays can be suppressed quite reliably, since they contain two photons with
(usually) substantial opening angle \cite{104a}. At higher energy, and higher
$p_T^{\gamma}$, this opening angle becomes much smaller, making it more
difficult to detect both photons individually. At \ppbar\ colliders a direct
photon signal could therefore only be detected \cite{105} if the photons were
isolated, i.e. after events were discarded if more than some maximal (small)
amount of energy was found in a cone around the photon. Such a cut reduces the
fragmentation contribution considerably; in particular, the contribution from
$g \to \gamma$ fragmentation becomes very small, since here most of the energy
usually goes into the accompanying jet rather than the photon. Nevertheless,
at hadron colliders contributions from $q \to \gamma$ fragmentation can still
be significant \cite{106}. Strictly speaking, the necessity to impose an
isolation cut at HERA has not yet been demonstrated. Indeed, the only
background study we are aware of \cite{107} reaches the conclusion that even
without isolation cut the background in the most interesting region of
positive rapidity can be suppressed to the 25\% level, which might be
tolerable. However, this study ignores resolved photon contributions to the
background (they are included for the signal); we saw in Sec.~3a that these
contributions will increase the total jet cross--section (and hence also the
cross--section for the production of high$-p_T$ particles) by a large factor
at positive rapidity. In the absence of a more complete background study we
are therefore inclined to believe that an isolation cut will indeed be
necessary at HERA.

In Fig.~17, which has been adapted from numerical results of ref.\cite{103},
we show the rapidity dependence of $\sigma(ep \to e \gamma X)$ after requiring
that the hadronic energy in a cone $\delta \equiv \sqrt{ (\Delta \eta)^2 +
(\Delta \phi)^2}/\cosh \eta_\gamma = 0.4$ around the outgoing photon be less
than 10\% of the energy of that photon. No cut on the energy of the incident
photon has been imposed. We see that, except at very large $\eta_\gamma$, the
cross--section is dominated by the direct contributions and resolved photon
contributions involving the quark content of the photon only. In particular,
in sharp contrast to the ``Born" cross--section of Fig.~16, after integration
over the incident photon spectrum the direct contribution has a very flat
rapidity distribution. The authors concluded that, while one might be able to
further constrain the quark densities in the photon from this measurement (see
the difference between the dotted curve, obtained from the GS parametrization,
and the solid one, which is for the GRV parametrization), there is very little
sensitivity to the gluon content of the photon.

However, we saw in Sec.~3a that even a rather modest lower cut on the energy
of the incident photon greatly suppresses direct photon contributions at
positive rapidity; see e.g. Fig.~10. Such a cut would also reduce
contributions $\propto q_i^{\gamma}$ substantially, but would have much less
effect on contributions $\propto \gga$. The sensitivity to the gluon content
of the photon would presumably be enhanced even more if the rapidity of the
jet balancing the direct photon can be measured as well
\cite{102}\footnote{Unfortunately, no NLO calculation for $\gamma+$jet
production exists as yet.}. We therefore believe that the conclusions of
refs.\cite{102,103} are probably too pessimistic. On the other hand, HERA will
have to accumulate a large body of data before a substantial number of direct
photon events with known (fixed) energy of the incident photon will become
available; only then can predictions like those shown in Fig.~16 be compared
with experiment.

So far no HERA data on direct photon production have been published. Data from
a fixed target photoproduction experiment exist \cite{104a}; they agree with
theoretical predictions \cite{107a}, but the energy is too low to extract a
signal for the resolved photon contribution.

Before closing this subsection we briefly mention some work on closely
related topics. In ref.\cite{108} the cross--section for $c+\gamma$ production
has been computed (ignoring fragmentation contributions, however), the main
motivation being that this might allow to constrain the charm content of the
photon. In ref.\cite{109} it has been suggested that one might learn something
about the parton content of polarized photons from the study of direct photon
production. Indeed, longitudinally polarized electron beams should become
available at HERA in a few years; part of this polarization will be passed on
to photons emitted from these electrons, if they carry a substantial fraction
of the electron's energy. However, one would also have to measure the
polarization of the {\em outgoing} photon, which seems quite difficult.

Finally, ref.\cite{110} is a first study of the production of {\em two}
direct photons. The cross--section is substantially smaller (by a factor
$\propto e_q^2 \aem/\alphas$) than the one for single direct photon
production. Apart from fragmentation contributions, only the resolved photon
process $\qqbar \to \gaga$ contributes in LO; processes like $\gamma q \to q
\gaga$ are formally of NLO, since $q^{\gamma}_i \propto \aem/\alphas$, and are
indeed found to be subdominant numerically. Due to the large charge factor,
this process might be useful for probing the up--quark density in the photon
at small $x$. If $p_T^{\gamma}$ values as low as 3 GeV are experimentally
accessible, one might even be able to extract the contribution from the
box diagram, $gg \to \gaga$; since it is proportional to the gluon density in
the photon, it should be concentrated at larger rapidities than the
contribution from \qqbar\ annihilation. This close analogue of the famous
light--by--light scattering process \cite{110a} has yet to be studied
experimentally.

\setcounter{footnote}{0}
\subsection*{3d) \jpsi\ Production in \gamp\ Scattering}
The inelastic production of \jpsi\ mesons in (virtual or real) \gamp\
collisions has long been regarded as one of the cleanest methods to constrain
the shape of the gluon distribution in the proton \cite{111}. ``Inelastic"
here means that the quantity
\be \label{e3.6}
Z \equiv \frac {p_p \cdot p_{J/\psi}} {p_p \cdot p_{\gamma}}
\ee
is significantly below unity. In the proton rest frame, eq.(\ref{e3.6})
reduces to $Z = E_{J/\psi}/E_{\gamma}$; $Z=1$ therefore means that the
incident photon transmits its entire energy to the \jpsi\ meson. The
cross--section for inelastic \jpsi\ production is nowadays usually computed
using the ``colour singlet" model of ref.\cite{112}, see Fig.~18. In this
model one computes the cross--section for $\gamma g \to \ccbar g$, and
projects out the contribution where the \ccbar\ system is in a
colour--singlet, $s-$wave, $J=1$ state. One further assumes that the relative
momentum of the $c$ and $\bar{c}$ in the \jpsi\ are negligible; the matrix
element for \jpsi\ production is then proportional to the wave function at the
origin $\Psi(0)$, which can be determined from the leptonic decay width of
\jpsi.

This model seems to describe the $Z$ distribution of fixed target data better
\cite{113} than the alternative ``dual model" \cite{114,84}. The agreement in
the high$-Z$ region becomes even better if one introduces a small but
nonvanishing relative momentum between the $c$ and $\bar{c}$ quarks
\cite{115}. Until recently the overall normalization of the cross--section was
not well reproduced by theoretical calculations, i.e. a sizable ``K--factor"
had to be introduced. Fortunately NLO corrections to the direct diagram of
Fig.~18 have recently been calculated \cite{116}. Together with the NLO
corrections to the leptonic decay width \cite{117} they give an overall
inelastic cross--section in agreement with fixed target photoproduction
experiments; however, the photoproduction cross--section extracted from fixed
target leptoproduction experiments seems to be somewhat higher \cite{113}. The
calculation of ref.\cite{116} predicts that the K--factor should be smaller at
high (HERA) energies.

As usual, resolved photon contributions are small at the energies of fixed
target experiments, but they could be quite important at HERA \cite{118,119}.
In addition to the process $gg \to \jpsi g$, one also has to consider $gg \to
\chi_c$ and $gg \to \chi_c g$, with subsequent decay of the heavier $\chi_c$
states into a \jpsi\ and a (soft) photon \cite{120}; the latter process has
to be considered if one imposes a cut on the transverse momentum of the \jpsi.
For $p_T(\jpsi) > 3$ to 5 GeV, the contribution from $b \to \jpsi$ decays
becomes important, and eventually even dominates the direct contribution
\cite{121}. There might also be sizable contributions from $cg \to \jpsi c$
\cite{122}. However, the typical momentum scale in most \jpsi\ events is too
small to reliably use charm distribution functions in the photon or proton;
we therefore neglect this contribution here.

As usual, resolved photon contributions are characterized by the presence of
the photonic remnant jet, and by a rapidity distribution that peaks at large,
positive values. In addition, direct and resolved photon contributions have
very different $Z$ distributions. The direct contribution is peaked at $Z
\simeq 1$, which corresponds to a small energy of the outgoing gluon. In
contrast, for fixed $x_{\gamma}$ the resolved photon contribution peaks at
$Z \simeq x_{\gamma}$; after convolution with the photon spectrum
(\ref{e2.10}) this leads to a $Z$ distribution that quickly rises with
decreasing $Z$.

This is shown in Fig.~19, which we adapted from numerical results of
ref.\cite{123}. Since only a mild cut on the $p_T$ of the \jpsi\ meson has
been applied, the contribution from $b$ decays is relatively small and has
been neglected. LO expressions have been used everywhere; neither NLO nor
non--relativistic corrections to the resolved photon contribution are as yet
known. The B1 parametrization of ref.\cite{124} has been used for the gluon
density in the proton, and DG \cite{39} for the photon. Finally, although the
cross--section has not been multiplied with any branching ratio, it is
clear that only the leptonic decays $\jpsi \to \epem, \mu^+ \mu^-$ are
detectable at HERA; the combined branching ratio for these modes is about
12\%. The results of Fig.~19 have been obtained under the condition that both
leptons can be reconstructed by the ZEUS tracking system. Unfortunately this
removes events at large rapidity, which greatly reduces the resolved photon
contribution. Including the leptonic branching ratio, it only amounts to about
7 pb after integration over $Z$, compared to a direct contribution of about
110 pb.\footnote{Recall that these are LO predictions, and hence somewhat
uncertain; the cross--sections also depend on the gluon densities in the
photon and proton, of course.} Requiring $Z > 0.2$ leaves a very pure direct
sample, which should allow to determine the shape of the gluon density in
the proton for $2 \cdot 10^{-4} \leq x \leq 0.1$ \cite{123}.

Unfortunately the extraction of \gga\ seems to be much less straightforward.
The simulation of ref.\cite{123} indicated that measurement errors would
smear out the direct contribution into the region of small $Z$, totally
obscuring the resolved photon contribution. It remains to be seen whether an
unfolding procedure, and/or the application of additional cuts (e.g., tagging
of the photon remnant jet) will improve the situation sufficiently to allow to
extract new information on \gga\ from \jpsi\ production at
HERA.\footnote{Previous analyses \cite{119,60a}, which had led to very
optimistic conclusions, had used much milder acceptance cuts on the leptons,
or none at all.}

Both H1 and ZEUS have published first results \cite{125} on \jpsi\
production. However, in both analyses events were rejected if any particle in
addition to the two leptons resulting from \jpsi\ decay was observed. This
cut removes most if not all inelastic contributions, but is sensitive to
elastic or diffractive \jpsi\ production. This allows to test Pomeron--based
models \cite{123,115,125a}, but teaches us nothing about the parton content of
the photon. Very recently, ZEUS has announced \cite{125b} preliminary results
on inelastic \jpsi\ production. However, they required $Z > 0.2$, which again
removes most of the resolved photon contribution.

Finally, we briefly mention associate $\jpsi + \gamma$ production. As first
pointed out by Fletcher et al. \cite{119}, in LO only the resolved process
$gg \to \jpsi + \gamma$ contributes, the direct contribution being forbidden
by colour conservation. In principle this process therefore allows a clean
determination of (the shape of) \gga\ \cite{126}. Unfortunately the
cross--section at HERA is quite small, very roughly of order 0.1 to 1 pb after
acceptance cuts and multiplication with the leptonic branching ratio.

\setcounter{footnote}{0}
\subsection*{3e) Production of Lepton Pairs in \gamp\ Collisions}
In this subsection we discuss the production of lepton pairs, either due to
the exchange of a virtual photon or from the decay of an on--shell $W$ or $Z$
boson. The corresponding cross--sections are quite small even at HERA
energies, so we will be brief here.

In the theoretical treatment of these reactions one has to distinguish the
cross--section integrated over the transverse momentum of the lepton pair (not
to be confused with the $p_T$ of the individual leptons) from the
cross--section for the production of a high$-p_T$ lepton pair. In the former
case, the only LO contribution comes from the resolved photon process $\qqbar
\to l^+l^-$, which produces a lepton pair with vanishing transverse momentum.
The corresponding $ep$ cross--section is ${\cal O}(\alpha_{\rm em}^3/
\alpha_s)$ since, as emphasized repeatedly, the parton distribution functions
in the photon are ${\cal O}(\aem/\alphas)$. In contrast, both the direct
process $\gamma q \to l^+ l^- q$ and the resolved photon process $g q \to l^+
l^- q$ contribute in LO to the production of a lepton pair with sizable $p_T$.
The corresponding $ep$ cross--sections are both ${\cal O}(\alpha_{\rm em}^3)$.

This distinction has been well understood in existing treatments of the
``Drell--Yan" production of $l^+ l^-$ pairs via the exchange of a virtual
photon; see ref.\cite{118} for a LO estimate, and \cite{89,127,128} for NLO
calculations. Compared to direct photon production this process has the
theoretical advantage that one does not have to worry about fragmentation
contributions. Moreover, backgrounds are low, and the final state can be
reconstructed cleanly even at rather low $p_T(l)$, allowing one to probe quite
small $x$ values \cite{128}. As usual, the region of positive rapidity
corresponds to small $x_{\gamma}$ and moderate $x_p$, while negative
rapidities correspond to large $x_{\gamma}$ and very small $x_p$. The
production of lepton pairs at positive rapidity and sizable $p_T$ is sensitive
to the gluon content of the photon \cite{127}. Unfortunately the expected
event rates are quite low: ${\displaystyle \frac{ d \sigma (ep \to l^+ l^- X)}
{d M_{l^+l^-}} \simeq 1}$ pb/GeV at $M_{l^+l^-}=4$ GeV after integration over
the transverse momentum of the pair and before any acceptance cuts have been
imposed \cite{128}.

In contrast, the fact that the {\em total} photoproduction cross--section for
$W$ and $Z$ bosons is, in LO, given by the resolved photon contribution {\em
only} has not been appreciated in the existing literature.\footnote{The total
$W$ and $Z$ cross--sections in $ep$ collisions also get contributions where
the heavy gauge boson is radiated off the electron line; in case of $Z$
production at HERA this contributes roughly 50\% of the total cross--section
\cite{129}.} The resolved photon contribution has first been estimated in
ref.\cite{130}, but here this contribution was considered to be
an addition to the direct process even at vanishing $p_T$ of the heavy gauge
boson. This is not correct, since the direct contribution has a $u-$channel
singularity which has to be absorbed into the resolved photon contribution;
simply adding both contributions implies double--counting. While current
calculations \cite{131,129} of the total $W$ and $Z$ photoproduction
cross--sections are in our view not entirely satisfactory, since they mix
LO and NLO contributions in an ill--controlled manner, they should predict
the production of high$-p_T$ gauge bosons quite accurately.\footnote{In
principle the resolved photon contribution from $gq \to W q'$ should be
included in a full LO treatment, but it is strongly suppressed at HERA
energies due to phase space constraints.} This high$-p_T$ region is sensitive
to the form of the $W^+W^- \gamma$ vertex \cite{131}. The total
cross--sections for $W^+$ and $W^-$ production at HERA amount to approximately
0.5 pb each. This cross--section is to be divided by another factor of five
if only the clean $e \nu_e$ and $\mu \nu_{\mu}$ final states are observable;
indeed, a first study \cite{132} has concluded that the detection of
hadronically decaying $W$ and $Z$ bosons at HERA is quite challenging,
although perhaps not impossible. Finally, we mention that recently the H1
collaboration has announced \cite{133} observation of one event that can be
interpreted as the production of a leptonically decaying $W$ boson at high
$p_T$; within the SM this interpretation is quite unlikely, due to the
smallness of the predicted cross--section, but all other interpretations seem
even less likely. Clearly much more data have to be analyzed before any
definite conclusion can be drawn.
\setcounter{footnote}{0}
\section*{4) Real \gaga\ Scattering at \epem\ Colliders}
In this section we discuss hard processes with two (quasi--)real photons in
the initial state and a hadronic final state. At present, and in the near
future, such reactions can only be studied at \epem\ storage rings, where the
effective photon flux is given by eq.(\ref{e2.10}). At future linear \epem\
colliders the photon spectrum might receive large additional contributions
from ``beamstrahlung" \cite{134}, which is emitted when a particle is
accelerated in the field produced by the opposite bunch; beamstrahlung photons
are exactly on--shell. The produced spectrum sensitively depends on the size
and shape of the electron and positron bunches; see ref.\cite{134a} for a
handy parametrization of the spectrum in terms of a few machine parameters.
Finally, in recent years the possibility has been discussed to convert a
linear \epem\ collider into a ``\gaga\ collider" by scattering laser photons
off the incident $e^{\pm}$ beams \cite{135}. The achievable luminosity is
predicted to be comparable to the (geometrical) \epem\ luminosity prior to
conversion of the beams; in contrast, at existing \epem\ storage rings one has
${\cal L}_{\gamma \gamma} \sim \left( \frac {\aem}{\pi} \ln \frac {s} {m_e^2}
\right)^2 \sim 10^{-3} {\cal L}_{ee}$. The photon spectrum of such a photon
collider again depends on the geometrical set--up, and also on the
polarization of the incident electron and laser photon beams. Most of the
results presented here will be for present and near--future colliders; these
are obviously of more immediate interest, and already allow to illustrate most
physics principles. For comparison we will also give a few results for the
more ``futuristic" colliders.

Since we now have two photons in the initial state, we have to distinguish
three physically distinct event classes \cite{37b}, see Fig.~20. Following the
notation of ref.\cite{136} we call reactions where the entire energy of both
photons goes into the hard subprocess ``direct", see Fig.~20a. If only one of
the photons couples in a pointlike manner while the other participates via
its quark and gluon content, as in Fig.~20b, the process is called ``once
resolved" (``1--res" for short). Finally, processes where both photons are
resolved into their partonic constituents are called ``twice resolved"
(``2--res"); an example is shown in Fig.~20c. Recall that each resolved
photon produces a remnant or ``spectator" jet, which goes approximately in the
direction of the incident $e^{\pm}$ beams; the three event classes are
therefore characterized by having zero, one or two of these photonic remnant
jets. Since the parton distribution functions in the photon are ${\cal
O}(\aem/\alphas)$, all three contributions are of the same order in coupling
constants, and have to be treated on the same footing \cite{37b}.

We saw in the previous section that direct and resolved photon contributions
to photoproduction processes mix in NLO QCD. Similarly, the three event
classes contributing to real \gaga\ scattering mix once higher--order QCD
corrections are included. Parts of the NLO direct (1--res) contributions have
already been included in the LO 1--res (2--res) terms; these parts therefore
have to be subtracted from the NLO contributions. Note that mixing between
direct and twice resolved contributions only occurs in NNLO in QCD.

Following our preceding discussion of \gamp\ scattering, we discuss different
final states in separate sub--sections. Jet production is treated in Sec.~4a,
open heavy flavour production in Sec.~4b, \jpsi\ production in Sec.~4c, and
direct photon production in Sec.~4d.

\setcounter{footnote}{0}
\subsection*{4a) Jet Production in \gaga\ Collisions}
As in case of \gamp\ scattering, the production of jets offers the largest
cross--section of all hard \gaga\ collisions that lead to hadronic final
states. In LO, the cross--section can be written as [see eq.(\ref{e3.1})]:
\be \label{e4.1}
\frac{d^3 \sigma(\epem \to \epem j_1 j_2)} {d p_T d \eta_1 d \eta_2} = 2 p_T
x_1 x_2 \sum_{i,j,k,l} f_{i|e} (x_1) f_{j|e} (x_2)
\frac { d \hat{\sigma}_{ij \to kl} ( \hat{s}, \hat{t}, \hat{u} )} {d \hat{t}}.
\ee
If $i$ or $j$ is a quark or gluon, $f_{i|e}$ can again be obtained by
convoluting the photon flux function \fgme\ with the quark or gluon density in
the photon, see eq.(\ref{e3.2}). However, antitag requirements at current
\epem\ collider experiments often allow larger photon virtualities than one
has in photoproduction events at HERA; at the same time, the scale $|\hat{t}|$
or $p_T^2$ of the hard process is usually smaller. This means that the
suppression of resolved photon contributions due to the reduced parton
content of virtual photons is usually larger at \epem\ colliders than at HERA.
In our numerical estimates to be presented below we have included this effect
using the formalism of ref.\cite{78}.

The scaling variables $x_{1,2}$ in eq.(\ref{e4.1}) are related to the jet
(pseudo)rapidities $\eta_{1,2}$ by:
\ben \label{e4.2} \beq
x_1 &= \frac {x_T}{2} \left( e^{\eta_1} + e^{\eta_2} \right); \label{e4.2a}
\\
x_2 &= \frac {x_T}{2} \left( e^{-\eta_1} + e^{-\eta_2} \right),  \label{e4.2b}
\eeq \een
with $x_T = 2 p_T / \rs$ as before. The Mandelstam variable $\hat{s}$ of the
hard scattering sub--process is again given by $\hat{s} = x_1 x_2 s$, and
$\hat{t}$ is as in eq.(\ref{e3.4}). Most of the sub--processes contributing
to jet production in \gaga\ collisions also contribute to jet production at
HERA. In particular, if both $i$ and $j$ are partons (twice--resolved
contribution), $\hat{\sigma}$ is the same as in resolved \gamp\ collisions,
and the case where either $i$ or $j$ is a parton while the other is a photon
(single--resolved contribution) corresponds to the direct contribution at
HERA. Finally, if $i=j=\gamma$ (direct contribution), then $k=q,\ l=\bar{q}$
and one has
\be \label{e4.3}
\frac{ d \hat{\sigma} (\gaga \to \qqbar)} {d \hat{t}} = 3 e_q^4
\frac {2 \pi \alpha^2_{\rm em}} {\hat{s}^2} \left( \frac {\hat{t}} {\hat{u}} +
\frac {\hat{u}} {\hat{t}} \right),
\ee
where $e_q$ is the electric charge of quark $q$ in units of the proton charge,
and the factor of 3 is due to colour.

The first quantitative estimate of high$-p_T$ jet production in (quasi-)real
\gaga\ scattering has been presented by Brodsky et al. \cite{137}. However,
gluon--initiated processes were omitted, and a rather crude parametrization
for the quark densities in the photon was used.\footnote{Ref.\cite{137} also
finds quite large higher twist contributions to the production of high$-p_T$
mesons and jets. It was shown later \cite{138} that the ``constituent
interchange model" used in ref.\cite{137} over--estimates the normalization
of these terms by as much as a factor of a thousand.} Gluon--initiated
contributions have been included in refs.\cite{139}, but again very simple
parametrizations for the parton densities in the photon were used. In spite of
their shortcomings, these early studies clearly demonstrated that resolved
photon contributions are quite important, and often even dominant, if $x_T
\leq 0.2$. Notice that jet production from two--photon collisions at present
\epem\ colliders cannot be studied experimentally if $x_T \geq 0.4$; the
cross--section becomes too small, and annihilation backgrounds too large.
Hence resolved photon contributions were predicted to be sizable for at least
half the experimentally accessible range of $x_T$.

Nevertheless data \cite{14} on multi--hadron production from \gaga\ collisions
taken at the PEP and PETRA storage rings were usually only compared to the
direct (QPM) contribution.\footnote{The PLUTO collaboration \cite{140}
attempted to include 1--res $qg$ or 2--res $qq$ final states in their
analysis. However, they made many simplifying assumptions, some of which are
incorrect. In particular, they assumed that the cross--sections for these
resolved photon contributions drops like the square of the \gaga\ invariant
mass $W$; in contrast, for fixed $p_T$ QCD predicts these cross--sections to
increase with $W$. Moreover, PLUTO did not attempt to include 1--res and
2--res contributions simultaneously.} Not surprisingly, a significant excess
of data over MC prediction was observed. Given that the importance of resolved
photon contributions had been emphasized quite early on \cite{137,139}, it is
surprising that it took ten years before the origin of this excess of data
over QPM prediction was clarified experimentally.

In ref.\cite{136} it was pointed out that the characteristics of these excess
events at least qualitatively agreed with those expected from resolved photon
contributions: Their $p_T$ spectrum is softer than that of the direct
contribution, just as in case of photoproduction events. In addition, resolved
photon events are less two--jet like, i.e. have smaller thrust $T$, than
direct events; in fact, the PLUTO collaboration had shown \cite{140} that the
excess could be removed by requiring $T \geq 0.9$. Ref.\cite{136} also
contains the prediction that, for fixed $p_T$, resolved photon events should
rapidly become more important as the beam energy is increased. The reason is
that raising \rs\ decreases $x_T$, and hence $x_{1,2}$ in eqs.(\ref{e4.2});
due to the additional convolution with photonic parton distribution functions,
the quark and gluon density in the electron is always considerably softer,
i.e. grows faster with decreasing $x$, than the photon flux in the electron.
Resolved photon contributions were therefore expected to play an even more
important role at TRISTAN and LEP than at PEP and PETRA. This was confirmed
soon afterwards by the AMY collaboration \cite{6}, who showed that their data
on multi--hadron production in antitagged \gaga\ collisions could be
reproduced by their QCD--based Monte Carlo program, in both shape and
normalization, if and only if it included the full set of resolved photon
contributions, including those with a gluon in the initial state. This was the
first unambiguous observation of resolved photon interactions other than in
deep--inelastic $e \gamma$ scattering, and the first direct experimental
evidence for a nonzero gluon density in the photon.

Early experimental analyses of multi--hadron production in \gaga\ collisions
\cite{14}, including the AMY analysis \cite{6}, were not amenable to direct
comparison with theoretical calculations, since the experiments defined a
``jet" as a thrust hemisphere. The production of actual, reconstructed jets
(using a cone algorithm) has been studied only quite recently, first by the
TOPAZ collaboration \cite{141} and then by AMY \cite{142}. This is an
important development, since it allows a much more direct comparison between
theory and experiment.

In the meantime theoretical estimates are also becoming more sophisticated, by
including NLO corrections. This process was actually already startd in 1979
with two calculations \cite{143} of $\gaga \to gg$ via a quark box diagram.
NLO corrections to the direct process $\gaga \to \qqbar$ (for massless quarks)
were calculated soon afterwards \cite{144}. However, the result contains
collinear divergencies, which have to be absorbed in the 1--res contribution;
in NLO it therefore makes little sense to consider the direct contribution in
isolation. This was recognized in ref.\cite{145}, which deals with the
production of high$-p_T$ hadrons. Here the direct contribution was treated in
NLO (including the contribution from $\gaga \to gg$, which is formally NNLO),
but at the time resolved photon contributions could only be included in
leading order.\footnote{Notice that one only needs the LO 1--res contribution
from $\gamma q$ scattering in order to absorb the collinear divergencies of
the NLO direct term. The procedure of ref.\cite{145} should therefore be quite
accurate at high $x_T$, where the direct contribution dominates.} This
shortcoming could be remedied only after the NLO corrections to the hard
partonic sub--process had been calculated \cite{68}. A full NLO calculation of
the single--jet inclusive cross--section has become available only recently
\cite{146}; it finds only modest NLO corrections to the total cross--section
($\leq 25\%$, for a jet cone size $\Delta R = 1$ used by both TOPAZ and AMY,
and renormalization and factorization scale $\mu = p_T$), but the relative
weights of direct, single-- and double--resolved contributions can change by
larger amounts.\footnote{However, there seems to be a discrepancy between the
LO result of ref.\cite{146} and our calculation. In particular, Aurenche et
al. find that 2--res processes still contribute about 25\% to the jet
cross--section measured by TOPAZ at $p_T =7.5$ GeV, while in our calculation
it amounts to less than 10\% at this rather large value of $x_T$ (unless we
use the LAC3 parametrization).} Finally, in ref.\cite{147} the 2--res
contribution has been studied in NLO, with emphasis on constraining the gluon
content of the photon.

Since NLO corrections appear to be quite modest, and since many more LO
parametrizations of parton densities in the photon exist, we will only present
results from LO analyses here. As mentioned earlier, we do include the
suppression of resolved photon contributions due to the finite virtuality of
the photons, using the simplest ans\"atze in ref.\cite{78}. (Quark and gluon
densities have to be treated separately.) We remind the reader that, unlike
ref.\cite{146}, these ans\"atze do not assume a power--like suppression of the
``hadronic" contribution to the parton densities of virtual photons; they
might therefore slightly under--estimate the (in any case only modest) size of
the correction.

In Fig.~21 we compare our LO calculation of the single--jet inclusive
cross--sections as measured by the TOPAZ (a) and AMY (b) collaborations at
TRISTAN ($\rs=58$ GeV). TOPAZ requires the jet to have pseudorapidity
$|\eta_{\rm jet}| \leq 0.7$, and rejects an event if it contains an electron
or positron at an angle $\theta > 3.2^{\circ}$ relative to the beam pipe and
with energy $E > 0.25 E_{\rm beam} = 7.2$ GeV. AMY accepts jets out to
$|\eta_{\rm jet}| = 1.0$; moreover, it can antitag an event only if it
contains an electron or positron with $E > 0.25 E_{\rm beam}$ at an angle
$\theta > 14.1^{\circ}$.

We see from Fig.~21 that the DG, LAC1 and WHIT1 parametrizations reproduce
the TOPAZ data at $p_T \leq 4.5$ GeV quite well, and also fit the AMY data
over the entire $p_T$ range. Notice, however, that we had to use $N_f=4$
active flavours already at $p_T({\rm jet}) = 2.5$ GeV in order to achieve this
agreement. This is not really justifiable for the DG and LAC parametrizations,
which treat the charm quark as massless; on the other hand, WHIT explicitly
includes some charm mass effects, which leads to a greatly reduced
contribution from charm in the photon, in agreement with expectations. (It
makes up for the shortfall by slightly larger light quark and gluon
densities.) The GRV parametrization seems to fall below the data at low $p_T$;
this is mostly because we have used $\Lambda_{\rm QCD}(N_f=3) = 0.4$ GeV for
DG and WHIT, as compared to 0.2 GeV for GRV and LAC, as is implicit in these
parametrizations. NLO corrections have been found to be most important at low
$p_T$ \cite{146}; they might very well bring the GRV prediction into agreement
with the data. In contrast, the WHIT4 prediction lies above the data at low
$p_T$, since it assumes a quite large and hard gluon density in the photon.
This discrepancy might be reduced if the ``hadronic" piece of the parton
density of virtual photons is suppressed by a power of the virtuality, but it
seems unlikely that this will restore full agreement with the data. All the
other WHIT parametrizations seem acceptable, however.

Fig.~21 shows that the LAC3 parametrization clearly over--estimates the
cross--section, while the direct contribution by itself falls well below the
data. Notice that the resolved photon contributions remain non--negligible out
to the highest $p_T$ where data exist, although their importance clearly
diminishes with increasing $p_T$, as discussed above. Finally, we observe that
our calculation falls somewhat below the TOPAZ data for $p_T > 4.5$ GeV. This
is disturbing, since in principle our neglect of the charm mass should be
better justified if $p_T^2 \gg m_c^2$. Notice, however, that our calculation
agrees very nicely with the AMY data. This hints at an experimental
discrepancy between these two data sets. Indeed, from the increased acceptance
in rapidity as well as the much looser antitag requirements used by AMY, one
expects that the cross--section measured by AMY should exceed that measured by
TOPAZ by a factor of about 1.7 to 1.8; here we have made use of the fact
\cite{141,142} that $d \sigma / d \eta_{\rm jet}$ is quite flat over the
observed range, and have used eq.(\ref{e2.10}) to estimate the effect of the
tagging criteria on the photon flux. On the other hand, the cross--section
measured by AMY only exceeds that reported by TOPAZ by a factor of $1.36 \pm
0.26$ ($1.29 \pm 0.20$) at $p_T = 4.55$ (6.5) GeV, where we have assumed all
errors to be independent. Although this procedure probably over--estimates the
true error, since part of the systematics should be common to both
experiments, this comparison indicates that the discrepancy in any one $p_T$
bin is not very significant statistically. Due to bin--to--bin correlations
produced by the unfolding procedure, it is difficult to combine different bins
to arrive at an overall significance of the discrepancy. At the lowest $p_T$
bin, the ratio of the two experimental cross--sections is 1.55, which agrees
very well with expectations.\footnote{The effect of the looser antitag
requirement of AMY is smaller at low $p_T$, due to the dynamical upper bound
$P^2 \leq p_T^2$ that has to be imposed on the virtuality of the exchanged
photons in order to meaningfully speak of parton or photon densities in the
electron.} Moreover, both collaborations also publish di--jet cross--sections,
although with somewhat larger relative errors; our LO calculation reproduces
both these measurements over the entire $p_T$ range.

\setcounter{footnote}{0}
Note that the published cross--sections are actually partonic cross--sections,
which have been obtained by unfolding the observed $p_T$(jet) and $\eta$(jet)
distributions using a LO MC program. We will argue in Sec.~5 that such a
procedure might be more reliable at TRISTAN than at HERA, since events with
multiple partonic interactions should not pose much of a problem here.
Nevertheless this procedure does produce some dependence on the details of the
Monte Carlo program, e.g. via the predicted jet reconstruction efficiency.
It might also be dangerous to directly compare these extracted cross--sections
with NLO calculations, since the higher order corrections seem to change
\cite{146} the relative weights of the three classes of contributions compared
to what has been assumed in the LO MC.\footnote{It is not obvious how the
``theoretical" definition of the three classes of contributions used in
ref.\cite{146} relates to a more experimental definition relevant for event
reconstruction, which could e.g. be based on the presence of absence of
(remnant) jets close to the beam pipes. For example, the relative weights of
the three classes of contributions in the NLO calculation strongly depend on
the factorization scale, which is an unphysical parameter. Unfortunately it is
quite difficult to write an MC generator based on a full NLO calculation; no
such generator for jet events produced in hadronic or photonic collisions
exists as yet.} Both experiments also give the measured rate of jet events
as a function of $p_T$; however, at these rather low transverse momenta the
jet reconstruction efficiency is still rather small and $p_T-$dependent. A
fair amount of MC work is therefore (unfortunately) necessary before a
comparison with theoretical predictions can be made.

The published data are based on an integrated luminosity of about 90 pb$^{-1}$
for TOPAZ, and 27 pb$^{-1}$ for AMY. The total available data samples are more
than three times larger than this. Notice that the error bars in Fig.~21 are
already quite small; if the slight discrepancy between AMY and TOPAZ can be
resolved, the full data set should therefore allow a quite precise comparison
between theory and experiment.

The larger data sets should also allow to investigate more differential
distributions, using events with two reconstructed jets. As an example we show
in Fig.~22 the triple--differential cross--section for events with two jets
with equal (pseudo)rapidity $\eta_1 = \eta_2 \equiv \eta$. This distribution
is quite flat for the direct contribution, at least over the range covered by
TRISTAN detectors. For parametrizations with a relatively hard gluon density
(WHIT1, DG) the total single--resolved contribution also depends only weakly
on $\eta$; for the given choice of $p_T$, the decrease of the contributions
$\propto q_i^{\gamma}$ ($qg$ final state) is more or less balanced by the
increase of those $\propto \gga$ (\qqbar\ final state). Increasing $\eta$
means increasing $x_1$ but decreasing $x_2$ in eqs.(\ref{e4.2}). Since the
gluon content in the electron is always quite strongly peaked at small $x$
while the photon flux function \fgme\ is relatively hard, the increase of
$f_{G|e}$ more than compensates the decrease of \fgme\ as $\eta$ is increased.
(Note that $f_{q|e}$ increases much more slowly with decreasing $x$ than
$f_{G|e}$ does.) If the gluon density in the photon is large but strongly
peaked at small $x$ (WHIT6, LAC1), the rise of the cross--section for
$\gamma g \to \qqbar$ even leads to a total 1--res contribution that peaks
at $\eta \simeq 2$, rather than at $\eta = 0$.

Turning to the twice--resolved contributions shown in Fig.~22b, we observe
that the cross--sections for processes with two gluons in the final state,
which most of the time also have two gluons in the initial state, is quite
strongly peaked at $\eta=0$, even for the comparatively hard gluon density of
WHIT1. The same parametrization predicts the cross--section from $qg \to qg$
scattering to remain quite flat for $|\eta| \leq 1$, since the decrease of
$f_{q|e}(x_1)$ is compensated by the increase of $f_{G|e}(x_2)$.\footnote{Of
course, there is also a contribution $\propto f_{q|e}(x_2) f_{G|e}(x_1)$, but
it decreases very quickly with increasing, positive $\eta$.} Nevertheless,
parametrizations with relatively hard gluon density still predict a
significant decrease of the total 2--res cross--section as $\eta$ is increased
from 0 to 1. In contrast, parametrizations with a large and soft gluon content
(WHIT6, LAC1) predict the 2--res contribution to rise with increasing $\eta$,
reaching a maximum at $\eta \simeq 1.3$. This increase is entirely due to the
$qg$ final state, where the increase of $f_{G|e}$ now over--compensates the
decrease of $f_{q|e}$.

As discussed in the next subsection, the TOPAZ collaboration has proven
capable of detecting photon remnant jets with an efficiency of about 70\%;
this ability was already implicit in their observation \cite{141} of
considerable energy flow at small angles relative to the beam pipes. They are
now beginning to exploit this capability also in the jet analysis \cite{148}.
This shows that the 1--res, 2--res and direct contributions can indeed be
studied separately. On the other hand, Fig.~22 shows that even the sum over
all contributions should allow to discriminate between some existing
parametrizations of the parton content of the photon. In particular, at $p_T =
3$ GeV, parametrizations with a large but soft gluon density predict a flat or
even slowly rising jet cross--section as $\eta$ is increased away from zero,
in contrast to parametrizations with a small or hard gluon density.
Gluon--initiated processes are expected to contribute more than 50\% of the
total 2--res cross--section, and still some 25 to 35\% of the total di--jet
cross--section at small rapidity. This fraction, and hence the difference
between predictions using different parametrizations, will be smaller (larger)
for $p_T > (<) 3$ GeV.

Experiments at the LEP storage ring should also be able to contribute
significantly to our understanding of (almost) real \gaga\ collisions. At LEP1
the vicinity of the $Z$ peak means that multi--jet annihilation events are a
much more severe background than at TRISTAN; however, a parton--level
investigation \cite{149} concluded that this should not be much of a problem
as long as the invariant mass $M_{jj}$ of the two high$-p_T$ jet system is
below 15 to 20 GeV. Indeed, some experimental results have already been
published. The ALEPH collaboration repeated \cite{150} the AMY analysis
\cite{6} of multi--hadron production. Although some of the details differ, as
will be discussed in Sec.~5, ALEPH also finds that the description of the data
is greatly improved once resolved photon interactions are included in the
Monte Carlo model.

DELPHI \cite{151} reached similar conclusions regarding ``minimum bias"
multi--hadron production. In addition, they analyzed a subsample of events
containing two jets with $|\eta| \leq 1.0$ and $p_T > 1.75$ GeV, where the
jets were defined using a cluster algorithm. At these small transverse momenta
the jet reconstruction efficiency is rather low [$\sigma(2-{\rm jet}) \sim 0.1
\sigma($parton)], which introduces a strong dependence on the details of the
MC program. DELPHI finds that DG falls below the data for $p_T \leq 3$ GeV,
while the LAC1 and GS parametrizations work well. Note that their event sample
is mostly sensitive to direct events, as well as resolved photon events with
large $x$, i.e. soft remnant jets, since they require $W_{\rm vis} < 13$ GeV
and $E_{\rm vis} < 20$ GeV; the calculation of these quantities includes
information from the small--angle taggers, which cover angles down to
2.5$^{\circ}$ and should therefore see parts of the remnant jets. Fig.~3 shows
that the DG parametrization does indeed fall well below LAC1 for large $x$ and
small $Q^2$.

More recently, DELPHI has published \cite{152} a study of hadronic \gaga\
events where either the $e^+$ or the $e^-$ is tagged at a very small angle,
corresponding to a photon virtuality $P^2 \sim 0.1$ GeV$^2$. Since even
stronger cuts were imposed on the hadronic system than in the first DELPHI
analysis, the data sample was quite small (491 events); nevertheless it was
sufficient to once again prove the necessity to include resolved photon
contributions if the data are to be described by the MC program. This small
angle tagging technique holds much promise, since the tag helps to reconstruct
the kinematics of \gaga\ events by determining the energy of one of the
two photons. If (as at HERA) the energy of the second photon can be determined
using the Jacquet--Blondel method, a measurement of the jet rapidities in
di--jet events would allow to directly reconstruct the Bjorken$-x$ variables
of the partons in the photons. If, as TOPAZ results indicate, the DELPHI
small angle detectors (not to be confused with the very small angle taggers
used for the electron tagging) can be used for detecting the presence or
absence of remnant jets, DELPHI (and similar detectors) should be able to
perform quite detailed analyses of hard two--photon reactions. In order to
fully exploit this potential, the strong upper limits on the visible energy
and invariant mass used in refs.\cite{151,152} should be relaxed; this should
certainly be possible at LEP2 energies, where the annihilation background is
much smaller.

The LEP energy is expected to soon be increased to $\rs \simeq 180$ GeV; LEP
experiments will then have the unique opportunity to study two--photon
cross--sections over a wide range of energies. As illustrated in Fig.~23, the
energy dependence can be quite strong. In this figure we show the
triple--differential cross--section at $p_T=5$ GeV as a function of $\eta_1 =
\eta_2 \equiv \eta$, imposing an antitag condition similar to that used by
ALEPH \cite{150}. For $\rs = 90$ GeV, $x_T$ is similar to the value used in
Fig.~22, which leads to a similar shape of the pseudorapidity distribution;
the kinks at $\eta \simeq 1.5$ occur because the ALEPH antitag becomes
ineffective if the outgoing electron has less than half the beam energy, i.e
for scaled photon energy $z > 0.5$. Raising \rs\ from 90 to 180 GeV increases
the direct contribution at $\eta = 0$ only by a factor of 1.45, while the
1--res and 2--res contributions increase by factors of 2.0 and 3.2,
respectively. This again demonstrates the strong dependence of the
cross--section for resolved photon processes on the available phase space, or,
equivalently, on the Bjorken$-x$ variables of eqs.(\ref{e4.2}). Note that the
single--jet inclusive cross--section at central rapdity grows even faster with
energy, since the kinematical integration limits for the rapidity of the
second jet also increase with \rs. Finally, we remind the reader that the
shape and normalization of the solid curves in Fig.~22 could be quite
different if the photon has a large but soft gluon component; this would also
result in an even more rapid increase of the cross--section for resolved
photon events with \rs.

LEP will almost certainly be the highest energy \epem\ storage ring ever. In
order to reach even higher energies, one will need to build a linear collider.
In such a device each bunch can most likely only be used once; the repitition
rate (number of bunch collisions per second) will therefore almost certainly
be smaller than at LEP. At the same time the total luminosity must increase
like the square of the beam energy in order to maintain a roughly constant
rate of \epem\ annihilation events. The luminosity per bunch crossing will
therefore have to be much larger than at existing storage rings, forcing one
to use very dense bunches. The correspondingly large charge density gives rise
to strong electromagnetic fields. When particles inside one bunch enter the
field produced by the opposite bunch, they will be accelerated, and will
therefore radiate real photons. This is known as ``beamstrahlung" \cite{134}.

The flux of these beamstrahlung photons depends quite sensitively on the
design characteristics of the collider; this is not surprising, since this
radiation is due to the field of the entire bunch, not due to individual
\epem\ collisions (unlike the only quasi--real ``bremsstrahlung" photons we
have dealt with up to now). Beamstrahlung can be reduced by using flat beams,
and by splitting large bunches into ``trains" of smaller ones. On the other
hand, for a given class of designs, beamstrahlung increases quite rapidly with
increasing beam energy.

This is ilustrated in Figs.~24, which show the single--jet inclusive
cross--section as a function of $p_T$, where we have accepted jets with
$|\eta| \leq 2.0$. We have used the WHIT1 parametrization, and imposed the
antitag requirement $\theta < 10^{\circ}$ when computing the bremsstrahlung
contribution to the total photon flux. The beamstrahlung spectrum has been
calculated using the analytical expressions of ref.\cite{134a}, for the JLC
design as specified at the 1993 international linear collider conference
\cite{153}. Designs for linear colliders are still evolving; the results of
Fig.~24 should therefore be considered as indicative only.

At $\rs=0.5$ TeV (Fig.~24a) the beamstrahlung spectrum is considerably softer
than the equivalent bremsstrahlung spectrum (\ref{e2.10}). This enhances the
relative importance of the direct contribution, since the cross--section for
resolved photon contributions increases with the two--photon invariant mass
$W$ while that for the direct contribution decreases. For $p_T > 100$ GeV, the
total cross--section is dominated by directly interacting bremsstrahlung
photons. Notice that this design leads to a luminosity of about 50 fb$^{-1}$
per year; the two--photon cross--section should therefore remain measureable
for jets with $p_T$ well above 100 GeV. Moreover, one expects of the order
$10^8$ events per year with a jet with $p_T > 5$ GeV. This sounds like a large
number, but corresponds to a trigger rate of 10 Hz or less, which should be
easily manageable.

Increasing \rs\ to 1.0 TeV (Fig.~24b) greatly increases the flux of
beamstrahlung photons, and also makes it harder. This enhances the relative
importance of resolved photon contributions, as can be seen from the $x_T$
value where direct and resolved photon cross--sections are equal. Notice also
that now beamstrahlung increases the cross--section by about a factor of 5.5
even at $p_T = 200$ GeV. Finally, comparing Figs.~24a and b, we see that the
jet cross--section increases by a factor of about 7 (20) for $p_T = 5$ (100)
GeV; without beamstrahlung, the corresponding factors would ``only" have
been 3 and 5, respectively. We remind the reader that these results depend on
the specific machine design; see refs.\cite{154,154a} for further discussions
of this point.

As mentioned in the beginning of this section, one might be able to convert a
linear \epem\ collider into a \gaga\ collider by back--scattering laser
photons off the $e^+$ and $e^-$ beams \cite{135}. The spectrum and luminosity
of such a collider depend quite strongly on details such as the polarization
of the laser photons and incident electrons beams. The production of jets at
such a collider has first been discussed in ref.\cite{155}. In Fig.~25 we
present a result for the simple case of unpolarized beams and small distance
between the conversion and interaction points. This leads to a photon spectrum
which peaks at $z=0.828$, where it also cuts off; the two--photon luminosity
is then quite flat over a wide range of $W$. Most modifications that have been
discussed in the literature \cite{135} give even harder photon spectra. By
comparing Fig.~25 with Fig.~24a we see that, as expected, the much harder
photon spectrum of the \gaga\ collider has greatly increased the relative
importance of resolved photon contributions. Moreover, the hard photon
spectrum allows to efficiently access soft partons in the photon via 1--res
contributions at large rapidity. As a result, the cross--section increases
with $\eta$ even for parametrizations with rather modest gluon content, like
WHIT1; this is to be contrasted with the situation at present and future
\epem\ colliders, see Figs.~22 and 23. This also explains the large difference
between the predictions from the LAC1 and WHIT1 parametrizations close to the
kinematical maximum of $\eta$, inspite of the rather large value of $x_T$.

Finally, we should warn the reader that from LEP2 energies onwards, multiple
interactions could substantially increase the true jet cross-section, compared
to the simple parton--level estimates of Figs.~23 to 25; we already saw in
Sec.~3a that this phenomenon seems to play an important role in jet
production from resolved photons at HERA. This will be discussed in more
detail in Sec.~5.

\setcounter{footnote}{0}
\subsection*{4b) Heavy Quark Production in \gaga\ Collisions}
Apart from the production of jets discussed in the previous subsection, the
production of heavy quarks is the only hard QCD process that has been
studied experimentally in two--photon collisions. As discussed in Sec.~3b, the
main advantage of heavy \QQbar\ pair production is that perturbative QCD
should be applicable over the entire phase space. However, as we already saw
in the discussion of the photoproduction of heavy quarks, predictions for
total \ccbar\ production rates are at present still quite uncertain, due to
unknown higher order corrections (which lead to a strong scale dependence) and
the uncertainty in the value of $m_c$ to be used here.

Progress in the theoretical treatment of heavy quark production in two--photon
processes has been relatively slow. The first complete LO calculation,
including resolved photon processes, was only performed in 1989 \cite{136}.
The contribution from twice resolved processes was found to be very small at
TRISTAN energies, but the 1--res contribution (from $\gamma g \to \ccbar$) is
quite sizable. Further LO predictions, for newer parton densities in the
photon, were published in \cite{41}. A full NLO analysis of the direct and
1--res contributions has been performed in ref.\cite{156}.

In Fig.~26 we show updated predictions for the total \ccbar\ cross--section in
the PETRA to LEP2 energy range. We have included NLO corrections to the direct
contribution, using a simple parametrization.\footnote{Note that here direct
and 1--res contributions remain well--defined even in NLO, since there is no
LO 1--res contribution from the quarks in the photon.} In ref.\cite{156} the
direct contribution to the total cross--section is written as
\be \label{e4.4}
\sigma_{\rm dir}(\gaga \to \QQbar (g)) = \frac {\alpha^2_{\rm em} e_Q^4}
{m_Q^2} \left( c^{(0)}_{\gamma \gamma}
+ 4 \pi \alpha_s c^{(1)}_{\gamma \gamma} \right).
\ee
Here $c^{(0)}_{\gamma \gamma}$ describes the well--known \cite{157}
tree--level (QPM) prediction, while $c^{(1)}_{\gamma \gamma}$ can be
parametrized as:
\beq \label{e4.5}
 c^{(1)}_{\gamma \gamma} &= \frac{\pi}{2} - \sqrt{r} \left( \frac{5}{\pi}
- \frac{\pi}{4} \right), \ \ \ \ \ r < 2.637 \nonumber \\
& = 0.35 r^{-0.3}, \hspace*{2.44cm} r \geq 2.637
\eeq
where $r = W^2_{\gamma \gamma} / (4 m_Q^2) - 1$. This parametrization is
exact at threshold $r \to 0$, and describes the full NLO result to better than
10\% accuracy for all $r \leq 100$. On the other hand, the predictions for the
1--res contribution shown in Fig.~26 have been computed in LO only. One reason
is that here NLO corrections are considerably smaller than for the direct
contribution \cite{156}, largely because the threshold (``Sommerfeld")
corrections are negative for a colour--octet \QQbar\ state. Moreover, the
uncertainty of the prediction is much larger than for the direct
cross--section, making it less important to include 10\% corrections.

In Fig.~26 this uncertainty is given by the width of the bands defined by two
curves with the same pattern. Following ref.\cite{136} we have in all cases
required $\wgg > 2 m_D = 3.74$ GeV, but the ``dynamical" charm quark mass
appearing in the expressions for the hard cross--sections is less well
defined. In case of the direct cross--section we have varied $m_c$ between 1.3
GeV (upper dotted curve) and 1.6 GeV (lower dotted curve). Another uncertainty
arises from the choice of scale in \alphas\ and (for the 1--res contribution)
in \gga. In Fig.~26 we have estimated this uncertainty by varying this scale
between $\mcc/4$ (upper curve) to \mcc\ (lower curve). In case of the direct
contribution the combined uncertainty only amounts to slightly over 20\%,
almost independently of \rs. Note, however, that other authors \cite{95,156}
prefer to use an even wider range of values for $m_c$.

Unfortunately the uncertainty for the prediction of the 1--res contribution is
considerably larger than this. One reason is that now \alphas\ already appears
in the tree--level cross--section, leading to a stronger dependence on the
renormalization scale; there is also a factorization scale dependence in this
case. As usual, the inclusion of NLO corrections should reduce these scale
uncertainties. However, the biggest uncertainty comes from the choice of
$m_c$, and of the minimal allowed \mcc. In case of the direct contribution one
always has $\wgg = \mcc$, at least in leading order. On the other hand, 1--res
events have $\wgg > \mcc$; it is then not clear whether one has to require
$\mcc > 2 m_D$ in order to describe open charm production, or whether it is
sufficient to have $\wgg > 2 m_D$. If $\mcc < 2 m_D < \wgg$, some energy has
to be transferred from the remnant jet to the ``hard" \ccbar\ pair; it has to
be remembered that in any case colour needs to be exchanged between these two
systems in the hadronization step. It seems unlikely to us that this soft
energy exchange can exceed 1 GeV, however. We have therefore used $m_c = 1.4$
GeV, $\mcc > 2 m_c$ and scale $\mu = \mcc/4$ in order to estimate the upper
limit of the 1--res contribution, and $m_c = 1.6$ GeV, $\mcc > 2 m_D$ and $\mu
= \mcc$ for the lower limit.

Fig.~26 shows that this uncertainty makes it impossible to distinguish between
the DG and WHIT1 parametrizations based on the total \ccbar\ production rate
alone. Note that the uncertainty is larger for LAC1, which has a very soft
gluon density and therefore reacts very sensitively to changes of the lower
bound of \mcc. Even for the more conservative parametrizations, the
uncertainty amounts to roughly a factor of two. Nevertheless, some sensitivity
to the parton densities does remain; we will come back to this point shortly.
Finally, we note that even at LEP2 the 2--res contribution amounts to at most
5\% of the 1--res one \cite{156}; we have therefore not shown it in this
figure.

In the last few years several experimental studies of charm production in
\gaga\ collisions have been published. The first analysis, by the JADE
collaboration \cite{158}, looked for fully reconstructed charged $D^*$ mesons
in single--tag events. They reported a considerable excess over QPM
predictions, but this has not been confirmed by the TASSO \cite{159} or
TPC/2$\gamma$ \cite{160} collaborations, who took data at similar energies.
TASSO also measured the cross section for exclusive $D^0 \overline{D^0}$
production. The total \ccbar\ production cross--sections derived by these two
experiments agree with expectations \cite{156}.

The first study of charm production in two photon collisions at TRISTAN, by
the TOPAZ collaboration \cite{161}, also searched for fully reconstructed
$D^*$ mesons. Unfortunately the reconstruction efficiency is quite low,
leading to rather poor statistics (a few dozen events per experiment). TOPAZ
therefore also published a study \cite{162} based on $D^{\pm*} \to \pi_s^{\pm}
D^0$ decays, where only the soft pion needs to be detected; here the signal
after background subtraction consists of $372 \pm 54$ events. VENUS
\cite{163} and TOPAZ \cite{164} have also published analyses of hadronic
two--photon events containing an electron or positron (electron--inclusive
analysis); after subtracting backgrounds (mostly from Dalitz decays and photon
conversions), these experiments extract a charm signal with ${\cal O}(100)$
events each. Finally, very recently ALEPH presented \cite{165} results of an
analysis based on 33 fully reconstructed $D^{\pm*}$ mesons.

Unfortunately these experimental results are somewhat contradictory. All
TRISTAN experiments find some excess of events with high $p_T$; e.g., TOPAZ
\cite{162} reports a 2.9 $\sigma$ excess of events with $p_T(D^*) \geq 3.6$
GeV. This is actually not all that surprising, given that we had to include
contributions from the charm in the photon in order to reproduce the
production of central jets with $p_T \geq 2.5$ GeV; such ``charm excitation"
contributions are not included in present MC programs.\footnote{The NLO
correction to the direct process should describe $\gamma c \to g c$ scattering
accurately at these energies; however, it is not clear whether the
parametrized form of the NLO corrections used by TOPAZ \cite{161,162} treats
such contributions properly. Moreover, 2--res excitation contributions are not
included at all.} On the other hand, ALEPH \cite{165} finds a cross--section
for the production of charged $D^*$ mesons with $p_T \geq 2.0$ GeV that agrees
with the {\em lower} range of predictions; no excess is visible here.

Most of these experimental analyses are only sensitive to charmed hadrons with
significant transverse momentum. The notable exception is the TOPAZ inclusive
electron analysis \cite{164}; electrons with momentum as low as 400 MeV are
accepted, so that even charm quarks at rest can contribute to the signal.
Moreover, TOPAZ used their forward calorimeter, which covers angles down to
3.2$^{\circ}$ with respect to the beam pipe, to look for the presence of
photon remnant jets in the event; their MC predicts a jet tagging efficiency
of $73\pm2\%$. This allows them to study direct and resolved photon
contributions separately. The direct contribution agrees roughly with the
upper range of NLO predictions. The resolved photon contribution agrees with
NLO predictions using the LAC1 parametrization, while the DG prediction seems
to be at least two standard deviations below the data. This demonstrates that
this kind of analysis can lead to significant constraints on the gluon density
in the photon.

In order to compare experimental results on charm production with theoretical
calculations, one has to model the quark to hadron transition. This is more
difficult here than in the more familiar high--energy \epem\ annihilation
events, since for the low values of \mcc\ or $p_T(c)$ relevant for present
two--photon data the effect of hard gluon radiation can {\em not} be absorbed
into fragmentation functions; this is only possible if gluons are
predominantly emitted collinear to the quarks, which requires $p_T \gg m_c$.
In this context it is interesting to note that the CLEO collaboration
\cite{166}, working at values of \mcc\ only slightly above the upper range
probed by present two--photon experiments, found that popular fragmentation
functions could describe the observed $D^*$ meson spectrum {\em only} if the
emission of hard gluons was allowed explicitly. In particular, it was not
possible to describe the spectrum by simply convoluting the tree--level
\ccbar\ cross--section with the Peterson et al. fragmentation function
\cite{87}. Other fragmentation functions gave a better fit, but only if the
fragmentation parameters were chosen differently from those used for data at
higher energy; this is not surprising, since fragmentation functions are scale
dependent, just like parton distribution functions. It is therefore
encouraging to note that an NLO event generator for heavy quark production in
two photon collisions, written by J. Zunft, is now publicly available
\cite{167}.\footnote{This generator only allows the emission of a single hard
gluon; the generator used by CLEO to model \epem\ annihilation events included
$2 \to 4$ matrix elements.} Once hard gluon radiation has been included
explicitly, the fragmentation function only has to include nonperturbative
effects, which should indeed factorize to good approximation. Studies like
that by the TOPAZ collaboration \cite{164}, which are sensitive to charmed
hadrons at rest, can play an important role in testing this formalism, since
by definition fragmentation functions can only change the spectrum, but not
the total cross--section. Once the fragmentation process has been fully
understood, direct \ccbar\ pair production might be the best way to measure
the value of $m_c$ to be used in perturbative QCD calculations, since the
scale uncertainty is rather small here. A good knowledge of $m_c$ would help
to sharpen predictions for 1--res \ccbar\ production, as well as photo-- and
hadro--production of charm.

Fig.~26 shows that the total \ccbar\ pair production cross--section is
expected to grow quite rapidly with energy. However, at higher energies it
might be necessary to impose quite stringent cuts in order to extract a charm
signal; this will reduce the detectable cross--section significantly. For
example, in ref.\cite{156} it was found that requiring one of the charm quarks
to have rapidity $|y| \leq 1.7$ and $p_T \geq 5$ GeV reduces the
cross--section by almost a factor of 50. This still leaves us with at least
5,000 \ccbar\ events in 500 pb$^{-1}$ of data; however, at this point we have
not yet required anything that would actually identify these events as being
due to charm, e.g. a hard lepton or a reconstructed $D^*$ meson.

At a 500 GeV \epem\ linear collider even the DG parametrization predicts
\cite{154a,167a} the total \ccbar\ pair cross--section to reach a value
between 5 and 50 nb, depending on the amount of beamstrahlung generated.
However, requiring the event to contain at least one muon with rapidity $|y|
\leq 2$ and $p_T(\mu) \geq 5$ GeV reduces this cross--section by at least a
factor of 2,000.

As in case of jet production, cross--sections for the production of heavy
quarks can be boosted considerably by converting an \epem\ collider into a
\gaga\ collider by means of laser backscattering. However, at such a collider
QCD processes will likely be regarded primarily as backgrounds to ``new
physics" signals. In particular, one advantage of a \gaga\ collider is that it
can produce a Higgs boson as an $s-$channel resonance. If $m_H < 150$ GeV,
this Higgs boson will predominantly decay into into a \bbbar\ pair \cite{168}.
The direct (tree--level) background from $\gaga \to \bbbar$ can be reduced
significantly by chosing the two incident photons to be (predominantly) in
a $J_z = 0$ state \cite{169}. There is still a direct background from $\gaga
\to \bbbar g$, but this appears to be manageable \cite{170}. However, at least
if one uses a collider with a broad photon spectrum, which allows to search a
wide range of Higgs masses simultaneously, the dominant background will
actually come from resolved photon events. This has been pointed out in
ref.\cite{171}, where the 1--res $\gamma g \to \bbbar$ contribution is
studied. As emphasized in ref.\cite{172}, the exact size of this background
will also depend on the extent to which a gluon in a polarized photon is
itself polarized.

The to date most complete study of resolved photon backgrounds to the Higgs
signal at a broad--band \gaga\ collider has recently been performed by
Baillargeon et al. \cite{173}. They ignore polarization effects on the parton
densities in the photon (which amounts to the assumption $\Delta \gga = 0$),
but include all possible processes, including $b$ and $c$ excitation
contributions; these are very large in the relevant kinematical region. They
include production of $c$ quarks, as well as light partons, in order to
estimate the effects of imperfect $b-$tagging.

An example is shown in Fig.~27, which shows the signal for a 120 GeV Standard
Model Higgs boson \cite{168} (assuming a \bbbar\ invariant mass resolution of
5 GeV), as well as various backgrounds. The electron beams have 175 GeV
energy, and 90\% polarization; 100\% polarized laser beams have been assumed.
This suppresses the direct backgrounds by about an order of magnitude.
Moreover, a $p_T$ cut of 30 GeV has been applied. We see that the largest
contribution to the single$-b$ inclusive cross--section comes from 2--res $b$
excitation processes (labelled as ``$bX_{2-res}$"). Depending on details of
the $b-$tagging efficiency, it is therefroe often advantageous to require both
high$-p_T$ jets to be tagged as $b$ quarks, or at least as heavy flavours.

Another conclusion of ref.\cite{173} is that Higgs searches at a \gaga\
collider become easier with increasing Higgs mass and {\em de}creasing beam
energy. The reason is that, as we have seen several times already,
cross--sections for resolved photon processes increase rapidly with \rs, but
decrease equally quickly when the invariant mass of the hard system is
increased for fixed \rs. In particular, for the case shown in Fig.~27, even
with a $b-$tagging efficiency comparable to present LEP detectors, a 7
$\sigma$ Higgs signal could be extracted from 10 fb$^{-1}$ of data, which
roughly corresponds to one year's running; for the same parameters, one could
at best hope to extract a 3.5 $\sigma$ signal at a 500 GeV collider. Since
one here probes the photon at very high momentum scales these conclusions do
not depend very sensitively on the parametrization chosen \cite{173}.
Nevertheless, Fig.~27 clearly shows that a good understanding of heavy quark
production in resolved photon interactions is mandatory if we ever want to
look for Higgs bosons with mass below 150 GeV at a \gaga\ collider.

\setcounter{footnote}{0}
\subsection*{4c) \jpsi\ Production in \gaga\ Collisions}
\jpsi\ production in principle offers a clean method for constraining \gga\ in
two photon collisions \cite{136}. At least in the framework of the colour
singlet model \cite{112} and to leading order in \alphas, only resolved photon
processes can contribute to $\gaga \to \jpsi +$hadrons. Moreover, the 2--res
contribution is expected to be very small, just as in case of open heavy
flavour production discussed in the previous subsection. In LO the
cross--section is therefore essentially proportional to \gga; the hard
sub--process, $\gamma g \to \jpsi + g$, is the same as in direct inelastic
photoproduction of \jpsi\ mesons, see Sec.~3d. As mentioned in that
subsection, NLO corrections to \jpsi\ production in $\gamma g$ fusion have
recently been computed \cite{116}. We use the results of this calculation to
update our previous LO predictions \cite{136} for \jpsi\ production at TRISTAN
energies, and also extend the calculation into the LEP energy range.

In order to simplify this calculation, we again use a parametrized form of the
NLO corrections. According to ref.\cite{116} the total cross--section for
$\gamma g \to \jpsi + X$, integrated over the region $Z \leq 0.9$ where $Z$ is
the inelasticity defined in eq.(\ref{e3.6}), can be written as:\footnote{In
ref.\cite{116} the factor $8/m^3_{J/\psi}$ has been written as $1/m^3_c$.
However, in the framework of the colour singlet model it seems more natural to
us to have the partonic cross--section scale like $1/m_c^2$, rather than
$1/m^5_c$. In the latter case the uncertainty related to the value of $m_c$
discussed below would obviously be significantly larger.}
\be \label{e4.6}
\hat{\sigma}(\gamma g \to \jpsi +X) = \frac {\aem \alpha^2_s} {m_c^2}
\frac { 8 e_c^2 |\psi(0)|^2 } {m^3_{J/\psi}} \left\{ c^{(0)}(r) +
4 \pi \alphas \left[ c^{(1)}(r) + \bar{c}^{(1)}(r) \log \frac {\mu^2}{m_c^2}
\right] \right\},
\ee
where $r = \hat{s}/m^2_{J/\psi} - 1$. The wave function at the origin
$|\psi(0)|^2$ can be derived from the measured leptonic partial width
\cite{117,116}. The function $c^{(0)}(r)$ describes the tree--level
cross--section \cite{112}; we do not impose any cut on $Z$ for this part,
since it remains finite in the limit $Z \to 1$, unlike the NLO correction,
which diverges in this limit. These corrections are given by the two functions
$c^{(1)}(r), \ \bar{c}^{(1)}(r)$, which can be parametrized as:
\ben \label{e4.7} \beq
c^{(1)}(r) &= \sqrt{r-0.1} \left[ \frac{ (r-0.8)(r-6.2) } {0.85 + 1.1r -0.1
r^2 + 2.7 r^{2.5}} - \frac {1} {1.22 -0.90 r + 1.14r^{1.5}} \right];
\label{e4.7a} \\
\bar{c}^{(1)}(r) &= (r-0.1)^{0.7} \frac {2.6-r} {1.4 - 1.2r + 3 r^{1.7}}.
\label{e4.7b}
\eeq \een
Note that the NLO corrections (\ref{e4.7}) remain finite as $r \to \infty$,
while the tree--level cross--section drops like $1/r^2$ in this limit.

In Fig.~28 we show results of a partial NLO calculation of \jpsi\ production
at current \epem\ colliders based on eqs.(\ref{e4.6}) and (\ref{e4.7}). We did
not include the small NLO contribution from $\gamma q$ scattering \cite{116};
moreover, the NLO direct contribution from $\gaga \to \jpsi + gg$ is not yet
available. Note also that $\psi'$ production with subsequent $\psi' \to \jpsi$
decay should increase the total \jpsi\ yield by another 15\% or so. The
results of Fig.~28 are for a no--tag situation, and include the suppression of
\gga\ due to the virtuality of the photon \cite{78}. We checked that our
parametrization of the NLO cross--section shows significantly reduced scale
dependence, as expected, at least as long as the scale $\mu^2 \geq 1.5 m_c^2$;
the cross--section drops quickly for even smaller $\mu$. The same behaviour
has been observed in \cite{116} for the case of photoproduction of \jpsi. We
therefore estimate the scale dependence by varying $\mu^2$ between $m_c^2$
(lower curves) and $2 m_c^2$ (upper curves); the cross--section slowly
decreases again for larger values of $\mu$. In addition, we have varied $m_c$
in eq.(\ref{e4.6}) from 1.6 GeV (lower curves) to 1.3 GeV
(upper curves). Since we are now explicity producing a colour singlet state
already in the hard scattering process we have always required the $\gamma g$
cms energy to exceed $m_{J/\psi}$, even if $2m_c < m_{J/\psi}$.\footnote{In
ref.\cite{116} $r$ has also been written as a function of $m_c^2$, rather
than a function of $m^2_{J/\psi}$. However, this results in negative
cross--sections for $m_c > m_{J/\psi}/2$. It seems to us that the $m_c$
dependence of the $c$ functions should be treated on a par with corrections to
the non--relativistic approximation used in the colour singlet model.}

Fig.~28 shows that the uncertainty estimated in this way is still quite
sizable; most of this uncertainty, however, is due to the variation of the
factor $1/m_c^2$ in eq.(\ref{e4.6}). Notice also that the LO estimate (dotted
curve, for scale $\mu^2 = 2 m_c^2$ and $m_c$ = $m_{J/\psi}/2$) is quite close
to the lower range of NLO predictions. This indicates that the perturbative
result is relatively stable. Recall also that ref.\cite{116} finds good
agreement of their prediction with low--energy photoproduction data, although
leptoproduction data favour larger cross--sections. However, there might be
sizable corrections to the non--relativistic approximation intrinsic to the
colour singlet model \cite{115}; this has not been included in our estimate of
the theoretical uncertainty.

We see that the total expected cross--section for \jpsi\ production at TRISTAN
lies in the range from about 0.4 to 2.5 pb, depending on the parametrization
of \gga. This corresponds to several hundred events per experiment in the
total TRISTAN data sample. However, one will almost certainly have to demand
that the \jpsi\ meson decays into an \epem\ or $\mu^+ \mu^-$ pair; this
reduces the rate by about a factor of 7.5 even before any acceptance cuts have
been applied. The cross--section increases quite rapidly with energy, possibly
exceeding 10 pb at LEP2; however, the acceptance at these higher energies
might also be poorer. A detailed MC study is necessary to decide whether the
study of \jpsi\ production in two photon events at current colliders is
feasible.

As usual, the cross--section at future linear \epem\ colliders is expected to
be even larger, partly due to contributions from beamstrahlung photons. Since
the invariant mass of the produced final state is quite small, the
cross--section is now largest for designs giving a large number of rather soft
photons. For example, at the 500 GeV TESLA collider (we again use the 1993
design \cite{153}) one expects a total \jpsi\ production cross--section
between 0.3 and 2.0 $\mu$b, or some $10^7$ events per year. At the same
collider the $\Upsilon(1s)$ cross--section is expected to lie between 0.25 and
0.85 pb, which still gives around 10,000 events per year. It is at present not
clear what fraction of these events would be detectable in the considerably
``dirtier" environment of such linear colliders.

\setcounter{footnote}{0}
\subsection*{4d) Direct Photon Production in \gaga\ Scattering}
The production of a direct photon in the collision of two (quasi--)real
photons, $\gaga \to \gamma \ X$, is another reaction that in LO only receives
contributions from resolved photon processes: The direct $2 \to 3$ process
$\gaga \to \qqbar \gamma$ is ${\cal O}(\alpha^3_{\rm em})$, and thus of
next--to--leading order compared to the 1--res contribution from $\gamma q
\to \gamma q$, as well as the 2--res contributions from $g q \to \gamma q$ and
$\qqbar \to g \gamma$, which are formally ${\cal O}(\alpha^3_{\rm
em}/\alphas)$; the relevant hard scattering cross--sections can be found in
ref.\cite{173a}. There are additional LO contributions involving parton $\to$
photon fragmentation; as discussed in Sec.~3c, the corresponding fragmentation
functions are ${\cal O}(\aem/\alphas)$, which compensates for the relative
factor of $\alphas/\aem$ that appears in the hard sub--process cross--sections
for these contributions. However, these fragmentation contributions will in
general lead to softer photons, which are accompanied by a jet.

The only existing calculation \cite{173a} for this process omitted the
fragmentation contributions; this roughly corresponds to imposing stringent
isolation requirements for the direct photon. Ref.\cite{173a} uses LO
expressions; an NLO calculation is now under way \cite{38b}.

In Fig.~29 we show updated LO estimates\footnote{As discussed in Sec.~3c, care
has to be taken when computing the rapidities in this case. This has not been
described properly in ref.\cite{173a}, although the numerical results of this
paper are correct.}. We have required $p_T^{\gamma} \geq 1.5$ GeV, since QCD
perturbation theory would be very unreliable for even softer photons. We have
also applied the acceptance cut $|\eta_{\gamma}| \leq 1$ on the
(pseudo)rapidity of the emitted photon. In case of the WHIT1 parametrization,
2--res and 1--res contributions are shown separately, together with the sum;
for all other parametrizations only the total prediction is given. We see that
WHIT1 predicts the 2--res contribution to be very small in the PEP to TRISTAN
energy range, and to remain sub--dominant even at the highest LEP2 energy. On
the other hand, according to the LAC1 parametrization the 2--res contribution
should be dominant for $\rs > 130$ GeV, due to the rapid rise of the
cross--section for $gq \to \gamma q$, which profits from the steep gluon
density assumed in this parametrization. In contrast, the GRV prediction for
the 2--res contribution at $\rs =200$ GeV is about a factor of 1.6 below the
WHIT1 result, largely due to the smaller value of $\Lambda_{\rm QCD}$ that has
been assumed for GRV; note that we are probing QCD at rather small momentum
scales, where \alphas\ depends quite sensitively on $\Lambda_{\rm QCD}$.

The similarity of the various predictions shown in Fig.~29 is therefore
somewhat misleading; clearly one could gain more information if the 2--res and
1--res contributions can be separated experimentally, e.g. by remnant jet
tagging. Due to the large charge factor, the 1--res contribution mostly probes
the $u-$quark density in the photon. Its measurement might therefore yield
valuable information about the flavour structure of the photon at rather low
momentum scales, where nonperturbative contributions to the photonic parton
densities are still important. Whenever the 2--res contribution is sizable, it
is dominated by $gq$ scattering; it could therefore give us an additional
handle on the gluon content of the photon.

Unfortunately the cross--section is not large even if we allow $p_T^{\gamma}$
to be as small as 1.5 GeV; the NLO calculation now being performed \cite{38b}
will show how well QCD perturbation theory converges at such low momentum
scales.\footnote{When calculating the LAC1 prediction shown in Fig.~29 we had
to ``freeze" the parton densities, i.e. use $\min(4 \ {\rm GeV}^2,
(p_T^{\gamma})^2)$ as momentum scale, since this parametrization produces
negative parton densities if used at scales below the input scale $Q_0^2=4$
GeV$^2$.} Furthermore, the Bjorken$-x$ of the partons (or photon) ``in" the
electron can only be reconstructed if the rapidity of the jet balancing the
photon is also known. However, requiring $|\eta_{\rm jet}| \leq 1$ reduces the
cross--section by another factor of 1.7 (2.0) at $\rs = 30$ (200) GeV;
moreover, the reconstruction efficiency for such a soft jet might be quite
low. On the positive side, Fig.~29 shows that each TRISTAN experiment should
have seen about 100 \gaga\ events with an isolated hard central photon in the
final state; at LEP2, several hundred such events should be accumulated by
each experiment. They are well worth looking for.
\setcounter{footnote}{0}
\section*{5) Beyond Perturbation Theory}
In this section we will address several topics that go beyond QCD perturbation
theory. We will be rather brief here, due to lack of both space and expertise
on our part. Nevertheless, we feel that we should at least comment on these
issues. For one thing, when comparing perturbative QCD calculations with data,
one nearly always has to correct for nonperturbative effects, usually with
the aid of Monte Carlo event generators; we have already commented on the MC
dependence of certain results in the preceding sections. Secondly, as remarked
in the Introduction, the study of reactions involving real photons might
provide us with new insight into the interplay of soft and hard QCD; given
that perturbative QCD is by now generally accepted as the theory of hard
hadronic interactions, much future research in strong interactions will
presumably be focussed on this transition region.

nonperturbative effects clearly play an important role in so--called minimum
bias events. This name derives from the (idealized) definition that in a
minimum bias event sample, each inelastic scattering event should be included,
without any experimental (trigger) bias. This is very hard to achieve in
practice, in particular for final states with low particle multiplicity and
small energy deposition in the detector. Measuring total inelastic
cross--sections is therefore not easy.

One usually distinguishes three classes of events in $\gamp$ scattering. In
\ub{quasi--elastic} events, $\gamp \to Vp$, the proton remains intact while
the photon is transformed into a single vector meson $V$; usually only $\rho,
\ \omega, \ \phi$, and occasionally \jpsi, are included here. In
\ub{diffractive} events either the photon or the proton (or both) gets broken
up into several hadrons, but no colour is exchanged between the two systems.
In such events there is usually a gap in rapidity space between the two
diffractive systems (in double diffractive events), or between the hadronic
system and the single $p$ or $V$ (in single diffraction). In contrast, in
\ub{non--diffractive} events colour is thought to be exchanged between the
photon and the proton. Both get broken up (in case of direct interactions, the
photon is absorbed), and soft particles usually fill rapidity space more or
less uniformly.

All these components have to be modelled accurately for a complete
understanding of minimum bias events, and in particular if one wants to
extract the total \gamp\ cross--section from the observed event rate. Both
HERA experiments have performed \cite{174,175} such measurements.
H1 \cite{175} has so far only published an analysis of 1992 data, while ZEUS
has also published \cite{174} results based on the much larger 1993 sample.
Both experiments use different MC models to describe the various event
classes, and then try to fit the relative weights to the data. For example,
H1 determines that about 26\% of all events are quasi--elastic or diffractive;
their result for the total \gamp\ cross--section at $\rs=195$ GeV
is\footnote{We have increased the published number by 8\%, since H1 used an
inaccurate expression for \fgme\ when converting $ep$ into $\gamp$
cross--sections} $\sigma^{\rm tot}_{\gamma p}({\rm H1}) = (171 \pm 7 \pm 22)
\ \mu$b, where statistical and systematic errors are listed separately. The
1993 ZEUS data at $\rs \simeq 180$ GeV favour a slightly larger quasi--elastic
$+$ diffractive event fraction ($\simeq 35\%$), but a lower total
cross--section: $\sigma^{\rm tot}_{\gamma p}({\rm ZEUS}) = (143 \pm 4 \pm 17)
\ \mu$b. Very recently, H1 announced \cite{176} preliminary results from a
special 1994 run with a minimum bias trigger. Their analysis indicates an even
larger quasi--elastic $+$ diffractive event fraction (around 40\%), and
$\sigma^{\rm tot}_{\gamma p}({\rm H1, \ prelim.}) = (172 \pm 3 \pm 10) \
\mu$b.

Note that the systematical uncertainties of these measurements are
significantly larger than the statistical errors. This indicates that our
understanding of minimum bias photoproduction events still needs to be
improved. Note also that the best description of non--diffractive events found
by ZEUS \cite{174}, based on a superposition of ``soft" and ``semi--hard"
events (see below), still has $\chi^2/d.o.f. > 2$. The detection efficiencies,
and hence the extracted value of $\sigma^{\rm tot}_{\gamma p}$, determined
from the same MC program may therefore also not yet be entirely reliable.

So far little effort has been made to understand diffractive events in the
framework of QCD. A possible connection \cite{177a} might be derived from the
observation of jets in single--diffractive \ppbar\ events by the UA8
collaboration \cite{177}, as well as the recent observation of di--jet events
with rapidity gap between the jets by the D0 collaboration \cite{178}. Very
recently the ZEUS collaboration announced \cite{179} preliminary results
indicating the presence of such di--jet events with gap between the jets at
HERA; ZEUS interprets them as resolved photon events with large \xgam.

Much more effort has been devoted to the understanding of non--diffractive
inelastic reactions. This is necessary even for the study of hard processes,
since the ``underlying event" in events with high$-p_T$ jets is closely
related to minimum bias events without any (obvious) hard interactions.
nonperturbative effects again play an important role in the description of
such events; however, it by now seems quite likely that semi--hard QCD
interactions, with partonic transverse momenta of order 1 to 2 GeV, are also
important here. Such interactions are said to lead to ``minijets": Even though
a perturbative (hard) scattering took place, the resulting ``jet" might be too
soft to pass experimental jet identification cuts.

The idea that such minijets might play an important role in minimum bias
hadronic physics dates back to 1973, when Cline et al. \cite{7} proposed that
the rapid increase of the inclusive jet cross--section with energy might be
the root cause of the observed increase of total hadronic cross--sections.
This jet cross--section is given by
\be \label{e5.1}
\sigma_{ab}^{\rm jet} (s) = \int_{p_{T,{\rm min}}}^{\rs/2} d p_T
\int_{4 p_T^2/s}^1 d x_1 \int_{4 p_T^2/(x_1 s)}^1 d x_2 \sum_{i,j,k,l}
f_{i|a}(x_1) f_{j|b}(x_2) \frac { d \hat{\sigma}_{ij \rightarrow kl}(\hat{s})}
{d p_T},
\ee
where subscripts $a$ and $b$ denote particles ($\gamma, \ p, \dots$) and
$i, \ j, \ k, \ l$ are partons. $\hat{s} = x_1 x_2 s$ as usual, and
$\hat{\sigma}$ are hard partonic scattering cross--sections. Note that
$d \hat{\sigma} / d p_T \propto p_T^{-3}$; the cross--section defined in
eq.(\ref{e5.1}) therefore depends very sensitively on \ptmin, which is
supposed to parametrize the transition from perturbative to nonperturbative
QCD.

If $\rs \gg \ptmin$, the integral in eq.(\ref{e5.1}) receives its dominant
contribution from $x_{1,2} \ll 1$. The relevant parton densities can then be
approximated by a simple power law, $f \propto x^{-J}$. In case of $pp$ or
\ppbar\ scattering, $a=b$ and the cross--section asymptotically scales like
\cite{180}
\be \label{e5.2}
\sigma^{\rm jet} \propto \frac {1} {p^2_{T,{\rm min}}} \left( \frac {s}
{4 p^2_{T,{\rm min}}} \right)^{J-1} \log \frac {s} {4 p^2_{T,{\rm min}}},
\ee
if $J>1$. For $J \simeq 1.3$, as measured by HERA, the jet cross--section
will therefore grow much faster than the total \ppbar\ cross--section, which
only grows $\propto \log^2 s$ (Froissart bound \cite{181}), or,
phenomenologically \cite{182} for $\rs \leq 2$ TeV, $\propto s^{0.08}$.
Eventually the jet cross--section (\ref{e5.1}) will therefore exceed the total
\ppbar\ cross--section.

This apparent paradox is solved by the observation that, by definition,
inclusive cross--sections include a multiplicity factor. Since a hard partonic
scattering always produces a pair of (mini--)jets, we can write
\be \label{e5.3}
\sigma_{ab}^{\rm jet} = \langle n_{\rm jet \ pair} \rangle
\sigma_{ab}^{\rm inel},
\ee
where $\langle n_{\rm jet \ pair} \rangle$ is the average number of
(mini--)jet pairs per inelastic collision. $\sigma_{ab}^{\rm jet} >
\sigma_{ab}^{\rm inel}$ then implies $\langle n_{\rm jet \ pair} \rangle > 1$,
which means that, on average, each inelastic event contains more than one hard
partonic scatter. The simplest possible assumption about these multiple
partonic interactions is that they occur completely independently of each
other, in which case $n_{\rm jet \ pair}$ obeys a Poisson distribution. At a
slightly higher level of sophistication, one assumes these interactions to
be independent only at fixed impact parameter $b$; indeed, it seems natural
to assume that events with small $b$ usually have larger  $n_{\rm jet \
pair}$. This leads to the eikonal formalism \cite{183}, where one writes the
cross--section as an integral over the two--dimensional impact parameter $b$:
\be \label{e5.4}
\sigma_{ab}^{\rm inel}(s) = P_{ab}^{\rm had} \int d^2 b \left[ 1 - \exp \left(
- \frac { A_{ab}(b) \chi_{ab}(s) } { P_{ab}^{\rm had} } \right) \right].
\ee
Here $P_{ab}^{\rm had}$ is the probability that both initial particles
are in a hadronic state when they interact (see below), and $A_{ab}$ describes
the transverse overlap of the partons. In realistic analyses \cite{184} it is
necessary to introduce both nonperturbative (soft) and hard contributions to
the eikonal $\chi_{ab}$:
\be \label{e5.5}
\chi_{ab}(s) = \sigma_{ab}^{\rm soft}(s) + \sigma_{ab}^{\rm jet} (s),
\ee
where $\sigma_{ab}^{\rm jet}$ is given by eq.(\ref{e5.1}). The connection to
the Poisson distribution then becomes evident from the identity
\beq \label{e5.6}
\sigma_{ab}^{\rm inel} &= P_{ab}^{\rm had} \int d^2 b \left[ 1 - \exp \left(
- \frac { A_{ab}(b) \sigma_{ab}^{\rm jet} } { P_{ab}^{\rm had} } \right)
\right]
\nonumber \\
&+ P_{ab}^{\rm had} \int d^2 b \exp \left( - \frac { A_{ab}(b)
\sigma_{ab}^{\rm jet} } { P_{ab}^{\rm had} } \right) \left[ 1 - \exp \left(
-\frac { A_{ab}(b) \sigma_{ab}^{\rm soft} } {P_{ab}^{\rm had}} \right) \right]
\nonumber \\
&= P_{ab}^{\rm had} \sum_{n=1}^{\infty} \int d^2 b \exp \left(
-\frac { A_{ab}(b) \sigma_{ab}^{\rm jet} } { P_{ab}^{\rm had} } \right)
\cdot \left( \frac { A_{ab}(b) \sigma_{ab}^{\rm jet} } { P_{ab}^{\rm had} }
\right)^n \frac {1} {n!} \ \ + \cdots,
\eeq
where the dots stand for the second line in eq.(\ref{e5.6}). Each term in the
sum corresponds to the cross--section for events with exactly $n$ hard
partonic scatters; the second line gives the cross--section for events without
any hard scatter [probability $= \exp \left( - \frac { A_{ab}(b)
\sigma_{ab}^{\rm jet} } { P_{ab}^{\rm had} } \right)$, for fixed impact
parameter $b$], but with a soft interaction. Note that the inclusive jet
cross--section (\ref{e5.1}) can be recovered from eq.(\ref{e5.6}) by
introducing a multiplicity factor $n$ in the sum:
\beq \label{e5.7}
\sigma_{ab}^{\rm jet} &= P_{ab}^{\rm had} \sum_{n=1}^{\infty} n \cdot
\int d^2 b \exp \left( -\frac { A_{ab}(b) \sigma_{ab}^{\rm jet} }
{ P_{ab}^{\rm had} } \right) \cdot \left( \frac { A_{ab}(b)
\sigma_{ab}^{\rm jet} } { P_{ab}^{\rm had} } \right)^n \frac {1} {n!}
\nonumber \\
&= \int d^2b A_{ab}(b) \sigma_{ab}^{\rm jet},
\eeq
since the function $A_{ab}$ describing the transverse overlap of the partons
in the two projectiles is by definition normalized such that $\int d^2 b
A_{ab}(b) = 1$.

We have already emphasized that, for fixed impact parameter, the ansatz
(\ref{e5.4}) assumes hard scatters to occur independently of each other. This
cannot be strictly true, since energy conservation implies that the maximal
Bjorken$-x$ for the partons in the second hard interaction is less than 1, but
it might still be a good approximation at high energies. A second assumption
is that the dependence of the parton densities on Bjorken$-x$ and on the
impact parameter $b$ factorizes; only in this case can the dependence on $b$
be parametrized by a single function $A_{ab}(b)$. At present we are not (yet?)
able to derive these properties from first principles. The ansatz (\ref{e5.4})
therefore goes beyond perturbation theory, even though it postulates that
perturbative interactions play an important role in minimum bias events; it
clearly needs to be checked against experimental data to assess its validity.

It has been shown \cite{184} that eq.(\ref{e5.4}) with $P_{p \bar{p}}^{\rm
had} = 1$ can be brought into agreement with data on total \ppbar\
cross--sections, with \ptmin\ around 1.5 GeV. In these calculations $A_{p
\bar{p}}(b)$ is usually computed from the Fourier transform of the charge form
factor of the proton; this amounts to the assumption that colour charges track
the distribution of electric charge in the nucleon. Moreover, the soft
cross--section is usually parametrized as
\be \label{e5.8}
\sigma_{p \bar{p}}^{\rm soft} (s) = \sigma_0 + \frac {\sigma_1} {\rs},
\ee
where $\sigma_{0,1}$ are constants. Including \ptmin, one therefore has three
new free paramters (beyond those describing high$-p_T$ jet production) to fit
total and inelastic cross--sections (the elastic cross--section can be
calculated in this model using the optical theorem); the success of such a fit
is not entirely trivial. In particular, the rise of the total cross--section
appears very naturally here, although the rate of increase is determined by
\ptmin, and is thus fitted rather than predicted. These fits do have some
weaknesses, however. They basically ignore the existence of diffractive
events; more exactly, if the diffractive cross--section is supposed to be part
of $\sigma^{\rm soft}$, eq.(\ref{e5.8}), then eq.(\ref{e5.6}) leads to the
prediction that the fraction of diffractive events (which are part of the
second term, without hard scatters) decreases quickly with energy. A remedy
within the framework of the minijet model is possible \cite{185}, but only at
the cost of introducing additional free parameters. Secondly, existing
analyses typically assume that parton densities only increase slowly with
decreasing $x$, at least at scale $\mu^2 \simeq p^2_{T,\rm{min}}$; HERA DIS
data favour a much steeper behaviour \cite{13}. It is not clear whether the
faster increase of $\sigma^{\rm jet}$ predicted by such parton densities can
be compensated by re--fitting the free parameters of the model. We should also
keep in mind that total cross--section data can just as well be fitted in a
more conventional Pomeron picture \cite{182}.

Fortunately there is more direct evidence that an ansatz like eq.(\ref{e5.4})
can describe some features of hadronic interactions. Using eq.(\ref{e5.6}),
and the standard machinery to describe parton $\to$ hadron transitions, the
idea of having multiple partonic interactions within a single \ppbar\ reaction
can be built into an event generator \cite{186,187,188}. In ref.\cite{186} it
was shown that many features of \ppbar\ scattering as seen at the CERN SpS can
be understood in such a picture. These authors did not try to fit total
cross--section data; instead, \ptmin\ was determined from the measured
multiplicity distribution. It is encouraging that this again leads to a value
around 1.5 GeV. Multiple interactions then naturally lead to the observed
broad (non--Poissonian) multiplicity distribution. The perhaps greatest
success of this ansatz is that it reproduces the ``jet pedestal" effect. It
had been observed that events with hard jets ($E_T > 15$ GeV or so) also
contain more hadronic activity far away from the jet cores (in the
``pedestal") than minimum bias events do. Eq.(\ref{e5.4}) predicts such a
behaviour, since events with hard jets are likely to have small impact
parameter $b$, which significantly enhances the probability of having
additional semi--hard interactions (minijets) in the same events. Further
qualitative and quantitative successes of this ansatz are described in
refs.\cite{186,187}. However, it should be noted that these analyses again
assume a rather flat small$-x$ behaviour of the parton distribution functions.

The most direct confirmation of eq.(\ref{e5.4}) would obviously be the
observation of events with (at least) two independent jet pairs due to
multiple interactions. These jets should be pairwise back--to--back in the
transverse plane; the transverse opening angle should have a flat distribution
for these events. Both properties are quite distinct from QCD $2 \to 4$ events
(usually called ``double bremsstrahlung"). The cross--section for the
production of at least two independent jet pairs can be computed by inserting
the multiplicity factor $\left( \begin{array}{c} n \\ 2 \end{array} \right)$
into the sum in eq.(\ref{e5.6}), and summing all $n \geq 2$:
\beq \label{e5.9}
\sigma_{ab}(\geq 2 \ {\rm jet \ pairs}) &= \frac {1} {2 P_{ab}^{\rm had}}
\left[ \sigma_{ab}^{\rm jet}(s) \right]^2 \int d^2b A^2_{ab}(b)
\nonumber \\
&\equiv \frac{1}{2} \frac{ \left[ \sigma_{ab}^{\rm jet} (s) \right]^2 }
{\sigma_{\rm eff}}.
\eeq
Energy conservation will again necessitate a slight modification of this
simple expression.

The first experimental search for multiple interactions was performed by the
AFS collaboration \cite{189} at the CERN ISR ($pp$ collisions at $\rs=63$
GeV). They required $\sum E_T({\rm jet}) > 28$ GeV, which implies $\sum_i x_i
\geq 0.45$. It is doubtful whether multiple interactions at these rather large
Bjorken$-x$ can be treated as independent; indeed, implementing energy
conservation here reduces the expected event rate by about a factor of 4.
Moreover, no complete calculation of the QCD $2 \to 4$ background processes
was available at the time, so that this background had to be modelled using a
leading logarithmic approximation in a region of phase space where no large
logarithms occur. The large signal reported by the AFS collaboration,
corresponding to $\sigma_{\rm eff} \simeq 5$ mb, therefore should be taken
with a grain of salt.

More recently the UA2 collaboration \cite{190} found some indication for the
existence of multiple interactions, but due to its rather small statistical
significance ($\sim 3$ standard deviations) they prefer to only quote the
lower bound $\sigma_{\rm eff} > 8.3$ mb. Finally, the CDF collaboration
\cite{191} found a 2.7 $\sigma$ signal, corresponding to $\sigma_{\rm eff}
\simeq 12$ mb, in the range expected by model calculations.

In the first theoretical studies \cite{192} of minijet production in \gamp\
collisions, eq.(\ref{e5.4}) with $P^{\rm had}_{\gamma p} = 1$ was assumed to
hold also for the case of photoproduction. In this case eikonalization effects
are very small even at HERA energies, and a rather rapid increase of the total
cross--section was predicted. However, as first pointed out by Collins and
Ladinsky \cite{8}, $P^{\rm had}_{\gamma p}$ should be ${\cal O}(\aem)$. This
enhances the exponent in eq.(\ref{e5.4}) by a factor of order $1/\aem$, so
that eikonalization does become relevant at HERA energies if $\ptmin \leq 2.5$
GeV. The necessity to have $P^{\rm had}_{\gamma p} \sim {\cal O}(\aem)$ can
most easily be seen from eq.(\ref{e5.6}), once we recognize that
$\sigma_{\gamma p}^{\rm jet}$ is ${\cal O}(\aem)$; if $P^{\rm had}_{\gamma p} =
1$, the production of additional minijet pairs would therefore be suppressed
by powers of \aem, even though only strong interactions are involved once the
photon has been transformed into a hadronic state. Notice also that $P^{\rm
had}_{ab}$ cancels out in eq.(\ref{e5.7}).

In ref.\cite{8}, $P^{\rm had}_{\gamma p}$ was taken to be the $\gamma \to
\rho$ transition probability $\simeq 1/300$. In ref.\cite{193} it was
suggested to instead estimate it as the momentum fraction carried by partons
in the photon at scale $\mu^2 \simeq p^2_{T,{\rm min}}$; this gives slightly
larger values $P^{\rm had}_{\gamma p} \simeq 1/200$. One can (roughly)
reproduce the HERA measurements with either number.

One might argue that $P^{\rm had}_{\gamma p}$ should really be of order
$\aem/\alphas$; after all, in eq.(\ref{e5.6}) one wants the production of a
second jet pair to be suppressed by a factor $\alpha_s^2$, not just a single
power of \alphas.\footnote{Recall that even the resolved photon contribution
to $\sigma_{\gamma p}^{\rm jet}$ is ${\cal O} (\aem \alphas)$, since the
parton densities in the photon are ${\cal O} (\aem /\alphas)$.} However, it
should be recognized that $P^{\rm had}_{\gamma p}$ cannot be discussed
independently of $A_{\gamma p}(b)$. In refs.\cite{193,194,195} $A_{\gamma
p}(b)$ was computed from the Fourier transform of the pion form factor; it is
not at all clear whether this describes the transverse distribution of partons
in a vector meson properly, and it is certainly not applicable to the
perturbative (``anomalous") component of the photon structure functions.
Mathematically, eq.(\ref{e5.4}) only depends on the combination $A_{ab}/P^{\rm
had}_{ab}$; one can see this by using the substitution $b' = b \cdot
\sqrt{P^{\rm had}_{ab}}$:

\be \label{e5.10}
\sigma_{ab}^{\rm inel}(s) = \int d^2 b' \left[ 1 - \exp \left(
- A'_{ab}(b') \chi_{ab}(s) \right) \right],
\ee
with $A_{ab}'(b') = A_{ab}\left(b'/\sqrt{P_{ab}^{\rm had}}\right) /
P_{ab}^{\rm had}$. Note that $\int d^2 A(b) = 1$ implies $\int d^2 b' A'(b') =
1$ as well; nevertheless $b'$ cannot be interpreted as the physical impact
parameter. Eq.(\ref{e5.10}) shows that one can always compensate an increase
in $P_{ab}^{\rm had}$ by a steeper decrease of $A_{ab}(b)$ at large $b$, which
implies an increase of $A(b)$ at small $b$. One can therefore simply fix
$P^{\rm had}_{\gamma p}$ to some value and fit $A_{\gamma p}(b)$ to data, or
vice versa; this approach was followed in ref.\cite{195}.

However, in case of photons the ansatz (\ref{e5.4}) might in any case be too
simple. As already discussed in the Introduction, the hadronic structure of
the photon is generally assumed to have both perturbative and
nonperturbative components. The corresponding contributions to the parton
densities have quite different $x$ and $Q^2$ dependence. It therefore makes
sense to split the cross--section (\ref{e5.4}) into (at least) two terms,
characterized by different values of $P^{\rm had}$, different parton densities
[and hence different $\sigma^{\rm jet}$, see eq.(\ref{e5.1})], and presumably
also different overlap functions $A(b)$. Yet another term has to be introduced
to describe direct interactions; these cannot be eikonalized, since here the
photon is ``used up" after the first interaction.

Two different ans\"atze of this kind have been proposed so far. Both Honjo et
al. \cite{196} and Schuler and Sj\"ostrand \cite{197,36a} split the hadronic
photon into a discrete sum over vector mesons, and a perturbative
(``anomalous") contribution; symbolically
\be \label{e5.11}
| \gamma \rangle = | \gamma \rangle_{\rm bare} + \sum_{\rho,\omega,\phi}
\frac {e}{f_V} |V \rangle + \sqrt{P_{q \bar q}} | \qqbar \rangle,
\ee
where $| \gamma \rangle_{\rm bare}$ is the ``direct" photon\footnote{Strictly
speaking the coefficient of the first term in eq.(\ref{e5.11}) should differ
from unity by an amount of order \aem; this effect can safely be ignored.},
$e/f_V$ are the $\gamma \to V$ transition amplitudes, $|V \rangle$ is a vector
meson state, and $| \qqbar \rangle$ is a state that develops from a hard
$\gamma \to \qqbar$ splitting. The sum in eq.(\ref{e5.11}) is assumed to be
incoherent, so that each term also contributes separately to the parton
densities and to the total \gamp\ cross--section. In refs.\cite{196,197} the
vector meson contributions to the parton content of the photon were described
in terms of pion structure functions, while in ref.\cite{36a} the shapes of
these contributions were fitted from \f2gam\ data (SaS parametrization; see
Sec.~2b). The perturbative contribution to eq.(\ref{e5.11}) is really a
continuum of states; its contribution to the parton densities can be written
as \cite{36a}
\be \label{e5.12}
f_{i|\gamma}^{\rm pert} (x,\mu^2) = \frac {\aem}{\pi} \sum_q
\int_{k_0^2}^{\mu^2} \frac {d k^2}{k^2} f_{i|q \bar q} (x,\mu^2, k^2).
\ee
Note that the perturbative $\gamma \to \qqbar$ transition has been factored
out here; it gives rise to the factor $dk^2/k^2$. The ``state distributions"
$f_{i|q \bar q}$ therefore obey {\em homogeneous} evolution equations in the
factorizatioon scale $\mu^2$, with the (leading order) boundary conditions
\cite{36a}
\ben \label{e5.13} \beq
f_{q|q \bar q} (x,k^2,k^2) &= f_{\bar{q}|q \bar q} (x,k^2,k^2) = \frac{3}{2}
\left[ x^2 + (1-x)^2 \right]; \label{e5.13a} \\
f_{q'|q \bar q} (x,k^2,k^2) &= f_{G|q \bar q} (x,k^2,k^2) = 0. \label{e5.13b}
\eeq \een

At this point the treatment of refs.\cite{196} and \cite{197,36a} diverges.
Schuler and Sj\"ostrand do not attempt to predict the total \gamp\
cross--section. Rather, the emphasis of their work is on the properties of
photoproduction events, both minimum bias events and events containing hard
jets. They therefore assume a parametrization of the total cross--section as
in ref.\cite{182}. The overall normalization of the nonperturbative
contributions to the photon structure functions, given by the $f_V$, is fixed
by low energy data. This also determines the contribution of this component to
the total cross--section, since the $Vp$ cross--sections are assumed to be
identical to the $\pi p$ cross--section. Moreover, they assume that the
perturbative component (\ref{e5.12}) only contributes to hard scattering
events, which are very rare at $\sqrt{s(\gamp)} \simeq 10$ GeV. However, their
choice of the $f_V$ gives a contribution to $\sigma^{\rm tot}_{\gamma p}$ which
falls about 20\% below the data at these low energies. The difference then has
to come from direct interactions. This forces them to include direct $\gamp$
scattering with partonic transverse momentum as low as 0.5 GeV.\footnote{In
ref.\cite{197} Schuler and Sj\"ostrand modify the proton structure functions
for scales $\mu^2 < 5$ GeV$^2$ and/or small $x$, in order to enforce a smooth
transition between DIS and photoproduction. However, in ref.\cite{36a} they
use standard leading twist QCD to describe \f2gam\ for momentum scales down to
about 1 GeV$^2$.} The same value is also used for the cut--off parameter $k_0$
in eq.(\ref{e5.12}); this is in accordance with ref.\cite{196}, as well as
(approximately) with the GRV \cite{42} and AGF \cite{47} parametrization.

\setcounter{footnote}{0}
Ref.\cite{197} devotes much attention to the description of quasi--elastic and
diffractive events, which are entirely due to contributions from the
$|V \rangle$ states. These states also contribute to soft and semi--hard
(minijet) non--diffractive \gamp\ interactions, where the cut--off $\ptmin
\simeq 1.3$ GeV at $\rs=200$ GeV is fixed from multiplicity distributions of
\ppbar\ events at the same energy. Schuler and Sj\"ostrand allow for multiple
interactions in this sector, but assume them to be completely independent of
each other; in contrast, the ansatz (\ref{e5.4}) assumes independent
scattering {\em only} at fixed impact parameter $b$. We have seen above that
the correlation between the presence of hard jets and small $b$ offered a
natural explanation of the jet pedestal effect. Assuming completely
independent interactions means that every partonic interaction occurs with
probability
\be \label{e5.14}
P_{\rm jet}(p_T) = \frac {1} { \sigma^{\rm nd}_{Vp} } \frac
{d \sigma^{\rm jet}_{Vp} } {d p_T},
\ee
where the superscript ``nd" stands for non--diffractive. Finally, in this
model no multiple interactions are allowed to originate from the perturbative
contribution (\ref{e5.12}) to the photon structure functions. In order to
maintain the assumed $s^{0.08}$ behaviour of the total \gamp\ cross--section
it then becomes necessary to increase \ptmin\ (in this sector only!) linearly
with \rs, so that $\ptmin \simeq 2.2$ GeV for HERA ($\sqrt{s(\gamp)} \simeq
200$ GeV).\footnote{In an alternative version of this model, \ptmin\ is held
fixed in the perturbative sector, while the $f_V$ are assumed to decrease with
energy. Clearly this can at best be a temporary solution, since eventually the
minijet contribution from the perturbative sector alone will exceed the
assumed total \gamp\ cross--section unless it is unitarized in some way.
Schuler and Sj\"ostrand therefore favour the variant of their model with
constant $f_V$. They also discuss a few other versions, not of all of which
are meant to be potentially realistic.}

The great advantage of the model of refs.\cite{197,36a} is that is comes
pre--packaged in the successful LUND/PYTHIA Monte Carlo event generator
\cite{198}. This allows to test many aspects of the model against data from
HERA, as well as from experiments at lower energies. On the positive side, not
only does the total \gamp\ cross--section seem to follow the ``universal"
$s^{0.08}$ behaviour, but the relative sizes of the contributions from
quasi--elastic, diffractive, and non--diffractive events also match more or
less the model predictions. The recent observation by the ZEUS collaboration
\cite{62c} that the transverse momentum $k_T$ of the photon remnant jet in
events with at least two high$-p_T$ jets is much harder than the Gaussian with
width $\sim 0.4$ GeV characteristic for hadronic remnant jets can also be
regarded as a success of this model; indeed, it has been predicted more than
ten years ago \cite{200} that the perturbative contribution should have a
$k_T$ distribution $\propto d k_T^2/k_T^2$ for large $k_T$, which is directly
related to the ansatz (\ref{e5.12}). Moreover, the observation \cite{80} that
in the region below about 1 GeV, the shape of the $p_T-$distribution of
charged particles at HERA closely matches that of \ppbar\ events at the same
energy indicates the presence of a component in the hadronic photon that
behaves similarly to other hadrons. At higher $p_T$, the spectrum becomes
harder, in agreement with perturbative calculations \cite{81} that include
direct contributions as well as contributions from the perturbative part of
the photon structure.

There also appear to be some problems with this model in its present form,
however. We already mentioned that the properties of the ZEUS minimum bias
sample \cite{174} cannot be described properly by PYTHIA. Moreover, when
trying to extract partonic cross--sections (and, ultimately, \gga) from the
measured di--jet cross--section, the H1 collaboration observed \cite{60} that
switching on multiple partonic interactions in PYTHIA is not sufficient to
describe the energy flow in resolved photon events.\footnote{Since including
multiple interactions improved agreement with the data, H1 implicitly assumed
that the remaining difference is due to even more partonic interactions; since
these produce a jet pedestal that is uniform in $\phi$, they can then simply
subtract this pedestal from the measured $E_T$ of the jet to arrive at the
partonic $p_T$ (up to showering and fragmentation effects, which are
presumably described adequately by the model). However, this treatment would
not give the correct answer if the additional energy flow was (partly) due to,
e.g., enhanced initial state radiation, which is not uniform in $\phi$.} It is
at present not clear whether this problem can be solved by replacing the
simple ansatz (\ref{e5.14}) by something like eq.(\ref{e5.4}) (for the
contribution from the $|V \rangle$ states), or whether a more substantial
modification of the model will be necessary.

Finally, we find the introduction of three different scales that are meant to
separate perturbative and nonperturbative interactions not very appealing.
One might argue that \ptmin\ ought to be larger for the pertubative $| \qqbar
\rangle$ component than for the $|V \rangle$ components, since the state
that develops from a hard $\gamma \to \qqbar$ splitting may have a smaller
transverse size (as we will see shortly, this is assumed in ref.\cite{196}),
so that a larger momentum transfer becomes necessary to resolve individual
colour charges. (A similar argument is used in ref.\cite{195} for the photon
as a whole.) However, this does not explain why $k_0$, which is also the
minimal $p_T$ of direct interactions, is so much smaller than even the value
of \ptmin\ used for the semi--hard interactions of the $|V \rangle$
components. Alternatively, one might argue that quark exchange (which
dominates direct interactions and, obviously, $\gamma \to \qqbar$ splitting)
is perturbative down to significantly smaller momentum scales than gluon
exchange (which dominates minijet production in all resolved photon
interactions). This could explain why $k_0$ is smaller than \ptmin, but the
difference between the \ptmin\ values used for the $|V \rangle$ and $|\qqbar
\rangle$ components cannot be explained by such an argument. The different
energy dependence of the two \ptmin\ values is also difficult to understand
intuitively.

In this respect the model of ref.\cite{196} seems somewhat more appealing.
Here a similar value for the cut--off $k_0$ in eq.(\ref{e5.12}) is used, but
all minjet production, from direct photons as well as the different classes
of resolved photons, is assumed to be regularized by a single, energy
independent parameter \ptmin. A fit to low energy data, which show a small
but significant increase of the total \gamp\ cross section between 10 and
about 18 GeV (the highest energy that had been reached in fixed target
experiments), gives $\ptmin \simeq 1.5$ GeV. Since this is significantly
larger than $k_0$, Honjo et al. also allow the $|\qqbar \rangle$ states to
have nonperturbative interactions, which are assumed to scale $\propto
1/k^2$; the idea here is that the ``hardness" $k^2$ of the $\gamma \to \qqbar$
splitting determines the transverse size of the state that develops from this
splitting. Minijet production from this state is also eikonalized, where the
typical transverse size [which determines $A(b)$] again scales like $1/k^2$;
that is, eikonalization \`a la eq.(\ref{e5.4}) occurs for each value of $k$
independently. As written in ref.\cite{196} the model predicts a total \gamp\
cross--section at $\rs = 200$ GeV between 190 and 250 $\mu$b, the lower end of
which is compatible at least with the H1 measurement \cite{175}; presumably
the agreement could be improved if one allowed the $x$ and/or $b$ dependence
of the partons in the $|V \rangle$ states to differ from those in the pion.
However, given the assumptions involved, we do not think that total
cross--section data can be used to constrain parton distribution functions.
Note that this model does predict a jet pedestal effect; it might even be
stonger than in \ppbar\ collisions, since for the perturbative contribution to
the photon structure functions a very hard interaction not only implies small
$b$, but also (at least on average) a larger value of $k^2$ and hence a
narrower $A(b)$, which increases the chances for additional semi--hard
interactions. Unfortunately this model has not yet been built into an event
generator.

We finally mention two slightly different approaches to \gamp\ scattering. In
ref.\cite{201} it was observed that, at least in a single--component model of
the photon\footnote{A single component description was used here since in
ref.\cite{201a} the perturbative component of the photon structure functions
had been found to be too small to have much impact on minijet production.
However, comparison with the explicit calculation of ref.\cite{36a} shows that
the simple estimate of ref.\cite{201a}, which used the double scaling limit of
QCD, greatly under--estimates the perturbative contribution to \gga\ at small
$x$ and scale $\mu^2 \simeq p^2_{T,{\rm min}}$.}, where $P_{\gamma p}^{\rm
had}$ and $A_{\gamma p}(b)$ are determined from ``canonical" VDM ideas, it is
difficult to simultaneously describe the rise of $\sigma^{\rm tot}_{\gamma p}$
between $\rs=10$ and 18 GeV, and the rather small value of the total
cross--section measured at HERA (or at least the smaller ZEUS result
\cite{174}). These authors therefore allowed the {\em soft} contribution to
the cross--section to grow $\propto s^{0.058}$, similar to the behaviour found
earlier in ref.\cite{202}. Adding a minijet contribution estimated using the
DG parametrization with $\ptmin=3.0$ GeV (where eikonalization effects are
still quite small for HERA energies) then gives $\sigma^{\rm tot}_{\gamma
p}(\rs=200 \ {\rm GeV}) \simeq 160 \ \mu$b. A similar approach is taken in
ref.\cite{203}. This work attempts a comprehensive description of \gamp\
scattering, including quasi--elastic and diffractive events, in the framework
of the Dual Parton Model \cite{204}; it is now being implemented in the PHOJET
event generator. Since the rise of the soft cross--section is now also fitted
from data, in these models one can no longer claim to explain the rise of
total cross--sections in terms of minijets. Moreover, $\ptmin \simeq 3.0$ GeV
seems rather large to us; there is evidence from \jpsi\ decays, DIS, and
\epem\ annihilation that perturbative QCD is applicable at lower momentum
scales. This is also supported by analyses of \gaga\ data, to which we turn
next.

Once the description of the photon has been fixed, e.g. by fitting the free
parameters of models like those of refs.\cite{196,197,203} from \gamp\ data,
the properties of \gaga\ interactions can in principle be predicted
unambiguously \cite{205}. However, two--photon experiments have needed
comprehensive event generators well before the recent sophisticated models of
\gamp\ scattering were developed. We already discussed in Sec.~2b that the
determination of Bjorken$-x$ in deep--inelastic $e \gamma$ scattering
necessitates a model of the hadronic final state; in Sec.~4 we mentioned that
the event generators that were used to model quasi--real \gaga\ scattering
prior to the pioneering AMY study \cite{6} did not include resolved photon
processes at all.

The situation has certainly improved a great deal since then; however, there
are still some problems. At present most experiments use the same basic ansatz
to describe real \gaga\ scattering, which contains three separate
contributions. Soft interactions, characterized by an exponential $p_T$
spectrum, are modelled using VDM ideas.\footnote{This ``VDM component" is not
to be confused with the nonperturbative contribution to photon structure
functions, which is often also estimated from the VDM. As emphasized in
refs.\cite{197,205}, these ``VDM partons" also participate in hard resolved
photon interactions.} The second component is estimated from the QPM. At high
$p_T$ it coincides with the direct contribution introduced in Sec.~4, but the
$p_T \to 0$ divergence is here regularized by constituent quark masses
(typically $m_u=m_d \simeq 300$ MeV, $m_s \simeq 500$ MeV), rather than by a
momentum cut--off. This QPM contribution therefore extends to (partonic)
$p_T=0$, and is hence not entirely perturbative.\footnote{Recall that in the
model of refs.\cite{197,205} direct interactions with partonic $p_T$ as low as
500 MeV are allowed; in practice this should give a similar contribution as
the QPM prediction using constituent quark masses.} In ref.\cite{206} it has
been shown that data on total \gaga\ cross--sections and \f2gam\ can be fitted
from these two components only. However, the description of multi--hadron and
jet production in \gaga\ collisions is only possible if one also introduces
single and double resolved contributions, which form the third component of
current \gaga\ event generators.

So far all currently used generators agree. There are some differences in the
details, however. For example, the TRISTAN experiments \cite{6,141,142}, as
well as DELPHI \cite{151,152} use parameter values for the description of the
soft (``VDM") component that have been determined by experiments at lower
energies. This might be dangerous, since these earlier analyses did not allow
for resolved photon interactions; there is a strong correlation between
the assumed size of the soft component and the parameter \ptmin, which
essentially fixes the size of the resolved photon contribution to the minimum
bias event sample. The ALEPH collaboration \cite{150} therefore preferred to
fit both \ptmin\ and the parameters describing the soft contribution from
their own data. This might explain the rather large value of \ptmin, 2.5 GeV,
obtained by ALEPH for the DG parametrization, compared to 1.45 GeV for DELPHI
and 1.6 to 2.0 GeV for AMY and TOPAZ. The expression for the total \gaga\
cross--section derived by ALEPH also differs significantly from conventional
VDM expectations. It should be noted that the jet cross--sections derived by
TRISTAN experiments \cite{141,142} are based only on high$-p_T$ events; these
analyses are therefore not sensitive to the assumed value of \ptmin.
Furthermore, the experimental definition (trigger conditions) of the ``minimum
bias" sample differs quite significantly between the various experiments,
with ALEPH having the loosest cuts and largest visible cross--section, while
AMY uses quite stringent requirements. Of course, a complete event generator
should still give the same values (within errors) for its free parameters even
when quite different data samples are used for the fit. The discrepancies
between the values of \ptmin\ determined by different experiments therefore
indicates that (some) present generators are not complete.

One problem of these generators is that they do not include initial and final
state radiation (parton showers). ALEPH states \cite{150} that generators like
PYTHIA \cite{198} and HERWIG \cite{207} with parton showering switched on did
not describe the data well. However, QCD tells us that such radiation must
exist at some level. It might, for example, change the $p_T-$dependence of the
partonic cross--sections extracted by TOPAZ \cite{141} and AMY \cite{142},
although the string fragmentation scheme used in these analyses may mimick
some of these effects; a preliminary TOPAZ study \cite{148} finds that their
MC program gives a reasonable, but not perfect description of the distribution
of the transverse opening angle $\phi$ between the jets in di--jet events.
Note that initial state radiation should be treated differently for photons
than for real hadrons \cite{197,208}, due to the presence of the hard $\gamma
\qqbar$ vertex. Whenever one reaches this vertex in the standard backward
evolution \cite{209} of initial state radiation, the shower has to be
terminated, even if the parton's virtuality is still quite high at this point.
This feature is included in the latest versions of PYTHIA and HERWIG, at least
in an average ($x-$independent) sense.

\setcounter{footnote}{0}
Another shortcoming of present \gaga\ event generators is that they do not
allow for the hard transverse momentum distribution of the remnant jets that
has been predicted by theory \cite{200} and seen experimentally by ZEUS
\cite{62c}. This seems to have little impact on the $p_T-$spectrum of the hard
jets \cite{62c}, but might change the predicted efficiency for detecting
remnant jets, e.g. in the TOPAZ detector \cite{210a,141,164}.

Finally, these event generators at present do not allow for multiple partonic
interactions. Even if only the contribution from the nonperturbative part
of photon structure functions is eikonalized, multiple interaction effects
will play a significant role from LEP energies onward \cite{210}. In a
single--component model of the photon, eq.(\ref{e5.4}) with $P^{\rm
had}_{\gamma \gamma} \sim \alpha^2_{\rm em}$, four jet events due to multiple
parton scattering might already be detectable at TRISTAN\footnote{Even in this
case they are not expected to significantly affect the properties of events
with one or two high$-p_T$ jets; they do therefore not invalidate the
analyses of refs.\cite{141,142}}, and ought to become a prominent feature of
two--photon events at LEP2 energies \cite{211}. The study of \gaga\ collisions
at the highest available energies should therefore tell us how to unitarize
the contribution to minijet production from the perturbative (``anomalous")
component of the photon structure.

Apart from its intrinsic interest, a good understanding of minimum bias \gaga\
events is necessary for a realistic evaluation of possible hadronic
backgrounds at future linear \epem\ colliders. In ref.\cite{212} it was
pointed out that in designs with strong beamstrahlung, several hadronic \gaga\
collisions might occur in each bunch crossing, thereby effectively giving rise
to an underlying event. This conclusion has been criticized \cite{213} on the
grounds that the minijet contribution had not been eikonalized in
ref.\cite{212}. However, in refs.\cite{154,214} is has been shown that
eikonalization does not significantly change predictions for the next
generation of linear colliders, planned to operate at $\rs \simeq 500$ GeV.
The ``hadron crisis" has not been solved by improved calculations, but by
improved machine designs, which reduce beamstrahlung by employing a larger
number of flat bunches. It should be emphasized that hadronic backgrounds are
not the only, and perhaps not even the most severe, problem of designs with
high beamstrahlung; they also suffer from large backgrounds from soft \epem\
pairs \cite{215}, as well as from a poorly defined beam energy, which
complicates the study of thresholds \cite{216}.

So far only one Monte Carlo study of the properties of these soft hadronic
backgrounds has been published, by Chen et al. \cite{214}. They assume that
the total \gaga\ cross--section tracks the measured \ppbar\ cross--section at
high energies. It should be pointed out that at present even the value of the
total \gaga\ cross--section at low energies is only poorly determined
experimentally. Chen et al. then modelled minimum bias events using ISAJET
\cite{217}, which contains a parametrization of the underlying event as
determined in $pp$ and \ppbar\ collisions. Using the same model for \gaga\
collisions may not be a bad first approximation, given the similarities of
some features of HERA and \ppbar\ events discussed earlier. However, Chen et
al. did not attempt to compare their model with actual data. They also used a
rather large value for \ptmin, 3.6 GeV, this being (approximately) the
smallest partonic $p_T$ leading to observable jets at the UA1 experiment
\cite{218} at the SpS. However, TRISTAN experiments have demonstrated
\cite{141,142,148} that much softer jets can be reconstructed in \gaga\
collisions. The analysis by Chen et al., which uses the simplified
unitarization scheme (\ref{e5.14}), therefore probably under--estimates the
effect of minijets on event characteristics.\footnote{As noted above, ISAJET
simply fits the properties of the underlying event in \ppbar\ collisions,
including effects that can be explained in terms of minijets. However, it is
known that \gamp\ events are more ``jetty" than \ppbar\ events, and hard
partonic reactions are expected \cite{205} to play an even more prominent role
in \gaga\ scattering.} Chen et al. conclude that the average \gaga\ event only
deposits about 3.5 GeV of hadronic energy in the central detector ($|\cos
\theta | \leq 0.9$) at a typical design for a 500 GeV \epem\ collider. It
would be interesting to repeat their analysis, using an event generator that
has been tuned to describe real \gaga\ data.

Finally, we briefly mention that eikonalization effects \cite{213} do
certainly play an important role if the next \epem\ collider is converted into
a \gaga\ collider, which greatly increases the average \gaga\ cms energy. The
analysis of ref.\cite{214} indicates that even here one can stay well below
one hadronic event per bunch crossing, at least at a 500 GeV collider; the
problem rapidly gets worse at higher energies, mostly due to the need to
increase the luminosity $\propto s$ in order to achieve a constant rate of
hard events. Since at those colliders \gaga\ scattering occurs at energies
well beyond the reach of current colliders, and in view of the incomplete
description even of these accessible events, any prediction of event
properties at future \gaga\ colliders should be taken with a grain of salt.
\setcounter{footnote}{0}
\section*{6) Outlook}
In this article we have reviewed the present knowledge of resolved photon
interactions. Great progress has been made in the last few years, both
experimentally and theoretically. We now know for a fact that partons ``in"
(quasi--)real photons play an important role in both \gamp\ and \gaga\
interactions. Theory tells us that the hard $\gamma \qqbar$ vertex should lead
to (a component of the) photonic parton densities that have a much harder
$x-$dependence than the more familiar nucleonic parton distribution functions,
and which should grow approximately linearly with the logarithm of the
momentum scale at which the photon is being probed. Both properties have been
confirmed experimentally, at least for the quark densities. Studies of
photonic remnant jets also indicate that resolved photons do not always act
like ordinary hadrons.

At the same time many properties of minimum bias photoproduction and two
photon events do resemble those of purely hadronic collisions at similar
energies. Moreover, the total \gamp\ cross--section seems to show
approximately the same energy dependence as total hadronic cross--sections.
We therefore have good evidence for the existence of both perturbative and
nonperturbative contributions to the hadronic structure of the photon.

Qualitatively speaking this summarizes the present level of understanding of
resolved photons. In the preceding sections we have reviewed the various
pieces of information that contributed to this understanding; we have also
described attempts to cast this understanding in a quantitative form. It seems
rather futile to try and summarize this summary. We therefore decided to
conclude this article with a list of open problems, which might become the
foci of future work.

The most pressing problems in \ub{deep--inelastic $e \gamma$ scattering}
(Sec.~2) have to do with the description of the final state:
\begin{itemize}
\item
The measurement of Bjorken$-x$ has traditionally relied on a reconstruction of
the invariant mass $W$ of the hadronic final state from the measured quantity
$W_{\rm vis}$. An improvement of this procedure has recently been suggested
\cite{36b}. It might also be possible to determine the energy of the target
photon using the Jacquet--Blondel method, as is done by HERA experiments.
Detailed Monte Carlo work, using realistic detector resolutions and
acceptances, is needed to decide how well these ideas work in practice.
\item
There are some discrepancies between existing data on \f2gam. Improved
reconstruction methods might be of some help here. It is not clear whether
older PEP and PETRA data can be re--analyzed in this way, but new measurements
and/or new analyses of \f2gam\ at small $x_\gamma$ might help to resolve the
discrepancy between recent results by the TOPAZ and OPAL collaborations. Such
studies have the potential to discriminate between existing parametrizations
of photonic parton densities.
\end{itemize}

A list of open problems in \ub{hard \gamp\ scattering} (Sec.~3) includes:
\begin{itemize}
\item
No complete NLO treatment of di--jet production exists. Recall that one needs
to measure the rapidities of both high$-p_T$ partons/jets in a hard event in
order to reconstruct the Bjorken$-x$ variables.
\item
The measured jet cross--sections should be extended both in rapidity and in
$p_T$. The former increases the sensitivity to the interesting region of small
\xgam, while the latter should allow to test theory cleanly, since a detailed
understanding of the underlying event (see below) is less crucial at high
$p_T$, and differences between parametrizations of photonic parton densities
are small at large $x_\gamma$ and large momentum scale.
\item
It might be interesting to try to correlate properties of the photonic remnant
jet with those of the high$-p_T$ jets. In the usual ``nonperturbative +
anomalous" description of the hadronic photon, the nonperturbative component
should always have a remnant jet with very small $k_T$; this component is also
characterized by soft parton densities. In this picture one therefore expects
nontrivial correlations between \xgam\ and $k_T$.
\item
Studies of heavy flavour production hold great potential. We do not think it
very interesting to try to derive total cross--sections from measurements
covering only a limited region of phase space, which contains only a small
fraction of all produced heavy quarks. It might be more fruitful to attempt to
extract the resolved photon contribution, which is sensitive to the as yet
poorly constrained gluon density in the photon. At high $p_T$, ``excitation"
contributions from the charm in the photon have to be taken into account. An
important open problem is the fragmentation of rather soft (low$-p_T$) charm
quarks, which contribute most to the total charm cross--section. Theoretical
predictions are more reliable for $b$ production, but it might be difficult to
find a clean signal.
\item
The production of direct photons is by now quite well understood, although an
NLO calculation of photon + jet production would certainly be welcome.
Realistic background studies are also needed, but can presumably only be
performed by members of HERA experiments.
\item
In spite of the recent NLO calculation \cite{116} of direct \jpsi\ production
in the colour singlet model, much needs to be done here: The resolved photon
contribution is only known to leading order in the colour singlet model.
Nothing is known about the contribution from the colour octet component of the
wave function of the \jpsi\ \cite{219}, which is accessible to resolved
photons already in LO. In addition, there are contributions at high $p_T$
coming from charm and gluon fragmentation \cite{220}.
\item
It is important to test our understanding of the hadronic photon in as many
different channels as possible. The production of Drell--Yan lepton pairs,
two photon final states, and associate $\jpsi + \gamma$ final states are all
plagued by rather small cross--sections, but this should at least partly be
compensated by the cleanliness of the final states.
\end{itemize}
We should emphasize that many of the measurements proposed here need more
luminosity than HERA has delivered so far. Fortunately new data are being
collected at an ever accelerating pace.

Some open problems in \ub{hard \gaga\ scattering} (Sec.~4) are:
\begin{itemize}
\item
An NLO calculation of di--jet production is still lacking.
\item
There is some discrepancy in the high$-p_T$ end between the partonic
cross--sections published by AMY and TOPAZ. Also, it is not clear how these
cross--sections, which have been extracted with the aid of leading order event
generators, can be compared to NLO calculations.
\item
More detailed tests of theory should be possible if the kinematic
reconstruction of events with high$-p_T$ jets can be improved. One possibility
might be to determine the Bjorken$-x$ variables using small angle electron
taggers and/or the Jacquet--Blondel method. A study of the energy distribution
of the remnant jet(s), and possible correlations with the hard jet(s), might
also prove rewarding. All these measurements are probably quite difficult,
however.
\item
Measurements of the cross--sections for the production of high$-p_T$ particles
should be relatively straightforward; they test similar aspects of the theory
as studies of jet final states do.
\item
The excess of high$-p_T$ charm events seen by TRISTAN experiments might well
be explainable in terms of charm excitation from the photon; in the
kinematical regime that is being probed by these experiments, this should be
well described by the NLO event generator that has become available recently.
Recall, however, that ALEPH does not see any excess.
\item
\jpsi\ and direct photon production has so far only been treated in LO. No
data on these final states exists as yet.
\end{itemize}

Finally, some of the most challenging problems go \ub{beyond perturbation
theory} (Sec.~5):
\begin{itemize}
\item
We have to understand the energy flow in resolved photoproduction events. This
is crucial for the comparison of measured jet rates with theoretical
(partonic) calculations, especially for presently accessible values of $p_T$.
\item
The modelling of minimum bias photoproduction events needs to be improved.
Among other things, this is necessary for an accurate determination of total
\gamp\ cross--sections.
\item
The total hadronic \gaga\ cross--section is at present only determined very
poorly. An accurate measurement can probably only be made if at least one of
the outgoing electrons is tagged at a small angle. This is important for the
assessment of hadronic backgrounds at future linear \epem\ colliders, which
may play a role in the decision which of several competing designs should be
pursued.
\item
The role of multiple partonic interactions needs to be clarified. They will
probably (help to) explain the energy flow in hard \gamp\ events mentioned
above. Theoretically least understood is the treatment of contributions from
the perturbative component of photon structure functions. Studies of
multi--jet production in high energy \gaga\ collisions might yield important
clues here. This also has ramifications for a theory (as opposed to
parametrization) of total hadronic cross--sections.
\item
\gaga\ event generators in present use are known to be incomplete: They do not
include parton showers, and assume too soft a $k_T$ distribution of the
remnant jets. At higher energies multiple interactions might also affect the
properties of the underlying event substantially. Finally, some
standardization might help to decide whether putative differences in published
results signal real discrepancies in the data, or are merely a reflection of
different MC generators.
\item
It is important to test the (potentially) complete event generators for \gamp\
and \gaga\ scattering that are now being developed in as many different ways
as possible. Not only should things like the energy flow, multiplicity
distributions, particle $p_T$ spectra, and strangeness production be
investigated, there might also be interesting correlations between some of
these quantities.
\end{itemize}

This list is almost certainly not complete. Moreover, new questions are likely
to arise as soon as current problems are solved. Much needs to be done before
we can say with confidence that we understand the structure of light.

\subsection*{Acknowledgements}
We greatly benefitted from numerous discussions with many people working in
this field. In particular, we wish to thank P. Aurenche, G. B\'elanger, F.
Borzumati, F. Boudjema, J. Butterworth, R. Enomoto, J. Forshaw, L. Gordon, M.
Greco, H. Hayashii, M. Iwasaki, D. Miller, T. Nozaki, G. Schuler, M. Seymour,
T. Sj\"ostrand and J. Storrow for their replies to our questions.
The work of M.D. was supported in part by the U.S. Department of Energy under
grant No. DE-FG02-95ER40896, by the Wisconsin Research Committee with funds
granted by the Wisconsin Alumni Research Foundation, as well as by a grant
from the Deutsche Forschungsgemeinschaft under the Heisenberg program. R.M.G.
wishes to acknowledge research grant no. 03(0745)/94/EMR-II, from the Council
of Scientific and Industrial Research.

\clearpage
\section*{Figure Captions}
\renewcommand{\labelenumi} {Fig. \arabic{enumi}}
\begin{enumerate}

\vspace{3mm}
\item
Generic Feynman diagram for deep--inelastic $e \gamma$ scattering: A probing
photon with large virtuality $Q^2 \equiv -q^2$ scatters off the (quasi--)real
target photon (with virtuality $P^2 \simeq 0$) to produce a hadronic final
state $X$.

\vspace{3mm}
\item
Feynman diagrams that contribute to $\gamma^* \gamma$ scattering, i.e. the
blob in Fig.~1. (a) shows a tree--level (QPM) contribution, while (b) and (c)
are examples of perturbative QCD corrections in the flavour non--singlet and
flavour singlet sector, respectively, and (d) shows nonperturbative
contributions as estimated in the VDM. All these contributions are formally
summed in (e), where the quark densities in the photon $q_i^{\gamma}$ are
introduced.

\vspace{3mm}
\item
Predictions of various existing parametrizations of photonic parton densities
for the structure function \f2gam\ at $Q^2=15$ GeV$^2$ are compared with data
from the OPAL \cite{28} (squares) and TOPAZ \cite{32} (diamonds)
collaborations.

\vspace{3mm}
\item
Predictions of various existing parametrizations of photonic parton densities
for the gluon content in the photon (a), and for the ratio of strange to
non--strange light quark densities (b). Notice that we have used a logarithmic
scale for the $y-$axis in (a), while in (b) $x$ has been shown on a
logarithmic scale. Both figures are for $Q^2=15$ GeV$^2$.

\vspace{3mm}
\item
Feynman diagrams contributing to the photoproduction of high$-p_T$ jets.
Direct contributions from the QCD Compton scattering and photon--gluon fusion
are shown in (a), while (b) gives examples of resolved photon contributions
involving the scattering of two quarks, a quark and a gluon, or two gluons.

\vspace{3mm}
\item
The topology of photoproduction events. In (a) the incident photon couples
directly, leading to a final state with two high$-p_T$ jets (in leading
order), and a remnant jet from the proton. In resolved photon contributions
(b) the photon gives rise to a second remnant jet.

\vspace{3mm}
\item
Example of a diagram that can be interpreted as either a leading order direct
or resolved photon contribution, or an NLO direct contribution, depending on
the virtualities of the exchanged quark and gluon; see the text for a detailed
discussion.

\vspace{3mm}
\item
The ratio of resolved photon and direct contributions to the inclusive jet
cross--section at HERA, for jet pseudorapidity between --1.0 and +2.0. We
have used the MRSD-' parametrization for the proton, and various
parametrizations for the photon as indicated. These are leading order
predictions, for momentum scale $\mu^2=p_T^2$. While most curves have been
computed using the simple antitagging requirement $Q^2<4$ GeV$^2$, the
dot--dashed curve has been calculated demanding $Q^2 < 0.01$ GeV$^2$ and
scaled initial photon energy $z$ between 0.20 and 0.75.

\vspace{3mm}
\item
Resolved photon (a) and direct (b) contributions to the inclusive jet
cross--section at HERA. Contributions from $qq, \ gq$ and $gg$ final states
are shown separately, where $q$ stands for a quark or antiquark of any
flavour. We have used the same antitagging and jet rapidity cuts as in
Fig.~8.

\vspace{3mm}
\item
The rapidity distribution of photoproduced jets with transverse momentum
exceeding 8 GeV at HERA. The dotted curves show the contribution from directly
interacting photons only, while the other curves show the total prediction
using various parametrizations of photonic parton densities. The requirements
for the scaled initial photon energy $z$ differ in (a) and (b); requiring
larger $z$ clearly increases the differences between the predictions.

\vspace{3mm}
\item
The triple differential di--jet cross--section at HERA, for the case of equal
jet pseudorapidities. The direct and three subclasses of resolved photon
contributions are shown separately. In (a) no cut on $z$ is applied, while
in (b) only events with $0.3 \leq z \leq 0.8$ are accepted; in particular,
the requirement $z \geq 0.3$ removes the direct contribution for
$\eta_1=\eta_2 > 0.2$.

\vspace{3mm}
\item
The average energy of the photonic remnant jet for resolved photon events at
HERA with $p_T=10$ GeV, as a function of the common pseudorapidity of the
high$-p_T$ jets. No cuts on the incident photon energy have been applied.
The contributions from $qq, \ qg$ and $gg$ final states are again shown
separately.

\vspace{3mm}
\item
The scale dependence of the inclusive jet cross section at HERA, for jets with
transverse energy $E_T = 25$ GeV and pseudorapidity = +1.5, as predicted
using the GRV parametrization for the photon and MRSD0 for the proton. The
dotted, dot--dashed and long dashed curves have been obtained by varying the
photonic factorization scale $M_{\gamma}$ only, while the short dashed and
solid curves are for the case where both factorization scales and the
renormalization scale are identical. Adapted from B\"odeker et al. \cite{72}.

\vspace{3mm}
\item
The transverse momentum spectrum of \ccbar\ and \bbbar\ pairs, for the case
where both heavy quarks are produced centrally (rapidity $y=0$). Note that
at high $p_T$ the single charm (or bottom) inclusive cross--section will
receive substantial ``excitation" contributions involving the heavy flavour
content of the photon, which are not included here. Even though we have used a
parametrization with a hard \gga, the resolved photon contribution is clearly
very small.

\vspace{3mm}
\item
The triple differential cross--section for the photoproduction of \ccbar\
pairs at HERA, for $p_T(c) = 10$ GeV, as a function of the rapidities
of the two heavy quarks, which are taken to be equal. These are LO results;
MRSD-' has been used for the nucleonic parton densities. For the WHIT1
parametrization contributions from $gg$ fusion and \qqbar\ annihilation are
shown separately, while only the sum is shown for the WHIT4 and LAC1
parametrizations. Note that for the range of rapidities shown here, the cut on
the scaled incident photon energy $z$ removes the direct contribution
completely (in leading order).

\vspace{3mm}
\item
The rapidity dependence of the direct photon cross--section for incident
photon energy $E_{\gamma}=10$ GeV and transverse momentum of the final photon
$p_T^{\gamma}=5$ GeV, as predicted from a leading order calculation
\cite{101} using the ABFOW parametrization \cite{abfow} for the proton and AGF
for the photon. Note that the rapidity $\eta_\gamma$ is measured in the \gamp\
centre--of--mass frame here. The contribution from the gluon in the photon
clearly dominates for $\eta_\gamma \geq 1$.

\vspace{3mm}
\item
The rapidity dependence of the direct photon cross--section at HERA for
$p_T^{\gamma}=5$ GeV, as predicted from an NLO calculation \cite{103} using
the GRV parametrization \cite{45} for the proton. Note that $\eta_\gamma$ is
measured in the HERA lab frame. No cut on the energy of the incident photon
has been applied; the contribution from the gluon in the photon is
sub--dominant everywhere. Some sensitivity to the quark densities in the
photon still remains, as demonstrated by the difference between the solid
(GRV) and dotted (GS) curves.

\vspace{3mm}
\item
A Feynman diagram contributing to the photoproduction of \jpsi\ mesons in the
colour singlet model. For the resolved photon contribution, the incident
photon has to be replaced by another gluon.

\vspace{3mm}
\item
The \jpsi\ production cross--section at HERA is shown as a function of the
elasticity parameter $Z$ of eq.(\ref{e3.6}), as predicted from a leading order
calculation \cite{123}. Acceptance cuts have been applied on the transverse
momentum of the \jpsi\ meson and on the angle of both leptons that originate
from \jpsi\ decay, but the leptonic branching ratio has not been included.

\vspace{3mm}
\item
Typical Feynman diagrams (left) and topologies (right) of the three classes
of contributions to the production of high$-p_T$ jets in (quasi--)real \gaga\
collisions: direct (a), single resolved (b), and double resolved (c). Note
that each resolved photon gives rise to a remnant jet.

\vspace{3mm}
\item
Leading order predictions for the inclusive jet cross--section at TRISTAN. The
lower dotted curve in (a) shows the direct contribution only, while the other
curves show the total prediction using various parametrizations of photonic
parton densities, as indicated. The cuts on the pseudorapidity of the jet and
on the maximal scattering angle of the outgoing $e^\pm$ differ, (a)
corresponding to the cuts used by TOPAZ and (b) to those used by AMY; note
that the antitagging cut is only effective if $z \leq 0.75$. Data by
these two experiments \cite{141,142} are also shown.

\vspace{3mm}
\item
The triple differential di--jet cross--section at TRISTAN, as predicted in
leading order. (a) shows the direct and 1--res contributions, while (b) shows
2--res contributions. For the WHIT1 parametrization the contributions from
different final states are shown separately, while only the sum is given for
the DG and WHIT6 parametrizations. The results are for TOPAZ antitagging
conditions.

\vspace{3mm}
\item
The triple differential di--jet cross--section at LEP1 and LEP2, as predicted
in LO using the WHIT1 parametrization. The steps at pseudorapidities
$\eta_1=\eta_2 \simeq 1.5$ (for $\rs=90$ GeV) or 2.2 (for $\rs=180$ GeV) occur
since the ALEPH antitagging condition has been used, which limits the
scattering angle of the outgoing $e^\pm$ only if it carries more than 50\% of
the beam energy.

\vspace{3mm}
\item
The transverse momentum distribution of jets produced in \gaga\ events at two
stages of the planned JLC collider, as predicted in LO using the WHIT1
parametrization. The dotted curves show the total result in the absence of
beamstrahlung, while the dashed curves show different contributions as
indicated, and the solid curves their sum, for realistic beamstrahlung
spectra. The comparison between the solid and dotted curves show that the
beamstrahlung spectrum is expected to be significantly harder at $\rs=1$ TeV
(b) than at 0.5 TeV (a).

\vspace{3mm}
\item
The triple differential di--jet cross--section at a \gaga\ collider, based on
a 500 GeV \epem\ collider with unpolarized beams. Due to the hardness of the
photon spectrum, there are substantial differences between predictions using
WHIT1 and LAC1 even at the high value of $p_T$ chosen here.

\vspace{3mm}
\item
Predictions for the total cross--section for \ccbar\ pair production in \gaga\
collisions at \epem\ colliders. The direct contribution (dotted) includes the
parametrized form (\ref{e4.5}) of NLO corrections, whereas the resolved photon
contribution (dashed and solid curves) has been computed in LO only. Two
curves of a given pattern show the uncertainty of the theoretical prediction,
as described in the text.

\vspace{3mm}
\item
The signal for a Standard Model Higgs boson of mass $m_H=120$ GeV decaying
into a \bbbar\ pair, as seen at a \gaga\ collider based on a 350 GeV \epem\
collider, as well as various backgrounds. 90\% electron beam polarization has
been assumed in order to suppress the direct background. Both \ccbar\ and
\bbbar\ pair production processes and $c$ and $b$ excitation processes are
included in the background calculation. The width of the Higgs signal, and of
the $Z$ peak, is determined by the resolution, not the intrinsic widths of
these particles. From Baillargeon et al. \cite{173}.

\vspace{3mm}
\item
The total cross--section for the production of \jpsi\ mesons in \gaga\
collisions, as predicted by the colour singlet model. The dotted curve shows a
LO prediction for scale $\mu^2=m^2_{J/\psi}$, while for the dashed and solid
curves NLO corrections to the 1--res $\gamma g$ fusion contribution have been
included using the parametrization of eqs.(\ref{e4.6}), (\ref{e4.7}), but the
NLO direct and $\gamma q$ scattering contributions have been omitted. Two
curves of a given pattern show the theoretical uncertainty of our estimate, as
described in the text.

\vspace{3mm}
\item
Leading order predictions for the production of a hard, central photon in
no--tag two--photon collisions (no antitag condition has been imposed).
Fragmentation contributions are not included. The long and short dashed curves
show 2--res and 1--res contributions as predicted from the WHIT1
parametrization; they add up to the solid curve. For all other
parametrizations, only the total sum has been shown.

\end{enumerate}

\end{document}